\documentclass[aps,twocolumn,superscriptaddress,showpacs]{revtex4}
\usepackage{apjfonts}
\usepackage{bm}
\usepackage{amsmath}
\usepackage{rotating}

\newcommand\lsim{\mathrel{\rlap{\lower4pt\hbox{\hskip1pt$\sim$}}
        \raise1pt\hbox{$<$}}}
\newcommand\gsim{\mathrel{\rlap{\lower4pt\hbox{\hskip1pt$\sim$}}
        \raise1pt\hbox{$>$}}}
\newcommand\propsim{\mathrel{\rlap{\lower4pt\hbox{\hskip1pt$\sim$}}
        \raise1pt\hbox{$\propto$}}}
\newcommand{\Si}{\mathrm{Si}}
\newcommand{\D}{\mathrm{d}}

\newcommand{\fast}{\mathrm{fast}}
\newcommand{\slow}{\mathrm{slow}}
\newcommand{\spin}{\mathrm{spin}}
\newcommand{\spinang}{SA}
\newcommand{\spinmag}{SM}
\newcommand{\dL}{d_{\rm L}}
\newcommand{\yr}{\,\mathrm{yr}}
\newcommand{\dy}{\,\mathrm{days}}
\newcommand{\hr}{\,\mathrm{hr}}
\newcommand{\AU}{\,\mathrm{AU}}
\newcommand{\mHz}{\,\mathrm{mHz}}
\newcommand{\Hz}{\,\mathrm{Hz}}
\newcommand{\Mchirp}{\mathcal{M}}
\newcommand{\Msun}{\mathrm{M}_{\odot}}
\newcommand{\ii}{i}

\newcommand{\hc}{h_{\mathrm{c}}}
\newcommand{\hmod}{h_{\mathrm{m}}}
\newcommand{\ti}{t_{\mathrm{i}}}
\newcommand{\tf}{t_{\mathrm{f}}}
\newcommand{\tc}{t_{\mathrm{c}}}

\begin{document}
\bibliographystyle{apsrev}
\title{Pre-Merger Localization of Gravitational-Wave Standard Sirens
With LISA:\\Harmonic Mode Decomposition}

\author{Bence Kocsis}
\affiliation{Institute of Physics, E\"otv\"os University, P\'azm\'any P. s. 1/A, 1117 Budapest, Hungary} \affiliation{Harvard-Smithsonian Center for Astrophysics, 60 Garden Street, Cambridge, MA 02138}

\author{Zolt\'an Haiman}
\affiliation{Department of Astronomy, Columbia University, 550 West 120th Street, New York, NY 10027}

\author{Kristen Menou} \affiliation{Department of Astronomy, Columbia University, 550 West 120th Street, New York, NY 10027}

\author{Zsolt Frei}
\affiliation{Institute of Physics, E\"otv\"os University, P\'azm\'any P. s. 1/A, 1117 Budapest, Hungary}

\begin{abstract}
The continuous improvement in localization errors (sky position and
distance) in real time as {LISA} observes the gradual inspiral of a
supermassive black hole (SMBH) binary can be of great help in
identifying any prompt electromagnetic counterpart associated with the
merger. We develop a new method, based on a Fourier decomposition of
the time-dependent, LISA-modulated gravitational-wave signal, to study
this intricate problem. The method is faster than standard Monte Carlo
simulations by orders of magnitude. By surveying the parameter space
of potential LISA sources, we find that counterparts to SMBH binary
mergers with total mass $M \sim 10^5$--$10^7 \Msun$ and redshifts $z
\lsim 3$ can be localized to within the field of view of astronomical
instruments ($\sim \deg^2$) typically hours to weeks prior to
coalescence. This will allow a triggered search for variable
electromagnetic counterparts as the merger proceeds, as well as
monitoring of the most energetic coalescence phase.  A rich set of
astrophysical and cosmological applications would emerge from the
identification of electromagnetic counterparts to these
gravitational-wave standard sirens.
\end{abstract}

\maketitle

\section{Introduction}\label{s:intro}

One of the key objectives of the planned, low-frequency
gravitational-wave (GW) detector LISA (Laser Interferometric Space Antenna) is the
detection of supermassive black hole (SMBH) binary mergers at
cosmological distances. The observation of these chirping GW sources
would deepen our understanding of (i) general relativity, e.g. by
offering unique tests of spacetime physics in the vicinity of SMBHs
\cite{dre04,mil05,hm05,bcw06,aru06}, (ii) cosmology, by providing
additional constraints on the luminosity distance--redshift relation
\cite{sch86,hug02,hh05}, (iii) large-scale structure, by indirectly
constraining hierarchical structure formation scenarios
\cite{bbr80,bh92,mhn01,bbw05}, and (iv) black hole astrophysics,
e.g. by allowing accurate determinations of Eddington ratios, and
other attributes of black hole accretion, in systems with SMBH mass
and spin known independently, from the GW measurements
\cite{mp05,koc06,dot06}.

From a purely astronomical point of view, one of the most attractive
features of the LISA mission design is the possibility to constrain
the 3-dimensional location (i.e. sky position and distance) of GW
inspiral sources to within a small enough volume that the
identification of potential electromagnetic (EM) counterparts to SMBH
merger events can be contemplated seriously. Indeed, the accuracy of
such LISA localizations at merger are encouraging, with an error
volume $\delta \Omega \times \delta z = 0.3\deg^2 \times 0.1$ for SMBH
masses $m_1=m_2=10^6\Msun$ at $z=1$, for instance \cite{vec04}. In
Ref.~\cite{koc06}, we have shown that this accuracy may be sufficient to
allow an unique identification of the bright quasar activity that may
be associated with any such SMBH merger.

Another possibility, examined here in detail, is to monitor the sky
for EM counterparts in real time, as the SMBH inspiral proceeds. This
is arguably one of the most efficient ways to identify reliably
(prompt) EM counterparts to SMBH merger events, since the exact nature
of such counterparts is a priori unknown. Using the GW
inspiral signal accumulated up to some look--back time, $\tf$, preceding the final
coalescence, one already has a partial knowledge of where the source
of GWs is located on the sky.  Since the sky position is deduced
primarily from the detector's motion around the Sun, one anticipates
that angular positioning uncertainties will not change too
dramatically during the last few days before merger, so that a
targeted EM observation of the final stages of inspiral may be a
feasible task. Here, we present an in-depth study of the potential for
such pre-merger localizations with LISA, while we discuss various
astrophysical concepts and observational strategies for EM counterpart
identifications in a companion work \cite{paper2}.

The main purpose of the present analysis is thus to determine the
accuracy of SMBH inspiral localizations with LISA, as a function of
look--back time, $\tf$, prior to merger.  The LISA detector is not
uniformly sensitive to sources with different sky positions and
angular momentum orientations.  Results will thus generally depend on
the fiducial values of these angles. Our first objective is to
calculate the time-dependence of distributions of localization errors,
for randomly oriented sources, over a large range of values for the
SMBH masses and source redshift. A second objective of our analysis is
to estimate source parameter dependencies for these distributions of
localization errors, i.e. how the 3-dimensional (sky position and
distance) localization error distributions depend on the fiducial sky
position of GW sources.  This is useful to understand which regions of
the sky may be best suited for the identification of EM counterparts
to SMBH merger events. To the best of our knowledge, this angle
dependence has not been explored in detail before, not even in terms
of final errors at ISCO (i.e. at $\tf=t_{\rm isco}$, when using the
complete inspiral data-stream, up to the innermost stable circular
orbit, or ISCO).

Parameter estimation uncertainties for LISA inspirals have been
considered previously, under a variety of approximations
\cite{cut98,mh02,hug02,bc04,vec04,bbw05,hh05,aru06,lh06}. These
studies differ in the levels of approximation adopted for the GW
waveform, using various orders of the post-Newtonian expansion. The
LISA signal output for these waveforms are obtained through a linear
combination of the two GW polarizations, $h_{+}(t)$ and
$h_{\times}(t)$, with the beam pattern functions, $F_{+}$ and
$F_{\times}$. The beam patterns define the detector sensitivity for
the two polarizations. They are determined by the angles describing
the instantaneous orientation of the LISA constellation relative to
the GW polarizations. As the LISA detector constellation orbits the
Sun, with a one year period, $F_{+}$ and $F_{\times}$ are slowly
changing in time and this introduces an additional time dependence in
the LISA signal. As first shown by Cutler~\cite{cut98}, the source sky
position can be determined with LISA using this modulation. In the
formalism given by Cutler~\cite{cut98}, this modulation couples time and
angular dependencies in a complicated way, making the estimation of
localization errors numerically costly for a large set of SMBH binary
random orientations and parameters.

Using a different approach, Cornish \& Rubbo \cite{cr03} have derived
the orbital modulation in a much simpler form, in which the angular
parameter dependence and the time dependence can be decoupled. Here,
starting directly from the original Cutler~\cite{cut98} expression, we
give an independent derivation of the Cornish \& Rubbo \cite{cr03}
formula and write it in an equivalent form, from which decoupling is
more evident. We do this by expanding the LISA response function into
a discrete Fourier sum of harmonics of the fundamental frequency of
LISA's orbital motion, $f_{\oplus}=1\yr^{-1}$. Since LISA's orbit does
not include high frequency features, we expect this sum to be quickly
convergent. In fact, it is clear from the Cornish \& Rubbo
\cite{cr03} result that the expansion terminates at $4f_{\oplus}$ and
that there are no higher order harmonics due to the detector's
motion. The series coefficients in the expansion are independent of
time and only depend on the relative angles at ISCO. We then develop a
Fisher matrix formalism in which parameter error distributions can be
mapped independently of time, while the time dependence can be
computed independently of the specific SMBH binary orbital elements. A
Monte Carlo simulation for random binary orientations then becomes a
simple linear combination, without any integral evaluations. This
greatly reduces the numerical cost of estimating parameter uncertainty
distributions, even at fixed observation time (e.g. to map
distributions of errors at ISCO). We use this numerical cost advantage
\begin{enumerate}

\item to map the distribution of localization errors for the full
three dimensional grid of SMBH total mass ($M=10^5$--$10^8\Msun$),
redshift ($z=0.1$--$7$) and arbitrary look--back time ($\tf$) before
merger,

\item to study how source localization error distributions vary
systematically with sky position, and

\item to discuss implications, in terms of advance warning times, for
prompt electromagnetic counterpart searches with large field-of-view
astronomical instruments.
\end{enumerate}

We call this new approach the harmonic mode decomposition (HMD).  The
method verifies that the amplitude modulation, which is restricted to
frequencies less than $4 f_{\oplus}=1.3\times 10^{-7}\Hz$, is indeed a
very slow modulation when compared to the GW frequency of LISA SMBH
inspirals ($0.03\mHz$--$1\Hz$). One plausibly expects that physical
parameters which determine the amplitude modulation (like the source
sky position and orbital inclination relative to the detector) can be
estimated independently of the parameters which determine the GW
frequency (like masses, orbital phase, time to ISCO). In the HMD
method, the two sets of parameters are naturally separated and can be
estimated independently. In particular, parameters related to the
modulation can essentially be determined on a background of GW-cycle
averaged signal. In the present work, we compute LISA inspiral
localization errors with the approximation that high frequency signal
parameters have strictly no cross-correlations with parameters related
to the slow orbital modulation.
In addition to the numerical advantages mentioned above, the HMD
formalism offers a clear interpretation of the time evolution of
uncertainties for the slow modulation parameters. This can be used to
gain a better understanding of the general evolutionary properties of
localization errors. The following questions, that we address in
detail in our work, are particularly relevant.
\begin{enumerate}
\item[(i)] Under what conditions do the localization uncertainties scale simply with the measured
signal--to--noise ratio, and how do these uncertainties evolve during the final stages of
inspiral?
\item[(ii)] To what extent do the high and low frequency signal parameters decouple?
\item[(iii)] What are the best determined combinations of the angular
parameters?
\item[(iv)] How and why does the shape of the 3D localization error
ellipsoid change during the final week(s) of observation?
\end{enumerate}

In our analysis, we neglect the ``Doppler phase'' due to LISA's orbital motion, SMBH spin precession
effects and any finite SMBH binary orbital eccentricities. These approximations
are advantageous for the resulting simplicity, but the use of the HMD
method is not restricted to these approximations. We also outline a
generalized HMD method which remains numerically much more efficient
than standard methods. We leave a numerical implementation of this
general HMD method to future work. It will be particularly interesting
to determine how our approximate results for the evolution of LISA
localization errors are modified when spin precession effects are
included, since spin precession effects were shown to improve the
final localization errors by factors of $3$--$5$ at ISCO
\cite{vec04,lh06}.

The remainder of this paper is organized as follows. In
\S~\ref{s:conventions} we define our conventions and the assumptions
made in our analysis.
In \S~\ref{s:HMD} we expand the LISA GW signal in Fourier modes and
obtain the conversion from actual physical parameters to corresponding
Fourier amplitudes.
In \S~\ref{s:HMDFisher}, we incorporate these results into a Fisher
matrix formalism and derive the expressions necessary to estimate
correlation errors for HMD signals.
In \S~\ref{s:advantages:computation_times}, we quantify the
computational advantages of the HMD method.
In \S~\ref{s:results}, we present results from Monte Carlo
computations of the time evolution of localization errors and discuss
results in terms of advance warning times for prompt electromagnetic
counterpart searches.
In \S~\ref{s:discussion}, we develop toy models to interpret the
time-dependence of LISA localization errors and to answer questions
(i)--(iv) above.
We summarize our results and conclude in \S~\ref{s:conclusions}.

\section{Assumptions and Conventions}\label{s:conventions}

This section is divided into three parts. First, we list the
definitions of physical quantities used in this paper, in particular
the variables describing a SMBH inspiral. Second, we give the
equations which determine the LISA inspiral signal. Third, we state
all the assumptions made in this work.

\subsection{Definitions}\label{s:definitions}

In general, an SMBH inspiral is described by a total of 17
parameters. These include 2 redshifted mass parameters, $(\Mchirp_z,
\eta_z)$, 6 parameters related to the BH spin vectors, ${\bm
p}_{\spin}$, the orbital eccentricity, $e$, the source luminosity
distance, $\dL$, 2 angles locating the source in the sky, $(\theta_N,
\phi_N)$, 2 angles that describe the relative orientation of the
binary orbit, $(\theta_{NL}, \phi_{NL})$, a reference time, $t_{\rm
merger}$, and a reference phase at ISCO, $\phi_{\rm ISCO}$, and the
orbital phase, $\phi_{\rm orb}$. Throughout this work, we restrict
ourselves to circular orbits by omitting the orbital eccentricity,
$e$, and instead of the orbital phase, $\phi_{\rm orb}$, we use the look--back time
before merger, $t$, as our evolutionary time parameter.
The LISA signal for a GW inspiral is determined by the
above set of parameters and two additional angular parameters
describing the orientation of LISA, $(\Xi,\Phi)$. We elaborate on the
definitions of our mass and angular parameters below.

\subsubsection{Mass Parameters}
For component masses $m_1$ and $m_2$, the total mass is $M=m_1+m_2$,
the reduced mass is $\mu=m_1m_2/M$, the symmetric mass ratio is
$\eta=\mu/M$ and the chirp mass is defined as $\Mchirp=M\eta^{3/5}$
\cite{Gravitation}.  Throughout this work, we use geometrical units:
${\rm G} \equiv {\rm c} \equiv 1$. In this case, the mass can be
expressed in units of time: $10^6 \Msun \equiv 4.95\sec$. The measured
GW waveforms are insensitive to the cosmological parameters, if they
are expressed in terms of the luminosity distance and the redshifted
mass parameters, e.g. $m_z=(1+z)m$ (same for redshifted chirp and
reduced masses).

\subsubsection{Time Parameters}\label{s:definitions:time}

We write a generic look--back time (or ``observation time'') before
merger as $t$, and a generic redshifted GW frequency (or ``observation
frequency'') as $f$\footnote{Note that, contrary to our convention for
redshifted mass parameters, we drop the $z$ index for $f$ and $t$
because we never consider comoving frequencies or times.}.
We use the leading order (i.e. Newtonian) approximation for the
frequency evolution. Therefore, the observed frequency at look--back
time $t$ before merger is (e.g. eq. 3.3 in ref.~\cite{pw95})
\begin{align}
f_0(\Mchirp_z,t)&= \frac{5^{3/8}}{8\pi}t^{-3/8}\Mchirp_z^{-5/8},\nonumber\\
      &= 2.7\times 10^{-4}\Hz\left( \frac{t}{\rm day}\right)^{-3/8}
      \eta_{0.25}^{-3/8}M_{6z}^{-5/8}\label{e:f0(t)},
\end{align}
or equivalently
\begin{align}
t_0(\Mchirp_z,f)&= 5 (8\pi f)^{-8/3}\Mchirp_z^{-5/3}\nonumber\\
      &=6.7{\,\rm min}\left(\frac{f}{f_c}\right)^{-8/3}
      \eta_{0.25}^{-1}M_{6z}^{-5/3}\label{e:t0(f)},
\end{align}
where $M_{6z}$ is the redshifted total mass in units of $4\times
10^6\Msun$, $\eta_{0.25}=\eta/0.25$ is the symmetric mass ratio
($\eta_{0.25}=1$ for equal component masses,
\S~\ref{s:conventions:assumptions}), $f_{c}= {\rm c}/{\rm
R}_{\oplus}=c/(1\AU)=2.00\mHz$ is the inverse light-travel time across
the radius of the LISA orbit, and the null index stands for the order
of approximation.  The inspiral phase extends until the innermost
stable circular orbit (ISCO), at $6 M$, is reached
\begin{eqnarray}
\label{e:fisco}
f\leq f_{\rm ISCO} &=& 6^{-3/2}\pi^{-1}M_z^{-1} = 1.1\mHz\times M_{6z}^{-1},\\
t\geq t_{\rm ISCO} &=& 5 (3/2)^4 \eta^{-1} M_z= 33{\,\rm min}\times
\eta_{0.25}^{-1} M_{6z}.
\end{eqnarray}
where $t_{\rm ISCO}$ is the (observer-frame) look--back time before
merger corresponding to the ISCO, and $f_{\rm ISCO}$ is the
(observer-frame) frequency at ISCO.

In the present work, we fix the start of the observation (i.e. when the
source first enters LISA's frequency band) at look--back
time $\ti$, and examine how the value of an end-of-observation time,
$\tf$, prior to merger affects the precision on source
localization. We restrict ourselves to pre-ISCO inspiral signals,
corresponding to $\tf\geq t_{\rm ISCO}$. Note that any instantaneous
look--back time $t$ associated with an observation lasting from look--back
times $\ti$ to $\tf$ must obey $t_{\rm ISCO}\leq \tf\leq t\leq \ti$ in
our notation.

\subsubsection{Angular Parameters}

LISA is an equilateral triangle-shaped interferometer with an
arm-length of $5\times 10^6\,$km, orbiting around the Sun. The
constellation trails $20^{\circ}$ behind the Earth and is tilted
$60^{\circ}$ relative to the ecliptic. The detector plane precesses
around the orbital axis with the same one-year period as the orbital
period \cite{Danz04}.

Following closely Refs.~\cite{cut98} and \cite{vec04}, including in
notation, we define two coordinate systems. The barycentric frame is
tied to the ecliptic, with ${\bf\hat x}, {\bf\hat y}$ lying in the
ecliptic plane and $\bf\hat z$ normal to it. The detector reference
frame tied to the detector, with ${\bf\hat z'}$ normal to the detector
plane, while ${\bf\hat x'}, {\bf\hat y'}$ are in the plane and
co-rotating with the detector so that the arms are described by
time-independent vectors. We refer to the barycentric frame with
normal coordinates and to the detector frame with primed
coordinates. The unit vectors defining the source location on the sky,
${\bf\hat N}$, and the SMBH binary orbital angular momentum, ${\bf\hat
L}$, are described by polar angles $(\theta_N,\phi_N)$ and
$(\theta_L,\phi_L)$ in the ecliptic frame, $(\theta'_N,\phi'_N)$ and
$(\theta'_L,\phi'_L)$ in the detector frame:
\begin{align}\label{e:angles_first} {\bf\hat N}(\theta_N,\phi_N) &= {\bf\hat z}\cos\theta_N + {\bf\hat
x}\sin\theta_N\cos\phi_N + {\bf\hat y}\sin\theta_N\sin\phi_N,\\
{\bf\hat L}(\theta_L,\phi_L) &= {\bf\hat z}\cos\theta_L + {\bf\hat x}\sin\theta_L\cos\phi_L + {\bf\hat
y}\sin\theta_L\sin\phi_L.
\end{align}
Since we assume no SMBH spins, orbital angular momentum is conserved
and the $(\theta_N,\phi_N,\theta_L,\phi_L)$ coordinates are
time-independent properties of the sources.

Let $(\Xi,\Phi)$ be the two angles specifying the orientation of the LISA system
in the ecliptic: $\Phi$ describes its orbital phase during its motion
around the Sun, while $\Xi$ describes the rotation of the triangle
around its geometrical center. If their values at merger are written
$\Xi_0$ and $\Phi_0$, then at an arbitrary look--back time $t$:
\begin{eqnarray}
\Xi(t) &=& \Xi_0 - \omega_{\oplus}t,\label{e:Xi(t)}\\
\Phi(t) &=& \Phi_0 - \omega_{\oplus}t,\label{e:Phi(t)}
\end{eqnarray}
where $\omega_{\oplus}\equiv 2\pi/\yr$ is the orbital angular velocity
around the Sun.

The time dependence of the detector normal vector $\bf \hat z'$ can be
expressed as
\begin{equation}
{\bf\hat z'} = \frac{1}{2}{\bf\hat z} - \frac{\sqrt{3}}{2}{\bf\hat x}\cos\Phi -\frac{\sqrt{3}}{2}{\bf\hat y}\sin\Phi.
\end{equation}
The detector angles are given by
\begin{align}
\cos \theta'_N &= \frac{1}{2}\cos \theta_N -
\frac{\sqrt{3}}{2}\sin\theta_N\cos(\Phi - \phi_N),\\\label{e:phi'_N}
\phi'_N &= \Xi + \tan^{-1}\left[ \frac{\frac{\sqrt{3}}{2}\cos
\theta_N + \frac{1}{2}\sin\theta_N\sin(\Phi-\phi_N)} {\sin\theta_N\sin(\Phi-\phi_N)} \right].
\end{align}
Let us also define $\psi'$, the {\it polarization angle} of the GW waveform, as
\cite{vec04}
\begin{equation}
\label{e:angles_last}
\tan \psi' = \frac{{\bf\hat L}\cdot{\bf\hat z'} - ({\bf\hat
L}\cdot{\bf\hat N})({\bf\hat z'}\cdot{\bf\hat N})}{{\bf\hat
N}\cdot({\bf\hat L}\times {\bf\hat z'})}.
\end{equation}

Note that there are only 6 independent angular parameters
$(\theta_N,\phi_N,\theta_L,\phi_L,\Xi,\Phi)$. Other detector specific
quantities like $\theta'_N$, $\phi'_N$, $\theta'_L$, $\phi'_L$, and
$\psi'$ can be expressed in terms of these 6 independent parameters using
eqs.~(\ref{e:angles_first}--\ref{e:angles_last}).

Let us introduce a new set of 6 independent angles,
\begin{equation}
\Omega=(\theta_N,\phi_N,\theta_{NL},\phi_{NL},\alpha,\gamma),
\end{equation}
with the following definitions:
\begin{itemize}
 \item $\theta_{NL}$ is the relative latitude of $\bf \hat L$ and $\bf \hat
 N$ (i.e. the inclination of the binary orbit to the line of sight),
 \item $\phi_{NL}$ is the relative longitude of $\bf \hat L$ and $\bf
 \hat N$,
 \item $\alpha \equiv \Xi-\Phi +\phi_{N} - \frac{3\pi}{4}$,
 \item $\gamma(t) \equiv \Phi(t)-\phi_{N}$.
\end{itemize} The explicit definitions are given in Appendix~\ref{app:angles}.

Let us refer to the angles at the reference time $t=t_{\rm merger}=0$ as
$\Omega(0)$. Although $\Phi\equiv\Phi(t)$ and $\Xi\equiv\Xi(t)$ are
time-dependent, as given by (\ref{e:Xi(t)},\ref{e:Phi(t)}) $\alpha$ is
a time-independent combination, unlike the time-dependent
$\gamma\equiv\gamma(t)$. The angles at $t=0$ are thus given by
$\Omega(0)=(\theta_{N},\phi_{N},\theta_{NL},\phi_{NL},\alpha,\gamma_{0})$.

These angles have the interesting property that they possess isotropic
{\it a priori} distributions, like the original $\Omega(0)$ variables,
but the measured GW waveforms expressed in terms of these new
variables are much simpler than when they are expressed in terms of the original set
eqs.~(\ref{e:angles_first}--\ref{e:angles_last}).

Two additional quantities which are useful to describe the sensitivity
of the detector in various directions are the antenna beam patterns
\cite{cut98}:
\begin{align}\label{e:Fx+} F_{\times,+}(\Omega) =& \frac{1 + \cos^2 \theta'_N}{2}\cos 2\phi'_N\cos 2\psi'_N \nonumber\\
&\;\pm \cos \theta'_N\sin 2\phi'_N\sin 2\psi'_N,
\end{align}
where the sign $\pm$ is defined to be positive for $F_{\times}$, and
negative for $F_{+}$. Equation~(\ref{e:Fx+}) and the transformation
rules eqs.~(\ref{e:angles_first}--\ref{e:angles_last}) define the time
evolution of the antenna beam patterns for a given set of final angles
$\Omega(0)$ as the LISA system orbits around the Sun. Note that the
LISA system is equivalent to two independent orthogonal-arm
interferometers which are rotated by $45^{\circ}$ relative to each
other \cite{cut98}. Both data-streams are given by the same equations (see eq.
[\ref{e:h_I,IIdef}] below), modulo a change of one of the angles for
the second detector: ${\phi'}^{\rm II}_N={\phi'}^{\rm I}_N-\pi/4$ (or
equivalently $\alpha^{\rm II}=\alpha^{\rm I}-\pi/4$ using our
time-independent angular variables). Thanks to this simple
relationship between the two data-streams, it is possible to carry out
all the calculations for the first data-stream, and later include the
second data-stream in the final expression by varying the fiducial
angle $\alpha$.

\subsubsection{Grouping the Parameters}
We group the most important parameters describing the inspiral as
follows:
\begin{eqnarray}
{\bm p}_{\slow}&\equiv &\{\dL,\Omega\},\\
{\bm p}_{\fast}&\equiv &\{\Mchirp_z,\mu_z,t_{\rm merger},\phi_{\rm ISCO}\},\\
{\bm p}_{\spin}&\equiv &\{2~\rm{spin~magnitudes},
4~\rm{spin~angles}\}.
\end{eqnarray}
This organization of parameters has fundamental importance in our
formalism. As we show in \S~\ref{s:waveform}, the parameters ${\bm
p}_{\fast}$ and ${\bm p}_{\spin}$ relate to the high frequency GW
signal, while the parameters ${\bm p}_{\slow}$ relate to the
distinctly slow orbital modulation.

\subsection{LISA Inspiral Signal Waveform}\label{s:waveform}

For a circular binary inspiral, the two polarizations of GW signal are
well approximated by the restricted post-Newtonian expressions
\begin{eqnarray}\label{e:h_+x}
h_{+}(t)&=& 2\frac{{\cal M}^{5/3}(\pi f)^{2/3}}{\dL}(1+\cos^2 \theta_{NL})\cos\phi_{\rm GW}(t),\\
h_{\times}(t)&=& -4\frac{{\cal M}^{5/3}(\pi f)^{2/3}}{\dL}\cos
\theta_{NL} \sin\phi_{\rm GW}(t).
\end{eqnarray}
The GW phase $\phi_{\rm GW}(t)\equiv \phi_{\rm GW}({\bm
p}_{\fast},{\bm p}_{\spin};t)$, which is twice the orbital phase ,
$\phi(t)=2\phi_{\rm orb}(t)$, can be expanded into the series
\begin{align}
\phi_{\rm GW}({\bm p}_{\fast},{\bm p}_{\spin};t) \approx &\phi_{\rm
ISCO} + \phi_{0}(\Mchirp_z;t) + \phi_{1}(\Mchirp_z,\mu_z;t)\nonumber\\
\label{e:PNexpansion} & + \phi_{2}(\Mchirp_z,\mu_z,{\bm
p}_{\spin};t) + \dots,
\end{align}
where $\phi_{0}(\Mchirp_z;t)$ is the leading order Newtonian solution
to the phase evolution, successive terms correspond to small general
relativistic corrections, $\phi_{\rm ISCO}$ is the reference phase at
ISCO and $\phi_{n}(t_{\rm ISCO})=0$ for all $n\geq 0$. The
instantaneous GW frequency is defined as the time derivative of the GW
phase (\ref{e:PNexpansion}), i.e. $f=f(t)\equiv \D \phi_{\rm GW}/\D
t$, which changes very slowly compared to the GW phase itself,
$\phi_{\rm GW}(t)$. In practice we use the Newtonian approximation
(\ref{e:f0(t)}), $f_0(t)=\D \phi_{0}/\D t$. Note that equation
(\ref{e:PNexpansion}) depends implicitly on the reference time,
$t_{\rm merger}$, since our time variable $t$ is the look--back time
before $t_{\rm merger}$ (see \S~\ref{s:definitions:time})

The signal measured by LISA is a linear
combination of the two polarizations (\ref{e:h_+x}), weighted by the
antenna beam patterns $F_{+}^{\rm I,II}$ and $F_{\times}^{\rm I,II}$
for each of the two equivalent interferometers,
defined by (\ref{e:Fx+}), resulting in the two observable data--streams
\begin{equation}\label{e:h_I,IIdef}
h^{\rm I,II}(t)= \frac{\sqrt{3}}{2}[F_{+}^{\rm I,II}h_{+}(t) +
F_{\times}^{\rm I,II}h_{\times}(t)],
\end{equation}
where the factor $\sqrt{3}/2=\sin(60^{\circ})$ comes from the opening angle
of the LISA arms. The beam patterns are determined by the relative
orientation of the source polarizations and the detector. Their
time-dependence is due to the following three main effects: LISA
changes its orientation as it orbits the Sun, LISA changes its
relative distance to the source as it orbits the Sun, and the orbital
plane of the SMBH binary can precess because of spin-orbit coupling
effects.  Substituting (\ref{e:h_+x}) in (\ref{e:h_I,IIdef}) and
expressing it in complex form, we get
\begin{equation}\label{e:h_I,II}
h^{\rm I,II}(t)=\frac{{\cal A}(\Mchirp_z,f)}{\dL}G^{\rm
I,II}(\Omega,f) e^{i\phi_{\rm GW}({\bm p}_{\fast},{\bm
p}_{\spin};t)},
\end{equation}
where ${\cal A}(\Mchirp_z,f)/\dL$ defines the overall amplitude
scale, with
\begin{equation}\label{e:A}
{\cal A}(\Mchirp_z,f)= 2\sqrt{3}(\pi f)^{2/3}{\cal
M}_z^{5/3}.
\end{equation}
The $G(\Omega,f)$ factor defines the angular dependence of the signal,
\begin{equation}\label{e:G}
G^{\rm I,II}(\Omega,f)=G_{\rm A}^{\rm I,II}(\Omega){e}^{\ii
\varphi_{\rm D}(\Omega,f)},
\end{equation}
where $G_{\rm A}(\Omega)$, the {\it amplitude modulation}, captures
the varying detector sensitivity with direction and polarizations of
the GWs,
\begin{equation}
\label{e:amplitude_mod} G_{\rm A}^{\rm I,II}(\Omega) =
\frac{1+\cos^2\theta_{NL}}{2}F_{+}^{\rm I,II}(\Omega) - \ii
\cos\theta_{NL} F_{\times}^{\rm I,II}(\Omega).
\end{equation}
The additional $\varphi_{\rm D}(\Omega,f)$ modulation is the {\it
Doppler phase modulation}, which is the difference between the phase
of the wavefront at the detector and at the barycenter \cite{cut98}:
\begin{equation}
\label{e:Doppler_phase} \varphi_{\rm D}(\Omega,f)=2\pi
\frac{f}{f_{c}} \sin\theta_N \cos\gamma.
\end{equation}
There is a non-negligible number of Doppler phase cycles only for a GW
frequency satisfying $f\geq f_{c}$ (see definition of $f_{c}$ below
eq.~[\ref{e:t0(f)}] above). However, equation~(\ref{e:fisco}) shows
that $f\leq f_{\rm ISCO} < f_c$, hence the $f_c$ frequency is reached
only after ISCO for typical SMBH component masses of
$m_1=m_2=10^6\Msun$ and redshift $z=1$. Even for smaller $10^5\Msun$
component masses, the total number of cycles, $N_{\rm pm}$, remains
$<1$ until the final $5\hr$ of inspiral.  Therefore the Doppler phase
(\ref{e:Doppler_phase}) is practically negligible for SMBH
inspirals. In fact, estimating localization errors without accounting
for the Doppler phase affects results by less than a factor of
$10^{-3}$ (for $m_1=m_2=10^6\Msun$ at $z=1$; S. A. Hughes, private
communication). Therefore, in eq.~(\ref{e:G}), we neglect
$\varphi_{\rm D}(\Omega,f)$ and restrict ourselves to the
approximation
\begin{equation}\label{e:G=G_A}
G^{\rm I,II}(\Omega,f) \equiv G_{A}^{\rm I,II}(\Omega).
\end{equation}
The explicit frequency-dependence dropped out, and the time evolution
of the signal $G_A$ is now fully determined by the time evolution of the angles $\Omega$.

Note that the amplitude modulation (\ref{e:amplitude_mod}), $G^{\rm
I,II}_{A}(\Omega)$, is traditionally expressed in complex polar
notation (e.g. \cite{cut98}), where the magnitude and argument of the
complex number are called {\it polarization amplitude} and {\it
phase}. As we will show, the mode decomposition is simplest in the
original Cartesian complex form (\ref{e:amplitude_mod}), which already
includes both the polarization amplitude and phase; thus, we do not
distinguish these two quantities in the following. The function
$G^{\rm I,II}(\Omega,f)$ given in (\ref{e:G}) also accounts for
spin-orbit precession if the orbital orientation $(\theta_{NL},\phi_{NL})$ in $\Omega$ is chosen to be time-dependent, to
satisfy the equations for spin-orbit precession, and if an extra
precession phase shift, $\exp(i\delta_{P}(\theta_{NL},\phi_{NL}))$, is
introduced (see eq. 2.14 in Lang \& Hughes \cite{lh06}) in addition to the Doppler
phase in (\ref{e:G}). In our calculations, we neglect spin precession
but discuss how the HMD method can be extended to include that effect
in \S~\ref{s:HMDFisher:general}.

Finally, we express the measured signal (\ref{e:h_I,II}) as
\begin{equation}
\label{e:schematic} h^{\rm I,II}({\bm p};t) =\hc({\bm
p}_{\fast},{\bm p}_{\spin};t) \times \hmod^{\rm I,II}({\bm
p}_{\slow};t),
\end{equation}
where $\hc$ is the {\it high frequency carrier signal} and $\hmod$
is the {\it slow modulation}:
\begin{eqnarray}
\label{e:PNrestricted} \hc({\bm p}_{\fast},{\bm
p}_{\spin};t)&=&{\cal A}(\Mchirp_z,f(t)) e^{\ii\phi_{\rm GW}({\bm
p}_{\fast},{\bm p}_{\spin};t)}
\\
\label{e:h1} \hmod^{\rm I,II}({\bm p}_{\slow};t)&=&\frac{G_A^{\rm
I,II}(\Omega(t))}{\dL}.
\end{eqnarray}

Equation (\ref{e:schematic}) shows that the two sets of parameters
${\bm p}_{\slow}$ and $\{{\bm p}_{\fast},{\bm p}_{\spin}\}$ are
exclusively determined by the low frequency modulation and the high
frequency carrier, respectively. For this reason, we only expect a low
level of cross-correlation between these sets of parameters:
parameters associated with very different timescale components should
essentially decouple. In Sec.~\ref{s:advantages:simple} and
Appendix~\ref{app:simple}, we consider several toy models which allow
us to understand the necessary conditions, and the extent to which,
parameters associated with high and low frequency components
decorrelate in the course of an extended, continuous observation.

\subsection{Simplifying Assumptions}\label{s:conventions:assumptions}

In the present work, we make the following assumptions:
\begin{enumerate}
\item We assume that the amplitude modulation can be used to determine
the luminosity distance and angular parameters, ${\bm
p}_{\slow}=\{\dL,\theta_N,\phi_N,\theta_{NL},\phi_{NL}\}$, while the
other parameters, ${\bm p}_{\fast}=\{\Mchirp_z,\mu_z,t_{\rm
merger},\phi_{\rm ISCO}\}$, are determined from the high frequency GW
phase. We assume no cross-correlations between these two sets of
parameters. This is supported by the results listed in Table~1 of
Hughes~\cite{hug02}, which shows the full covariance matrix of a Monte Carlo
realization of 2PN waveforms.
The correlation coefficients are $\sim 0.1$ for the above quantities,
and the absolute scale of the second set of parameters is very low in
the first place.  Berti~et~al.~\cite{bbw05,bbw05b} also report that the
sets ${\bm p}_{\fast}$ and ${\bm p}_{\slow}$ are relatively uncorrelated for
general relativity and even for alternative theories of gravity.  In
the latter case, the carrier $\hc(t)$ in the signal
(\ref{e:schematic}) is modified but not the slow modulation,
$\hmod(t)$, so that the general expectation of decoupling is
maintained.

\item We assume that there are no additional errors on the detector
orientations $\Phi(0)$ and $\Xi(0)$. These parameters are given by
$t_{\rm merger}$ via eq. (\ref{e:Phi(t)}) and (\ref{e:Xi(t)}), and
$t_{\rm merger}$ itself is determined by the high frequency carrier
signal to high precision. Using the full data-stream up to ISCO,
$\delta t_{\rm merger}\sim 2\,{\rm sec}$ \cite{hug02,aru06}.  Using
(\ref{e:Phi(t)}) and (\ref{e:Xi(t)}), we estimate $|\delta
\Phi(0)|=|\delta \Xi(0)|\equiv\omega_{\oplus}\delta t_{\rm merger}= 4
\times 10^{-7} \,{\rm rad} = 0.08''$. This is so small that we expect
the errors $\delta \Phi(0)$ and $\delta \Xi(0)$ to be negligible at
any relevant end-of-observation times $\tf>t_{\rm ISCO}$, even if the
$\tf$-dependence of these errors scale as steeply as $(S/N)^{-1}$ (see
also Appendix~\ref{app:simple}).

\item We use the circular, restricted post-Newtonian (PN)
approximation for the GW waveform, keeping only the leading order
(i.e. Newtonian) term in the signal amplitude. Higher order
corrections to the GW amplitude introduce additional structure to the
waveform. They improve the parameter estimation uncertainties for high
mass binaries \cite{bs07a,bs07b} and introduce additional correlations
between the parameters. It will be important to consider these
corrections to the amplitude in future investigations. Arbitrary PN
corrections to the GW phase only enter via $\hc$ in the signal given
by eq.~(\ref{e:schematic}).  Since we neglect correlations between
the sets of parameters ${\bm p}_{\slow}$ and $({\bm p}_{\fast},{\bm
p}_{\spin})$, all the restricted PN corrections to the phase drop out
and become irrelevant for the ${\bm p}_{\slow}$ parameter estimations.

\item We neglect the effects of Doppler phase modulation. This is
plausible for SMBH binaries with component masses $m>10^5\Msun$, since
the GW wavelength in this case is generally greater than LISA's
orbital diameter and $N_{\rm pm}<1$ (see eqs.~[\ref{e:Doppler_phase}]
and [\ref{e:f0(t)}]).

\item We neglect SMBH spins and, in particular, neglect the
spin--orbit precession for angular determinations. This assumption is
useful in simplifying our equations and in focusing on the behavior of
pure angular modulation. Future studies can incorporate spin--orbit
precession by convolving the angular modulation decomposition with
spin--orbit effects.

\item We fix the start of LISA observations at a look--back time
$\ti\equiv \min\{t_0(f_{\min}),1\yr\}$ prior to merger.  This
corresponds to the time when the GW inspiral frequency $f$ crosses the
low frequency noise wall of the detector at $f_{\min}=0.03\mHz$, but
we limit the initial look--back time to a maximum of $1\yr$ before
merger. Note that LISA's effective mission lifetime is estimated to be
$3\yr$. Integrated observation times longer (but also shorter) than
our assumed $\ti$ values are possible in principle, depending on
source specifics.  In a more elaborate treatment, one could define
$\ti$ as an a priori random variable. We fix $\ti$ here mostly for
simplicity and focus on the effects of varying the values of $\tf$
$(<\ti)$. In \S~\ref{s:advantages:simple} we show that localization
errors are primarily determined by the end-of-observation time, $\tf$,
and that values of $\ti>1\yr$ do not significantly change the
evolution or final localization error estimates. If, however, $\ti\ll
1\yr$ (that is, if $t_{\rm merger}$ is within a few months of the
beginning of observation), then localization errors can become
significantly worse than in our results with $\ti=1\yr$.

\item We neglect finite arm-length effects and we do not make use
of the three independent observables of the time delay interferometry
\cite{pri02}. This is a valid assumption for SMBH inspirals since
here $f\ll c/L=0.01\Hz$.

\item We neglect any finite orbital eccentricities. We note that, for
eccentric orbits, higher order harmonics appear in the GW phase. In
principle, since these harmonics affect the high frequency GW phase,
but not the slow modulation, including finite eccentricities should
not significantly affect localization error estimates. For rather
eccentric orbits, high-order harmonics with $f\gg f_c$ can have a
non-negligible Doppler phase (\ref{e:t0(f)}), which would lead to an
improvement in the determination of $\theta_N$ and $\phi_N$. Although
eccentricity is efficiently damped by gravitational radiation reaction
\cite{Peters64}, the presence of gaseous circumbinary disks could lead
to non-zero eccentricities for at least some LISA inspiral events
\cite{ArmNat05,PapNelMas01}.

\item We follow Barack~\&~Cutler~\cite{bc04} in calculating the LISA root spectral
noise density, $S_{n}(f)$, which includes the instrumental noise as
well as galactic/extra-galactic backgrounds. For the instrumental
noise \cite{bbw05}, we use the effective non-angularly averaged
online LISA Sensitivity Curve Generator\footnote{{\tt
www.srl.caltech.edu/$\sim$shane/sensitivity/}}, while we use the
isotropic formulae for the galactic and extra-galactic background
\cite{bc04}.

\item Our analysis focuses on statistical errors and does not account
for possible systematic errors. For example, waveform templates might
be inaccurate either due to the imprecision of the theory if the true
waveform is not the one predicted by general relativity, or due to
practical limitations from necessarily finite realizations of the
large template space. Such inaccuracies can introduce new systematic
errors.
\end{enumerate}

\section{Harmonic Mode Decomposition}\label{s:HMD}

In our formalism, the angular information of the LISA inspiral signal
is contained exclusively in the periodic modulation due to the
detector motion around the Sun, which adds an amplitude modulation to
the high frequency waveform. This modulation has a fundamental
frequency, $f_{\oplus}=1/\yr$, along with upper harmonics
$jf_{\oplus}$, where $j$ is an integer. Although it is intuitively
clear that the high frequency harmonics will tend to have a vanishing
contribution, it is hard to establish this just by looking at
eqs.~(\ref{e:angles_first}--\ref{e:angles_last}), which define the
time evolution. In this section we show that it is possible to derive
surprisingly simple analytical expressions for the amplitude of each
harmonic. We provide an outline of the derivation starting from the
commonly used Cutler~\cite{cut98} formulae
(\ref{e:angles_first}--\ref{e:angles_last}) and alternatively from
those in Cornish \& Rubbo \cite{cr03}. We show that the derivation is much simpler in
the latter case, in the sense that the Cornish \& Rubbo \cite{cr03}
expression is almost already in the desired form.

\subsection{Derivation using Cutler~\cite{cut98}}\label{s:HMD:Cutler}

We expand the modulating signal (\ref{e:amplitude_mod},\ref{e:h1}) in
a Fourier series
\begin{equation}
\label{e:Gexpansion} \hmod({\bm p}_{\slow}(0);t) =
\frac{G_{A}(\Omega(t))}{\dL(z)} = \sum_{j=-\infty}^{\infty} g_j({\bm
p}_{\slow}(0)) {e}^{{\ii} j \omega_{\oplus} t},
\end{equation}
where $g_j({\bm p}_{\slow}(0))$ are the mode amplitude coefficients
and ${\bm p}_{\slow}(0)$ are the distance and angle variables at $t=0$
(see \S~\ref{s:definitions}). The coefficients can be obtained as
\begin{equation}
\label{e:g_jdef} g_j({\bm p}_{\slow}(0)) =
\frac{1}{2\pi\dL}\int_0^{1 \yr} \D t \,G_{\rm A}(\Omega(t))
{e}^{-{\ii} j \omega_{\oplus} t}.
\end{equation}
Substituting the definition of $G_{\rm A}(\Omega(t))$ from
eq.~(\ref{e:amplitude_mod}), using the time evolution of $\Omega(t)$,
eqs.~(\ref{e:angles_first}--\ref{e:angles_last}), integral
(\ref{e:g_jdef}) can be evaluated.

Although conceptually simple, a direct analytical evaluation of
integral (\ref{e:g_jdef}) is overly cumbersome. Thus, for practical
reasons, we follow an alternative path. We start with the original
Cutler~\cite{cut98} formulae, given by eqs. (\ref{e:Fx+}) and
(\ref{e:amplitude_mod}). First, using general trigonometric
identities, we can express $\cos 2x$ and $\sin 2x$ with $\tan(x)$ for
$x=\phi'_{N}$ and $x=\psi'$. In the second step, we express and
substitute for $\tan\phi'_{N}$ and $\tan\psi'_{N}$ with ecliptic
variables using (\ref{e:phi'_N}) and (\ref{e:angles_last}). In the
third step, we express the trigonometric functions in complex form.
After this step, each term in the beam pattern (\ref{e:Fx+}) is of the
form
\begin{equation}
\frac{\sum_n a_n e^{i n \gamma}}{\sum_m b_m e^{i m \gamma}},
\end{equation}
where the sums over $n$ and $m$ integers are finite, containing only a
few terms, and $a_n$ and $b_n$ depend only on the angles
$(\theta_N,\phi_{NL},\alpha)$. In the fourth step we simplify the
product of fractions. It turns out that, after combining terms, the
denominators drop out exactly, leaving a formula just like
(\ref{e:Gexpansion}), except that the largest element in the sum is
$|j|=4$. In the fifth step, we substitute in (\ref{e:amplitude_mod}),
and finally, change back from complex to trigonometric notation for
the coefficients, using the half-angles $\theta_N/2$ and
$\theta_{NL}/2$. Finally we arrive to the remarkably simple form:
\begin{equation}\label{e:hmod}
\hmod({\bm p}_{\slow}) = \dL(z)^{-1}\sum_{j=0}^{4}(L N_{j} D_{j} +
L^{*}N^{*}_{j}D_{j} + L^{*}N_{j}D^{*}_{j} + LN^{*}_{j}D^{*}_{j}),
\end{equation}
where the functions $L$, $N$, and $D$ depend only on the angular
momentum angles, sky position angles, and detector angles,
respectively:
\begin{eqnarray}
L(\theta_{NL},\phi_{NL}) &\equiv& \sin^4\left(\frac{\theta_{NL}}{2}\right)e^{-2i\phi_{NL}},\\
N_{j}(\theta_{N}) &\equiv& w_{j}\cos^{j}\left(\frac{\theta_N}{2}\right)\sin^{4-j}\left(\frac{\theta_N}{2}\right),\\
D_{j}(\alpha,\gamma) &\equiv& i e^{2i\alpha}e^{ij\gamma},
\end{eqnarray}
where $w_{j}=1/16\times(9,12\sqrt{3},18,4\sqrt{3},1)$ for
$j=(0,1,2,3,4)$, respectively, and we have defined asterisks to refer
to the following conjugates:
\begin{eqnarray}
L^{*}(\theta_{NL},\phi_{NL}) &\equiv& L(\pi - \theta_{NL},-\phi_{NL}),
\\ N^{*}_{j}(\theta_{N}) &\equiv& (-1)^{j} N_{j}(\pi-\theta_{N}),\\
D^{*}_{j}(\alpha,\gamma) &\equiv& - D_{j}(-\alpha,-\gamma)\equiv
\overline{D_{j}(\alpha,\gamma)}.
\end{eqnarray}
Note that using these conjugate functions, only the non--negative terms $0\leq j\leq
4$ remain in the sum (\ref{e:hmod}).

Substituting the time dependence implicit in $\gamma\equiv \gamma(0) +
\omega_{\oplus}t$, equation (\ref{e:hmod}) becomes
\begin{equation}\label{e:hmodres}
\hmod({\bm p}_{\slow}(0),t)=\dL(z)^{-1}\sum_{j=-4}^{4} g_j({\bm
p}_{\slow}(0)) {e}^{{\ii} j \omega_{\oplus} t},
\end{equation}
where the coefficients are
\begin{equation}\label{e:g_jres}
g_j({\bm p}_{\slow}(0)) = \left\{\begin{array}{l}
LN_jD_{|j|}(0) + L^{*}N^{*}_jD_j(0) \text{~~~if~} j\geq0 \\
L^{*}N_{|j|}D^{*}_{|j|}(0) + LN^{*}_{|j|}D^{*}_{|j|}(0) \text{~~~if~} j\leq0
\end{array}\right.
\end{equation}
and the detector functions $D_j(0)$ and $D^{*}_j(0)$ refer to their
values at $t=0$, $\gamma(0)$. (Note that $L,N_{j},L^{*}, N^{*}_{j}$
are all time-independent.) Since the decomposition
(\ref{e:Gexpansion}) is unique, the coefficients (\ref{e:g_jres}) that
we read off from our result also satisfy eq.~(\ref{e:g_jdef}).

\subsection{Derivation using Cornish \& Rubbo \cite{cr03}}\label{s:HMD:CornishRubbo}

Our result in (\ref{e:hmod}) can also be derived from the Cornish \&
Rubbo \cite{cr03} formulae for the LISA response function. In their
paper, these authors use a different set of angles, which relate to
ours as follows: $\theta^{CR}=\theta_{N}$, $\phi^{CR}=\phi_{N}$,
$\psi^{CR} \equiv \phi_{NL}-(\pi/2)$, $\lambda^{CR}\equiv -\alpha +
\phi_{N} \equiv \Phi - \Xi - (3\pi/4)$, and $\alpha^{CR}\equiv\gamma +
\phi_{N}\equiv\Phi$. Note that our set of angles is very similar to
theirs, except that we measure the detector angles relative to the
source, $\phi_N$. This is advantageous given the rotational symmetry
around the Earth orbital axis, making angles relative to the source
the only ones that should have an effect on the measured GWs; we
expect $\phi_N$ to drop out of the equations when using $\alpha$ and
$\gamma$. Note, once again, that the variables
$(\theta_{N},\phi_{N},\theta_{NL},\phi_{NL}, \alpha)$ are time
independent, while $\gamma\equiv\gamma(t)$. Writing the Cornish \&
Rubbo \cite{cr03} beam patterns for low frequencies, which is fully
equivalent to Cutler \cite{cut98}, with our angular
variables\footnote{Cornish \& Rubbo \cite{cr03} combine the
$\sqrt{3}/2$ factor with the beam patterns $F^{CR, \rm
I,II}_{+}=\frac{\sqrt{3}}{2} F^{\rm I,II}$, but we follow the original
definition, where $\sqrt{3}/2$ appears only when taking the linear
combination of GW polarizations (\ref{e:h_I,IIdef}).}, we get
\begin{eqnarray}
F^{\rm I,II}_{+} &=& -\frac{1}{2}[\cos(2\phi_{NL}) D_{+}^{\rm I,II}(t) - \sin(2\phi_{NL}) D_{\times}^{\rm I,II}(t)],\\
F^{\rm I,II}_{\times} &=& -\frac{1}{2}[\sin(2\phi_{NL}) D_{+}^{\rm I,II}(t) + \cos(2\phi_{NL}) D_{\times}^{\rm I,II}(t)],
\end{eqnarray}
where
\begin{eqnarray}
D_{+} = &\frac{1}{32}&\{ -36\sin^2 \theta_N \sin(2\gamma + 2\alpha) + (3+\cos 2\theta_N)\nonumber\\
&\times& \{\cos(2\phi_{N}) [ 9\sin (-2\alpha + 2\phi_N)-\sin(4\gamma +2\alpha + 2\phi_N) ]\nonumber\\
&+& \sin(2\phi_{N})[ \cos(4\gamma + 2\alpha - 2\phi_N) -9\cos(-2\alpha + 2\phi_N)  ]\}\nonumber\\
&-& 4\sqrt{3}\sin(2\theta_N) [ \sin(3\gamma + 2\alpha) - 3\sin(\gamma+2\alpha)]\},
\end{eqnarray}
and
\begin{eqnarray}
D_{\times} = &\frac{1}{8\sqrt{3}}&\{ \sqrt{3} \cos\theta_{N} [9\cos(-2\alpha)-\cos(4\gamma +4\alpha)] \nonumber\\
&-& 6\sin\theta_N [\cos(3\gamma +2\alpha) + 3\cos(\gamma + 2\alpha )]\}.
\end{eqnarray}

One notices instantly that the time dependence here is much simpler
than in the original Cutler~\cite{cut98} formula, as it is inscribed only in
the various harmonics of $\gamma$. We can identify the highest
harmonic present to be $4\gamma$. Expanding the trigonometric
functions using standard identities, we obtain
\begin{equation}\label{e:D_+}
D_{+} =
-\frac{1}{32}\left(\begin{array}{c}
9(3+\cos2\theta_N) \\
-12\sqrt{3}\sin 2\theta_N \\
36\sin^2\theta_N \\
4\sqrt{3}\sin 2\theta_N  \\
3+\cos2\theta_N
\end{array}\right)\cdot
\left(\begin{array}{c}
\sin(2\alpha) \\
\sin(2\alpha + \gamma) \\
\sin(2\alpha + 2\gamma) \\
\sin(2\alpha + 3\gamma) \\
\sin(2\alpha + 4\gamma)
\end{array}\right),
\end{equation}
and
\begin{equation}\label{e:D_x}
D_{\times} =
-\frac{1}{8}\left(\begin{array}{c}
-9\cos\theta_N \\
6\sqrt{3}\sin \theta_N \\
0 \\
2\sqrt{3}\sin \theta_N \\
\cos\theta_N
\end{array}\right)\cdot
\left(\begin{array}{c}
\cos(2\alpha) \\
\cos(2\alpha + \gamma) \\
\cos(2\alpha + 2\gamma) \\
\cos(2\alpha + 3\gamma) \\
\cos(2\alpha + 4\gamma)
\end{array}\right),
\end{equation}
where ${\bm a}\cdot{\bm b} = \sum_n a_n b_n$ is the usual dot
product. Now, the second sets of elements carry the time dependence
and the detector orientation information, while the first sets
describe the sky position. Note that the explicit $\phi_N$ dependence
dropped out, as expected. Next, we manipulate equations
(\ref{e:D_+},\ref{e:D_x}), substituting complex expressions for the
trigonometric ones, and substituting these into
eq.~(\ref{e:amplitude_mod}). We finally arrive at eq.~(\ref{e:hmod}) after
changing to half--angles $\theta_N/2$ and $\theta_{NL}/2$.

We note that eqs.~(\ref{e:hmod}) or (\ref{e:hmodres},\ref{e:g_jres})
are fully general expressions, equivalent to the standard
LISA inspiral signal in eqns.~(\ref{e:Fx+}) and (\ref{e:amplitude_mod}). The
two data-streams are obtained by substituting $\alpha=\alpha^{I}$ and
$\alpha^{II}$, corresponding to the two independent LISA-equivalent
Michelson interferometers (see \S~\ref{s:definitions}). To verify our
final result, we compare numerically the signals computed using
eqs.~(\ref{e:Fx+},\ref{e:amplitude_mod}) with the signals computed using
eqs.~(\ref{e:hmodres},\ref{e:g_jres}), for a large set of random choices of
angles. Agreement is achieved at machine precision levels.

The main utility of eq.~(\ref{e:hmod}), is that it can be used to
``deconstruct'' parameter error histograms, i.e. to understand how the errors
depend on the fiducial values of the parameters.
As compared to Cornish \& Rubbo \cite{cr03}, our result leads to an
explicit decoupling of the signal angular dependence into simple
products of one-dimensional functions. In particular, the dependence
on sky position, angular momentum, and detector angles are separated.
Using the special conjugate functions $L^{*}$, $D^{*}$, $N^{*}$,
eq.~(\ref{e:hmod}) displays the symmetry properties of the signal.
Finally, one angular variable, $\phi_N$ is eliminated exactly.

\section{Estimating Parameter Uncertainties in the HMD formalism}
\label{s:HMDFisher}

Parameter estimations for LISA GW inspiral signals are possible with
matched filtering and the expected uncertainties can be
forecast using the Fisher matrix formalism \cite{fin92,cf94}. In
this section, we apply this approach to the LISA signal derived
in \S~\ref{s:HMD}, with an angular dependence of the signal decomposed
into harmonic modes. In \S~\ref{s:HMDFisher:approximate}, we consider
the simple case of a high frequency carrier signal that is modulated
by a low-frequency function, without any cross-correlation between the
two sets of relevant parameters. We derive a simple formula for the
estimation of modulating parameter uncertainties. In
\S~\ref{s:HMDFisher:general}, we consider a more general
post-Newtonian signal and show that parameters related to source
localization can still be decoupled from the time evolution and the
other source parameters.

\subsection{Fisher Matrix Formalism}

Let us consider a generic real signal described by the function
$h(x)$, which depends on $N$ parameters $\{p_a\}_{a\in[1,N]}$. The
measured signal is $y(x)=h(x)+n(x)$, where $n(x)$ is a realization of
the noise specified by a probability distribution. Let us assume that
the noise is Gaussian, is statistically stationary with respect to
$x$, has zero mean value, $\langle n(x)\rangle = 0$, where $\langle
\rangle$ represents an ensemble average, and has known variance,
$\sigma(x)^2=\langle n(x)^2\rangle$. The parameter estimation errors
for $p_a$ can then be calculated using the Cramer-Rao bound~\cite{fin92}
\begin{equation}
\label{e:papb}
\langle\delta p_a\,\delta p_b\rangle \geq \langle \Gamma^{-1}
\rangle_{a\,b},
\end{equation}
where equality is approached for high $S/N$ signals. Here
$\Gamma_{a\,b}$ is the Fisher matrix defined by
\begin{equation}
\label{e:Gamma0}
\Gamma_{a\,b} = \int_{x_{\min}}^{x_{\max}} \frac{
\partial_a h(x)\;
\partial_b h(x)}{\sigma^2(x)}\D x,
\end{equation}
where $ \partial_a$ is the partial derivative with respect to the
parameter $p_a$.  Note that $\sigma(x)$ here is defined as the noise
per unit $x$.  In eq.~(\ref{e:Gamma0}), $x$ denotes time $t$ for
time-domain samples, or $f$ for frequency-domain samples. The purpose
of this work is to study how an arbitrary end of the observation, at
$x_{\max}$ (or $\tf$ below, for time samples) affects the resultant
correlation errors $\langle\delta p_a\,\delta p_b\rangle$, for a fixed
start-of-observation at $x_{\min}$ (or $\ti$ below, for time samples).

An important quantity for the evolution of parameter estimation errors
is the signal-to-noise ratio, $S / N$, defined by
\begin{equation}
\label{e:snr}
\left(\frac{S}{N}\right)^{2} = \int_{x_{\min}}^{x_{\max}}
\frac{h^2(x)}{\sigma^2(x)}\D x.
\end{equation}

For LISA, the noise varies with signal frequency. In this case, the
Fisher matrix can be evaluated in Fourier space \cite{fin92,cf94},
\begin{equation}
\label{e:GammaLISAf}
\Gamma(\tf)_{a\,b} = \Re\left\{4\int_{f_{\min}}^{f(\tf)}
\frac{\overline{
\partial_a\tilde h(f)}\;
\partial_b\tilde h(f)}{S_n^2(f)}\D f\right\},
\end{equation}
where $\tilde h(f)$ is the Fourier transform of $h(t)$, the GW signal
(\ref{e:schematic}), $ \partial_a$ is the partial derivative with
respect to parameter $p_a$, bars denote complex conjugation, and
$S_n^2(f)$ is the one-sided spectral noise density
(\S~\ref{s:definitions}).

\subsection{Approximate solution}\label{s:HMDFisher:approximate}

We seek an alternative equivalent form of eq.~(\ref{e:GammaLISAf})
specific to GW inspirals for which, as in eq.~(\ref{e:schematic}), the
high frequency carrier signal is decoupled from the slow
modulation. In case of SMBH inspirals, with a high frequency signal
$\hc(t)$ changing its frequency slowly as $f_0(t)$ given in eq.~(\ref{e:f0(t)}), and
further modulated by a slowly varying function $\hmod(t)$ as given in
eq.~(\ref{e:schematic}), the integral in eq.~(\ref{e:GammaLISAf}) can be evaluated in
the stationary phase approximation, by substituting
\begin{equation}
\label{e:h(f)} \tilde h(f) = \hmod[t_0(f)]\times\tilde \hc(f),
\end{equation}
where $\tilde \hc(f)$ is the Fourier transform of the carrier signal
and $\hmod[t_0(f)]$ is the modulating function evaluated at the time
when the carrier frequency is $f$. This can be converted to the
time domain, by simply changing the integration variable to $t=t_0(f)$ using
the frequency evolution in eq.~(\ref{e:t0(f)}):
\begin{equation}
\label{e:GammaLISAt1}
\Gamma(\tf)_{a\,b} = \Re\left\{4\int_{\tf}^{\ti}
\frac{\overline{
\partial_a\tilde h(t)}\;
\partial_b\tilde h(t)}{S_n^2[f_0(t)]}
\left|\frac{\D f_0(t)}{\D t}\right|\D t\right\},
\end{equation}
and
\begin{equation}
\label{e:h(t)} \tilde h(t) = \hmod(t)\times\tilde \hc[f_0(t)].
\end{equation}

We are only interested in estimating uncertainties for the ${\bm
p}_{\slow}$ variables (\S~\ref{s:definitions}), which are determined
exclusively by $\hmod(t)$. Recall from eq.~(\ref{e:PNrestricted}) that
$|\hc(t)|={\cal A}$ so that, for the Fourier transform\footnote{The
reason for the factor 4 is that the mean squared of $\cos(x)$ or
$\sin(x)$ is $1/2$ in (\ref{e:h_+x}), and since we use one-sided
signals in frequency domain ($f>0$), responsible for another factor of
$1/2$ in comparison.}, we have $|\hc(t)|^2=4|\tilde \hc(f)|^2(\D f/\D
t)$. Using these relationships, let us define the instantaneous
relative noise amplitude per unit time $\sigma(t)$ as
\begin{align}\label{e:sigma}
\sigma^{-2}(t) &= 4\frac{{\tilde \hc}^2[f_0(t)]}{S_n^2[f_0(t)]}
\left|\frac{\D f_0(t)}{\D t}\right| = \frac{{\cal A}^2[f_0(t)]}{S_n^2[f_0(t)]}\\
&=
\frac{3\sqrt{5}}{4}\frac{\Mchirp_z^{5/2}t^{-1/2}}{S_n^2[f_0(\Mchirp_z,t)]}.\nonumber
\end{align}
The last equality follows from the Newtonian waveform and frequency
evolution, eqs.~(\ref{e:A}) and (\ref{e:f0(t)}). We point out that the
mass dependence is captured entirely by $\sigma(t)$ and does not
appear anywhere else in what follows.

By combining eqs.~(\ref{e:GammaLISAt1}), (\ref{e:h(t)}), and
(\ref{e:sigma}), we arrive at
\begin{equation}
\label{e:GammaLISAt2}
\Gamma(\tf)_{a\,b} = \int_{\tf}^{\ti} \frac{\Re[\overline{
\partial_a \hmod(t)}\;
\partial_b \hmod(t)]}{\sigma^2(t)}\D t.
\end{equation}
Equation~(\ref{e:GammaLISAt2}) is the special case of
(\ref{e:GammaLISAf}), where the carrier signal-to-noise ratio and
modulation, $\hmod$, are conveniently isolated.

We are now ready to make use of the harmonic mode decomposition. Substituting eq.~(\ref{e:Gexpansion}) into (\ref{e:GammaLISAt2})
gives
\begin{equation}
\label{e:GammaLISAt3} \Gamma(\tf)_{a\,b} =
\Re\left\{\sum_{j_1,j_2=-4}^{4} \overline{
\partial_a g_{j_1}}\;
\partial_b g_{j_2}P_{j_2-j_1}(\tf)\right\},
\end{equation}
where
\begin{equation}
\label{e:Pj(t)} P_{j_2-j_1}(\tf)=\int\limits_{\tf}^{\ti}
\frac{{e}^{\ii (j_2-j_1)\omega_{\oplus}t}}{\sigma^2(t)} \D t.
\end{equation}
\begin{figure}
\includegraphics[width=8.5cm]{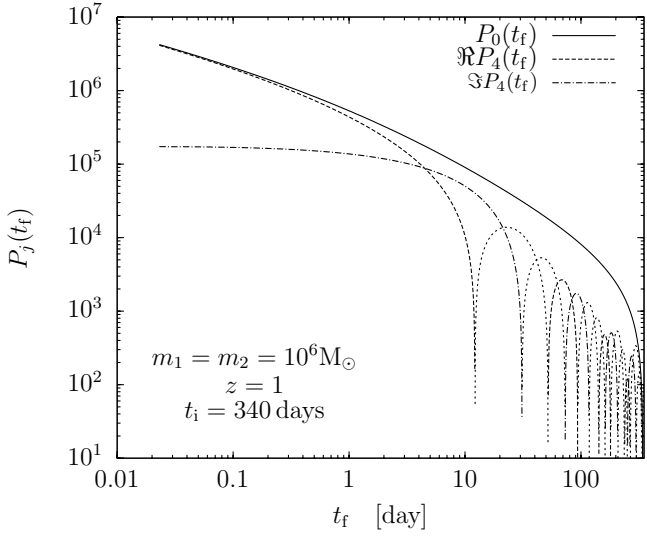}
\caption{\label{f:Pj(t)} The time evolution of the fundamental
functions $P_j(\tf)$, used to construct the Fisher matrix to forecast
localization errors by LISA.  The dependence of the Fisher matrix on
the look--back time $\tf$ is obtained from the 9 fundamental functions
with $0 \leq j \leq 8$. The curves show $P_0(\tf)$, as well as the
real and imaginary parts of $P_4(\tf)$, for $m_1=m_2=10^6\Msun$ and
$z=1$.  Thin dotted lines represent negative values. Note that
$P_0(\tf)\equiv (S/N)^{2}$, which is the simple scaling of inverse
squared errors, neglecting correlations. The signal-to-noise ratio
scales steeply, approximately as $S/N\propto \tf^{-1}$. The curve
$P_{4}(\tf)$ illustrates how all the other similar $P_{j}(\tf)$
functions vary, with a relative number $|j|$ of oscillations, and
$P_{-j}(\tf)\equiv \overline{P_{j}(\tf)}$.  }
\end{figure}

The function $P_j(\tf)$ is shown in Figure~\ref{f:Pj(t)} for $j=0$,
together with real and imaginary parts for the $j=4$ case, for
$m_1=m_2=10^6\Msun$ at $z=1$. Since the accumulated signal-to-noise
ratio is $S/N = P_0(t)$, the figure shows that the instantaneous
signal-to-noise ratio is $[\D/\D t](S/N) = [\D/\D t] P_0(t)\approx
t^{-2}$. The extrapolated signal-to-noise blows up at ``merger''
($t=0$). Data analysis for such a non-stationary signal-to-noise ratio
evolution has several interesting implications, which we study further
with toy models in Appendix~\ref{app:simple}. We find that, for such a
signal-to-noise ratio evolution, specific combinations of parameters
can always be measured to very high accuracy.

The time dependence in eq.~(\ref{e:Pj(t)}) couples only to the combination
$j=j_2-j_1$. This allows us to rearrange the double sum on $(j_1,j_2)$
and evaluate one of them independent of time:
\begin{equation}
\label{e:GammaLISAt4} \Gamma({\bm p}_{\slow},\tf)_{a\,b} =
\Re\left\{\sum_{j=-8}^{8} [{\cal F}_j({\bm p}_{\slow}(0))]_{a\,b}
P_j(\tf)\right\},\\
\end{equation}
where
\begin{equation}
\label{e:Fj}
 [{\cal F}_j({\bm p}_{\slow}(0))]_{a\,b} = \sum_{j'=-8}^{8}\overline{
\partial_a g_{j+j'}({\bm p}_{\slow}(0))}\; \partial_b g_{j}({\bm
p}_{\slow}(0)).
\end{equation}

Our parameters in the correlation matrix are
$p_a=(\dL,\theta_N,\phi_N,\theta_{NL},\phi_{NL})$ since we assume that
the other parameters, i.e. $\{\Mchirp_z,\eta_z,\phi_{\rm ISCO},t_{\rm
merger},\alpha,\gamma(0)\}$, are known from the high frequency carrier
signal (\S~\ref{s:conventions:assumptions}). It is straightforward to
compute the derivatives of $g_j(\dL,\Omega)$ using eq.~(\ref{e:g_jres})
for all parameters $p_a$, except $\phi_{N}$.  The $\phi_{N}$
dependence in $g_j$ in eq.~(\ref{e:g_jres}) is implicit in $\alpha\equiv
\alpha(\Xi(0),\Phi(0),\phi_N)$ and $\gamma(0)\equiv
\gamma(\Phi(0),\phi_N)$ (see \S~\ref{s:definitions}). Since we assume
that $\Xi(0)$ and $\Phi(0)$ are measured to very high precision with
the high frequency carrier (\S~\ref{s:conventions:assumptions}), we
can use the chain rule to express the $\phi_N$ derivative as
$\partial_{\phi_{N}} g_j = \partial_{\alpha} g_j -
\partial_{\gamma_{0}} g_j$.

Up to this point we did not make use of the fact that the LISA signal
is equivalent to two orthogonal arm interferometers rotated by $\Delta
\alpha=\pi/4$ with respect to each other. To account for both
data-streams being measured simultaneously, the Fisher matrix is
written as the sum of the two Fisher matrices corresponding to each
individual interferometer.  Writing out only the $\alpha$ dependence,
we have $\Gamma^{\rm
tot}_{a\,b}(\alpha)=\Gamma_{a\,b}(\alpha)+\Gamma_{a\,b}(\alpha-\pi/4)$.
Finally, according to eq.~(\ref{e:papb}), the parameter error covariance
matrix is the inverse of this total Fisher matrix:
\begin{align}
\langle\delta p_a\,\delta p_b\rangle \geq&
[\Gamma_{a\,b}(\dL,\theta_N,\theta_{NL},\phi_{NL},\alpha,\gamma(0))\nonumber\\
&\;+\Gamma_{a\,b}(\dL,\theta_N,\theta_{NL},\phi_{NL},\alpha-\pi/4,\gamma(0))]^{-1}.\label{e:covarianceresult}
\end{align}
Equation (\ref{e:covarianceresult}) along with (\ref{e:GammaLISAt4})
is our final expression, describing the time evolution of parameter
estimation uncertainties. We note that after combining both
data-streams, the matrices $[{\cal F}_j({\bm p}_{\slow}(0))]_{a\,b}$
for $4<|j|\leq 8$ modes vanish exactly for all ${\bm p}_{\slow}(0)$.

Let us emphasize the most important features of
eq.~(\ref{e:GammaLISAt4}):
\begin{itemize}
 \item The parameter dependence is separated from the time
 dependence. The Fisher matrix, $\Gamma_{a\,b}$, is written as a
 linear combination of matrices ${\cal F}_j({\bm p}_{\slow})$ weighted
 by the scalars $P_j(t)$, where ${\cal F}_j({\bm p}_{\slow})$ is
 independent of time and $P_j(t)$ is independent of the parameters
 $p_a$. Evaluating ${\cal F}_j({\bm p}_{\slow})$ requires the
 computation of the parameter derivatives $\partial_a g_{j}$.

 \item The evaluation of all integrals $P_j({\tf}_n)$ for different
 $n=1,2,\dots,N_{\tf}$ can be done in the same amount of time as
 needed for a single integration, since the $\tf$ dependence enters
 only in the integration bound in eq.~(\ref{e:Pj(t)}),

 \item Large Monte Carlo (MC) simulations can easily be performed
 since the time evolution is given by a small number of functions,
 $P_j(t)$, which can be calculated a priori and pre-saved. No
 integrations at all are necessary during the MC simulation for
 calculating distributions of correlation matrices.
\end{itemize}
In \S~\ref{s:advantages:computation_times} below, we
estimate the improvement in the computation time provided by the HMD method
for calculating distributions of parameter errors and their
time evolution.

\subsection{Generalization to the exact PN signal}\label{s:HMDFisher:general}

So far, we only considered the simplest case, assuming no
cross-correlation between parameters ${\bm p}_{\slow}$ and $({\bm
p}_{\fast}, {\bm p}_{\spin})$, for a restricted post-Newtonian
waveform. Moreover, we assumed the Doppler-phase (\ref{e:G}) to be
negligible. Including cross-correlations and the Doppler phase would
allow us to examine the range of validity of our approximations, and
it would allow us to extend computations to the lower component mass
regime, $m<10^5\Msun$, where the Doppler phase becomes
important. Furthermore, including spin precession effects can modify
angular localization errors by a factor of $ \sim 3$, at least for
the final errors at $\tf\approx t_{\rm ISCO}$
\cite{vec04,lh06}.  While we continue to use our initial set of
approximations in later sections, we outline here how the HMD
formalism could be used to decouple the time--dependence from the
angular parameter--dependence, even in the case of the most
general (arbitrary order) restricted post--Newtonian waveform. Source
sky position angles $(\theta_N,\phi_N)$, detector angles at ISCO
$(\alpha,\gamma(0))$ and luminosity distance ($\dL$) can all be
decoupled even if spin-orbit and spin-spin precessions are included in
the waveform, which is potentially a great advantage over the
traditional calculation methods (see
\S~\ref{s:advantages:computation_times} for a detailed discussion).

Consider a general restricted post-Newtonian signal given by
eq.~(\ref{e:h_I,II}), for which we substitute the harmonic mode expansion
\footnote{Note that this expression includes both the polarization
amplitude and the polarization phase, as both of these terms are
accounted for by the complex harmonic mode coefficients $g^{\rm
I,II}_j({\bm p}_{\slow})$.},
\begin{equation}\label{e:h_I,IIgeneral}
\tilde h^{\rm I, II}(f) = {\cal \tilde A}\sum_{j=-4}^{4} g^{\rm
I,II}_j f^{-7/6}e^{i[j\omega_{\oplus}t(f)+\varphi_{\rm D} +
\phi_{\rm GW}]},
\end{equation}
where ${\cal \tilde A}\equiv{\cal \tilde A}({\bm p}_{\fast})$ is the
amplitude in the frequency-domain (eq. 68 in ref.~\cite{vec04}),
$g_j\equiv g_j({\bm p}_{\slow})$ is the modulation amplitude in
eq.~(\ref{e:g_jres}), $\varphi_{\rm D}\equiv\varphi_{\rm D}({\bm
p}_{\slow},f) = c_{\rm D}({\bm p}_{\slow})\,f$ is the Doppler phase
(see eq.~\ref{e:Doppler_phase} above), $c_{\rm D}({\bm
p}_{\slow})=2\pi f_{c}^{-1}\sin \theta_N\cos \gamma$), $\phi_{\rm
GW}\equiv\phi_{\rm GW}({\bm p};f)$ is the GW phase
(\ref{e:PNexpansion}) (e.g. eq. 3.2 in ref.~\cite{pw95}),
\begin{equation}\label{e:phi_GW}
\phi_{\rm GW}({\bm p};f) = \sum_{n=0}^{N} c^{\rm GW}_{n}({\bm
p}_{\fast},{\bm p}_{\spin}) u^{\rm GW}_n(f),
\end{equation}
and time-frequency relationships $t(f)\equiv t({\bm p};f)$ can be
written as (e.g. eq. 3.3 in ref.~\cite{pw95})
\begin{equation}\label{e:t(f)}
t({\bm p};f) = \sum_{n=0}^{N} c^{\rm T}_n({\bm p}_{\fast},{\bm
p}_{\spin}) u^{\rm T}_n(f).
\end{equation}

Here, the $c^{\rm GW}_n$ and $c^{\rm T}_n$ coefficients are
frequency-independent, while $u^{\rm GW}_n(f)$ and $u^{\rm T}_n(f)$
are parameter-independent. They correspond to the various
post-Newtonian terms in the post-Newtonian expansion, and $N$
corresponds to the highest order term. The frequency functions are
very simple powers of $f$, i.e. $u^{\rm GW}_n(f)=f^{n-(5/3)}$ and
$u^{\rm T}_n(f)=f^{n-(8/3)}$. (Note that neither the $c^{\rm GW}_n$
and $c^{\rm T}_n$ coefficients nor the $u^{\rm GW}_n(f)$ and $u^{\rm
T}_n(f)$ functions are complex. Every term in
eq.~(\ref{e:h_I,IIgeneral}) is real except for the $g_j\equiv g_j({\bm
p}_{\slow})$ coefficients and the complex argument.)

Equations (\ref{e:h_I,IIgeneral}-\ref{e:t(f)}) can be combined into
\begin{equation}
\tilde h(f)^{\rm I,II} = \sum_{j=-4}^{4} A^{\rm I,II}_j({\bm p})
f^{-7/6}e^{i \Psi_{j}({\bm p};f)},
\end{equation}
where
\begin{equation}\label{e:A_j}
A^{\rm I,II}_j({\bm p}) = {\cal \tilde A}({\bm p}_{\fast}) g^{\rm
I,II}_j({\bm p}_{\slow}),
\end{equation}
and
\begin{eqnarray}\label{e:Psi}
\Psi_{j}({\bm p};f) &=& c_{\rm D} f
+ j\omega_{\oplus}\sum_{n=0}^{N} c^{\rm T}_n u^{\rm T}_n(f)
+ \sum_{n=0}^{N} c^{\rm GW}_n u^{\rm GW}_n(f)\\
&\equiv& \sum_{k=0}^{2N+1} c_{j k} u_k(f),
\end{eqnarray}
where in the last step we introduced $u\equiv\{f,u^{\rm T},u^{\rm
GW}\}$ and $c_j\equiv\{c_{\rm D},j\omega_{\oplus}c^{\rm T},c^{\rm
GW}\}$ to collect all $f$-functions and coefficients in one vector
and one matrix, respectively.

To compute the Fisher matrix, we need to obtain the partial
derivatives of the signal with respect to the parameters:
\begin{equation}\label{e:dh}
\tilde h_{,a}(f) = \sum_{j=-4}^{4} \left(A_{j,a} + i
\sum_{k=0}^{2N+1}A_jc_{j k,a} u_k(f) \right)
f^{-7/6}e^{i\Psi_{j}(f)},
\end{equation}
where commas in indices denote partial derivatives with respect to the
parameter following the index. Note, that the parameter index $a$ spans
all parameters ${\bm p}_{\slow}$, ${\bm p}_{\fast}$, and
${\bm p}_{\spin}$, and the Fisher matrix accounts for correlations
between these parameters.

Equation (\ref{e:dh}) can now be substituted in the Fisher matrix
in eq.~(\ref{e:GammaLISAf}). We get
\begin{eqnarray}\label{e:Fishergeneral}
\Gamma(\tf)_{a\,b} &=&\Gamma^{(0)}_{a\,b}(\tf) + \Gamma^{(1)}_{a\,b}(\tf) +\Gamma^{(2)}_{a\,b}(\tf),
\end{eqnarray}
where
\begin{align}\label{e:Fishergeneral0}
\Gamma^{(0)}_{a\,b}(\tf) =&\Re\left\{
\sum_{j_1,j_2=-4}^{4} A_{{j_1},a}\overline{A_{{j_2},b}}P^{(0)}_{j_1\,j_2}(\tf)
\right\},\\
\Gamma^{(1)}_{a\,b}(\tf) =&\Re\left\{\-i\sum_{j_1,j_2=-4}^{4} A_{{j_1},a}\overline{A_{{j_2}}}\sum_{k=0}^{2N+1}c_{{j_2}k,b}
P^{(1)}_{j_1\,j_2\,k}(\tf)\right.\nonumber\\
&\left.-i\sum_{j_1,j_2=-4}^{4} A_{{j_1}}\overline{A_{{j_2},b}}\sum_{k=0}^{2N+1}c_{{j_1}k,a}
P^{(1)}_{j_1\,j_2\,k}(\tf)\right\},\label{e:Fishergeneral1}\\
\Gamma^{(2)}_{a\,b}(\tf)=&
\Re\left\{-\sum_{j_1,j_2=-4}^{4} A_{{j_1}}\overline{A_{{j_2}}}\sum_{k_1,k_2=0}^{2N+1}c_{{j_2}k_1,a} c_{{j_2}k_2,b}
P^{(2)}_{j_1\,j_2\,k_1\,k_2}(\tf)
\right\},\label{e:Fishergeneral2}
\end{align}
and where
\begin{eqnarray}
P^{(0)}_{j_1\, j_2}(\tf)&=&4\int_{f_{\min}}^{f(\tf)}\frac{f^{-7/3}e^{i(j_1-j_2)\omega_{\oplus}t(f)}}{S_n^2(f)}\D f,\nonumber\\
P^{(1)}_{j_1\, j_2\, k}(\tf)&=&4\int_{f_{\min}}^{f(\tf)}\frac{f^{-7/3}u_{k}(f)e^{i(j_1-j_2)\omega_{\oplus}t(f)}}{S_n^2(f)}\D f,\label{e:Pgeneral}\\
P^{(2)}_{j_1\, j_2\, k_1\, k_2}(\tf)&=&4\int_{f_{\min}}^{f(\tf)}\frac{f^{-7/3}u_{k_1}(f)u_{k_2}(f)e^{i(j_1-j_2)\omega_{\oplus}t(f)}}{S_n^2(f)}\D f,
\nonumber
\end{eqnarray}
are the frequency dependent terms.

Equation (\ref{e:Fishergeneral}) is our final result, where the
localization parameters (i.e. angles and distance) are decoupled from
all other parameters (i.e. masses, spins, reference time and phase at
ISCO). The equation explicitly shows that, contrary to the traditional
methods usually adopted for Monte Carlo computations of random binary
orientations and sky positions
\cite{cut98,mh02,hug02,bc04,vec04,bbw05,hh05,aru06,lh06}, the
localization of a LISA inspiral event and its time--dependence can be
explored without the need to evaluate integrals for each realization
of the fiducial angles. Note that the only approximation made to
obtain eq. (\ref{e:Fishergeneral}) was to neglect of spin-orbit and
spin-spin precession in the general restricted post-Newtonian solution
for the Fisher matrix. The time--dependence is given by the $P(\tf)$
functions in eq.~(\ref{e:Pgeneral}) and the extrinsic parameter--dependence is
given by the coefficients, $A$. The $P(\tf)$ functions
in eq.~(\ref{e:Pgeneral}) can be computed a priori, independently of the
fiducial angles. Note that $P(\tf)$ depends implicitly on the
parameters $({\bm p}_{\fast},{\bm p}_S)$ through $t(f)$ in
eq. (\ref{e:t(f)}), and its inverse $f(t)$.  Generally, there are at
most $(2J_{\max}+1)\times[1+(2N+1)+(1/2)(2N+1)(2N+2)]$ such
independent functions.

From the general case, we can now deduce the special solution
in eqs.~(\ref{e:GammaLISAt4}) and (\ref{e:covarianceresult}) valid for a
Newtonian evolution, no Doppler phase, and no cross-correlations
between ${\bm p}_{\slow}$ and $({\bm p}_{\fast},{\bm p}_{\spin})$.
This approximation simply corresponds to the first term
$\Gamma^{(0)}_{ab}(\tf)$ in eq.~(\ref{e:Fishergeneral}), where in
eq.~(\ref{e:Pgeneral}) the $t(f)$ function is computed using the Newtonian
formula $t_0(f)$ given by eq.~(\ref{e:t0(f)}). Note also, that the next
term in eq.~(\ref{e:Fishergeneral}), $\Gamma^{(1)}_{ab}(\tf)$, corresponds
to the cross-correlation of the amplitude modulation with the ``high
frequency carrier signal'' (i.e. Doppler phase and GW phase). The last
term, $\Gamma^{(2)}_{ab}(\tf)$, corresponds to the cross-correlations
among parameters in the high frequency carrier.

Finally, we briefly consider extensions to include spin-orbit and
spin-spin precessions in the signal. Let us refer to the angular
momentum angles as ${\bm p}_{\rm
L}(t)\equiv(\theta_{NL}(t),\phi_{NL}(t))$, which are now
time-dependent. As we briefly show next, in the case of spin
precession, the $P^{0,1,2}(\tf)$ time-dependent integrals loose the
convenient property of being independent of ${\bm p}_{\rm L}$, but
nevertheless, the parameters describing the sky position and detector
orientation are still time-frequency independent and they are
decoupled in this prescription. Indeed, an extra precession phase
$\exp[i\delta_{\rm P}({\bm p}_{\rm L},p_{\spin},t)]$ has to be
included in eq.~(\ref{e:h_I,IIgeneral}) and $g^{\rm I,II}_j({\bm
p}_{\slow})$. We now have $\Psi_j^{\rm prec}({\bm p};f) = \Psi_j({\bm
p};f)+\delta_{\rm P}({\bm p}_{\rm L},p_{\spin};t)$ instead of
eq.~(\ref{e:Psi}). Thus, when taking the derivatives of the signal in
eq.~(\ref{e:dh}), we will get additional terms proportional to $A^{\rm
I,II}_j(\bm p)\partial_a \delta_{{\rm P},a}$ and the original
$A_{j,a}$ terms will have time variation due to the ${\bm p}_{\rm L}$
dependence of $g^{\rm I,II}_j$. Finally, after these modifications,
the Fisher matrix will be similar to that in eq.~(\ref{e:Fishergeneral}), except now
the ${\bm p}_{\rm L}$ terms cannot be moved outside of the
frequency-integral but have to be attached to the time--varying
$P^{0,1,2}(\tf)$ part \footnote{Note that if ${\bm p}_{\rm L}$ was
intricately coupled to the other angular parameters in $A^{\rm
I,II}_j$ then it would be impossible to detach the ${\bm p}_{\rm L}$
part from $A^{\rm I,II}_j$ and attach it to
$P^{(0),(1),(2)}(\tf)$. Fortunately, these terms were originally
included exclusively in the coefficients $A_j^{\rm I,II}(\bm p)$ in
eq.~(\ref{e:A_j}) and in a very simple way: the ${\bm p}_{\rm L}$ terms
are found only in the $L$ and $L^{*}$ coefficients in $g^{\rm
I,II}_j({\bm p}_{\slow})$ (see
eqs.~[\ref{e:hmodres},\ref{e:g_jres},\ref{e:A_j}]). The precession
phase terms, $\partial_a \delta_{\rm P}({\bm p}_{\rm L},p_{\spin},t)$,
can also be simply attached to $P^{(0),(1),(2)}(\tf)$. Due to the
precession phase, the index $k$ now spans the range $0\leq
k\leq2N+2$.}. The main advantage we retain is therefore that the
source position and detector angles $(\theta_N, \phi_N, \alpha,
\gamma(0))$ will still only be included in the coefficients $A^{\rm
I,II}_j(\bm p)$ and $c_{jk,a}$ and the time--evolution can be still be
computed independently of these parameters.
In~\S~\ref{s:advantages:computation_times:spinprec} we show that this
indeed reduces computation times by a large factor relative to the
traditional methods (e.g. \cite{vec04,lh06}).

We leave numerical implementations and explorations of parameter
distributions and their time--dependence, in this case of a general
inspiral waveform, to future work.

\section{Computation time}\label{s:advantages:computation_times}

One of the great advantages of introducing the HMD method is the
reduction in the computational time needed to evaluate error
distributions for the parameters which determine how efficiently LISA
can localize SMBH binary inspiral events: sky position, angular
momentum orientation and final detector orientation. In general, this
is a computationally very demanding task because of the large
dimensionality of the angular parameter space. Mapping the structure
of the distribution of correlations in the parameter space of mock
LISA measurements requires vast Monte Carlo simulations, which are
presently limited by computational resources. Currently, only a small
portion of this space has been explored
\cite{mh02,hug02,vec04,bc04,bbw05,hh05,lh06}. In this context, it is
desirable to tune methods to the specific problem at hand.  The HMD
method described above is specifically constructed to exploit the
structure of LISA inspiral signals.

\subsection{Approximate solution}\label{s:advantages:computation_times:approximate}

The computational time for parameter space exploration, using the HMD
method with the approximations described in
\S~\ref{s:HMDFisher:approximate}, is significantly reduced for the
following reasons. The standard approach for estimating parameter
errors requires the evaluation of an $N_{p}\times N_{p}$ symmetric
Fisher matrix, where each matrix element is an integral over the range
spanned by the GW frequency during inspiral (see Refs.~\cite{fin92,cf94};
and \S~\ref{s:HMDFisher}). Here, $N_{p}$ is the number of parameters
describing the signal. The number of independent elements in a
symmetric matrix is $(1/2)N_{p}(N_{p}+1)$. Let us assume that the
evaluation of a single integral requires to compute the waveform at
$N_{\rm int}$ separate instances. The computation of one integral is
sufficient also to trace the time evolution at $N_{\rm int}$ different
$\tf$ values, if one uses a single trapezoidal integral in frequency
from $f_{\rm ISCO}$ to $f_{\min}$ and stores results at each
intermediate value of $f$. Since the time evolution of the frequency
is known independently of the angles, we can already get the
integral for $N_{\rm int}$ different $\tf$ values. For randomly chosen
fiducial angles in a MC simulation of size $N_{MC}$ requires the
calculation to be repeated $N_{MC}$ times. To evaluate the evolution
of parameter errors as a function of $\tf$ for $N_{\tf}$ different
instances requires $N_{\tf}$ computations.  Therefore, the standard
method costs $\Delta T_{\rm standard}=(1/2)N_{p}(N_{p}+1) N_{\rm int}
N_{MC}$ computational time units.

In contrast, with our proposed HMD method
(\S~\ref{s:HMDFisher:approximate}), the ${\bm p}_{\slow}$ parameters
are decoupled from the ${\bm p}_{\fast}$ parameters
(\S~\ref{s:conventions:assumptions}) and from time. The $N_{p}\times
N_{p}$ Fisher matrix can be split into two smaller matrices, with
$N_{p1}\times N_{p1}$ and $N_{p0}\times N_{p0}$ components, where
$N_{p} = N_{p1} + N_{p0}$. The $N_{p1}\times N_{p1}$ matrix determines
the angular errors (which are deduced from the amplitude modulation of
the signal), while the other matrix determines the remaining
parameters (e.g. masses, phase and time at ISCO, using the high
frequency carrier only). Since we are only interested here in
parameters relating to the localization of the source, ${\bm
p}_{\slow}$, it is sufficient for us to consider the corresponding
$N_{p1}\times N_{p1}$ matrix only. Since it is symmetric, it has only
$(1/2)N_{p1}(N_{p1}+1)$ independent elements, but it turns out that
the computation of only $N_{p1}$ elements is sufficient for a single
harmonic (we need only the $N_{p1}$ derivatives of the $g_{j}$
functions, see eq.~[\ref{e:GammaLISAt3}]).  Using $2N_{J}+1$ harmonic
modes and a MC simulation with $N_{MC}$ random choices of fiducial
parameters costs $N_{p1}(2N_{J}+1)N_{MC}$ time units. In this method,
the time dependence is decoupled, so that parameter dependencies can
be taken outside of the integral (see eq.~[\ref{e:GammaLISAt3}] and
${\cal F}_{j}({\bm p}_{\slow})$ in eq.~[\ref{e:GammaLISAt4}]). The MC
sampling can therefore be evaluated independently of time. The time
evolution of the signal for each harmonic is known independently of
the angles, by construction ($P_{j}(\tf)$, eq.~[\ref{e:Pj(t)}]), and
this integration for each component can be evaluated {\it a priori},
independently of the fiducial parameter values. For each such mode, we
would like to evaluate a number $N_{\tf}$ of integrals. Fortunately,
since the different integrals differ only via the lower integration
bound in the time domain, all integrals can be obtained during a
computation of the integral with the largest time domain, $\tf=t_{\rm
ISCO}$. Therefore for a total of $(2N_{J}+1)$ modes, building the
time-evolution functions $P_{j}(\tf)$ takes of order
$(2N_{J}+1)N_{\tf}$ time units. This is generally much faster than
building the time-independent coefficients ${\cal F}_{j}({\bm
p}_{\slow})$. In summary, with the HMD method, one only needs $\Delta
T_{\rm HMD}=N_{p1}(2N_{J}+1)N_{MC} + (2N_{J}+1)N_{\tf}$ units of time.

Comparing methods, we find that the computational requirements of the
HMD method is lower by a factor of
\begin{equation}\label{e:comptime}
\frac{\Delta T^{\rm no~spin}_{\rm HMD}}{\Delta T^{\rm no~spin}_{\rm
standard}} = \frac{N_{p1}(2N_{J}+1)N_{MC} +
(2N_{J}+1)N_{\tf}}{(1/2)N_{p}(N_{p}+1) N_{\rm int} N_{MC}}.
\end{equation}
Recall from \S~\ref{s:definitions} that the number of parameters for a
no-spin case is $N_{p0}=4$, $N_{p1}=5$, so that
$N_{p}=N_{p0}+N_{p1}=9$. Choosing $N_{\rm int}=10^3$, $N_{MC}=10^4$,
$N_{\tf}=100$, and $N_{J}=4$ for the other parameters in
eq.~(\ref{e:comptime}) as a representative example, the gain in
computational efficiency is $\Delta T_{\rm standard}/\Delta T_{\rm
HMD}=(5 \times 10^{8})/(5\times 10^5)\approx N_{\rm
int}=1000$. Moreover, the Fisher matrix is much smaller, $5\times 5$,
which offers an important further advantage when performing the inversion to obtain the
error covariance matrix.  Using the HMD method, the inversion of the Fisher
matrix is computationally less expensive than generating the
matrix. Note that the second term in $\Delta T^{\rm no~spin}_{\rm
HMD}$, corresponding to the $P_j(\tf)$ functions, is negligible in
this case and the computation time is dominated by constructing the
coefficient matrices. A calculation of the representative MC example
above, with a non-optimized implementation of the HMD method, takes
less than a minute on a regular workstation.

The case for substantial improvement with the HMD method becomes even
more compelling when additional parameters (spins and higher order PN
terms) are included, as we discuss next.

\subsection{Post-Newtonian Signal without Spin Precession}\label{s:advantages:computation_times:nospinprec}

We now consider the general HMD method outlined in
\S~\ref{s:HMDFisher:general}, with $N_{\spin}\equiv 6$ spin
components. The spin parameters can be grouped as $N_{\spinmag}\equiv
2$ independent spin magnitudes and $N_{\spinang}\equiv 4$ independent
spin angles. Since the spins can be oriented arbitrarily, the spin
angular parameters have to be randomly chosen, in addition to the
other angular parameters in any Monte Carlo computation. This
enlargement of the parameter space of random parameters greatly
increases the computational cost, both for the standard method and the
HMD method. However, we show next that the incremental cost is much
less severe for the HMD method. The HMD method should be considered in
future work aimed at computing time-dependent parameter estimation
errors in the general case with spins. Here, we neglect the effects of
spin precession, so that the angles
$\Omega=(\theta_N,\phi_N,\theta_{NL},\phi_{NL},\alpha,\gamma(0))$ are
decoupled and, since the signal does not depend on $\phi_N$, there are
only $N_{\Omega}\equiv 5$ independent (spin-unrelated) angles.

The larger the parameter space, the larger the sample size must be in
a Monte Carlo computation.  Let us assume that the sample size is
chosen to be $N_{MC}=\left(N^{(1)}_{MC}\right)^d$, where
$N^{(1)}_{MC}$ is the effective number of samples for a single
parameter, and $d=N_{\Omega}$ when spanning the $\Omega$-space only,
$d=N_{\spinang}$ when spanning the spin-angle space only, and
$d=N_{\Omega}+N_{\spinang}$ when spanning both.

To compute the time-independent matrices, we have to evaluate
$N_{p}(2N_{J}+1)$ independent $A_{j,a}$ coefficients for the full
$\Omega$-space and we have to compute the $N_{p}(2N_{J}+1)(2N_{PN}+1)$
independent $c_{jk,n}$ matrices over a $d=N_{\rm D}+N_{\spinang}$
dimensional parameter space. (Here $N_{\rm D}\equiv 2$ denotes the
number of parameters on which the Doppler phase depends,
$(\theta_N,\gamma)$ in eq.~[\ref{e:Doppler_phase}], using the fact
that both $c_{\rm GW}$ and $c_{\rm T}$ also depend on all spin angles
in eqs. [\ref{e:phi_GW},\ref{e:t(f)}].)  In
\S~\ref{s:HMDFisher:general} we have shown that there are
$(2N_{J}+1)\times(2N_{PN}^2+7N_{PN}+6)$ independent integrals, where
$N_{PN}$ is the number of terms in the post-Newtonian expansion plus
the Doppler phase. We have to compute these integrals for all spin
angle orientations. Therefore, the computational cost scales as
\begin{eqnarray}
\Delta T^{\rm no~spin~prec}_{\rm HMD}&&=
N_{p}(2N_{J}+1)\left(N^{(1)}_{MC}\right)^{N_{\Omega}} \nonumber\\ &&+
(2N_{J}+1)(2N_{PN}^2+7N_{PN}+6)N_{\tf}\left(N^{(1)}_{MC}\right)^{N_{\spinang}}\nonumber\\ &&+
N_{p}(2N_{J}+1)(2N_{PN}+1)\left(N^{(1)}_{MC}\right)^{N_{\rm
D}+N_{\spinang}}.
\label{e:T_HMD_noprec}
\end{eqnarray}
For the standard method, reiterating the argument in
\S~\ref{s:advantages:computation_times:approximate}, we get
\begin{equation}
\Delta T^{\rm no~spin~prec}_{\rm standard}=
\frac{1}{2}N_{p}(N_{p}+1) N_{\rm int}
\left(N^{(1)}_{MC}\right)^{N_{\Omega}+N_{\spinang}},
\end{equation}
where now $N_{p}=N_{p1}+N_{p0}+N_{\spin}=15$. Taking
$N^{(1)}_{MC_1}=10$, $N_{PN}=4$, $N_{J}=4$, $N_{D}=2$, $N_{\Omega}=5$,
$N_{\spinang}=4$, $N_{\rm int}=10^3$, $N_{\tf}=100$ as a
representative example, we find that the HMD method is computationally
less expensive by a factor $\Delta T^{\rm no~spin~prec}_{\rm
standard}/\Delta T^{\rm no~spin~prec}_{\rm HMD}=(1\times
10^{14})/(2\times 10^9)=7\times 10^{4}$, as compared to the standard
method.

Note that the first term in eq.~(\ref{e:T_HMD_noprec}) corresponds to our
original approximations in \S~\ref{s:HMDFisher:approximate}, i.e. no
cross-correlations between ${\bm p}_{\spin}$ and ${\bm
p}_{\slow}$. This approximation indeed leads to much faster
computations, since $\Delta T^{\rm no~spin~prec}_{\rm standard}/\Delta
T^{\rm approximate}_{\rm HMD}=9\times 10^6$ for the same
representative example.

\subsection{Post-Newtonian Signal with Spin Precession}\label{s:advantages:computation_times:spinprec}

Accounting for spin precession, the $N_{L}=2$ angular momentum angles,
($\theta_{NL},\phi_{NL}$), and the $N_{\spinang}$ spin angles are now
changing with time. In this case, one has to solve a differential
equation for the evolution of these angles for each individual Monte
Carlo realization. We assume that this can be computed independently
of the Fisher matrices and that it would take $N_{DE}N_{\tf}$
computation time units to evaluate, at each of the $N_{\tf}$ time
instances, for each initial set of angles.

The HMD method costs
\begin{eqnarray}
\Delta T^{\rm spin~prec}_{\rm HMD}&&=
N_{p}(2N_{J}+1)\left(N^{(1)}_{MC}\right)^{N_{\Omega}} \nonumber\\ &&+
N_{p}(2N_{J}+1)(2N_{PN}+2)\left(N^{(1)}_{MC}\right)^{N_{\rm
D}+N_{\spinang}}\nonumber\\ && +
(2N_{J}+1)(2N_{PN}^2+7N_{PN}+6)N_{\tf}\left(N^{(1)}_{MC}\right)^{N_L+N_{\spinang}}\nonumber\\
&&+
N_{DE}N_{\tf}\left(N^{(1)}_{MC}\right)^{N_L+N_{\spinang}},
\label{e:T_HMD_prec}
\end{eqnarray}
where the first term involves constructing the $A_{j,a}$ coefficient
matrices, the second term involves constructing the $c_{jk,a}$
coefficient matrices, the third term is for computing all three time--evolution
quantities $P(\tf)$ in eq.~(\ref{e:Pgeneral}), and the fourth term is
for solving the precession equations.

In the standard method, we need first to solve the precession
evolution differential equation and then construct the Fisher matrix
\cite{lh06}. Following the assumptions made above, we estimate a cost
\begin{eqnarray}
\Delta T^{\rm spin~prec}_{\rm standard}&=&
N_{DE}N_{\tf}\left(N^{(1)}_{MC}\right)^{N_L+N_{\spinang}}\nonumber\\
&&+\frac{1}{2}N_{p}(N_{p}+1) N_{\rm int}
\left(N^{(1)}_{MC}\right)^{N_{\Omega}+N_{\spinang}}N_{\tf}.
\end{eqnarray}

Using $N_{L}=2$ and all other parameters as in
\S~\ref{s:advantages:computation_times:nospinprec}, we find that the
HMD method is computationally more efficient by a factor $\Delta
T^{\rm spin~prec}_{\rm standard}/\Delta T^{\rm spin~prec}_{\rm
HMD}=(2\times 10^{14})/(6\times 10^{10})=2\times 10^{3}$, as compared
to the standard method. Since the $(\theta_{NL},\phi_{NL})$ subspace
could no longer be decoupled, the efficiency of the HMD method
relative to the traditional method lost a factor of $30$, as compared
to the no spin precession case in
\S~\ref{s:advantages:computation_times:nospinprec}. Nevertheless, the
computational advantage remains very substantial.

\section{Results}\label{s:results}

Having described the HMD formalism in detail, we now apply it to build
MC simulations aimed at studying how RMS source localization errors
\footnote{The Fisher matrix method yields $\sqrt{\langle \delta
p_i^2\rangle}$ RMS error for each set of fiducial angles. As an
approximation, we identify the distribution of errors with the
distribution of RMS errors.} evolve as a function of look--back time,
$\tf$, before merger. The low computational cost of the HMD method
allows us to survey simultaneously the dependencies on source sky position,
SMBH masses and redshifts.  We carry out MC calculations with $3\times
10^3$ random samples for the angles $\cos
\theta_{N},\cos\theta_{NL},\phi_{NL},\alpha,\gamma(0)$.  Several
thousands values of $M$ and $z$ are considered, in the range
$10^5<M/\Msun<10^8$ and $0.1\leq z\leq7$.  In addition, we ran
specific MC calculations to study possible systematic effects with
respect to the source sky position, by fixing $\theta_N$ and $\phi_N$
(on a grid of several hundred values) and varying all the other
relevant angles.

In all of our computations, we calculate the error covariance matrix
for $\dL,\theta_N,\phi_N,\theta_{NL},\phi_{NL}$. Following
Lang~\&~Hughes~\cite{lh06}, we calculated the major and minor axes of the 2D sky
position uncertainty ellipsoid, $2a$ and $2b$, and the equivalent
diameter, $\sqrt{4ab}$.

We have verified our HMD implementation and the general validity of
our assumptions by comparing our results at ISCO with those of
Lang~\&~Hughes~\cite{lh06} (for $m_1=m_2=10^5, 10^6, 10^7\Msun$ at $z=1$ and
$m_1=m_2=10^5, 10^6\Msun$ at $z=3$, in the no spin precession
case). Depending on SMBH masses and redshifts, we found agreement at
the $5$--$30\%$ level for the mean errors on the luminosity distance,
major axis, and minor axis.  The small discrepancies may be due to
differences in the set of assumptions made. Lang~\&~Hughes~\cite{lh06} account for
the small cross-correlations between the ${\bm p}_{\slow}$ and $\{{\bm
p}_{\fast},{\bm p}_{\spin}\}$ parameters and they choose $\ti$ to be
uniformly distributed between merger time and LISA's mission
lifetime. Recently, Lang \& Hughes reported angular errors that are
a factor of 2--3 lower \cite{lh07}, which are inconsistent with our results
at this level. Nevertheless, these discrepancies are
still small relative to the typical width of error distributions or to the
systematic variations of mean errors with $t_f$, $M$, and $z$ (from a factor
of few to orders of magnitudes, see Fig.~\ref{f:errortau} below). This
successful comparison justifies the use of the HMD method to study the
dependence of localization errors on look--back time.

\subsection{Time dependence of source localization errors}

\begin{figure}
\centering{
\mbox{\includegraphics[width=8.5cm]{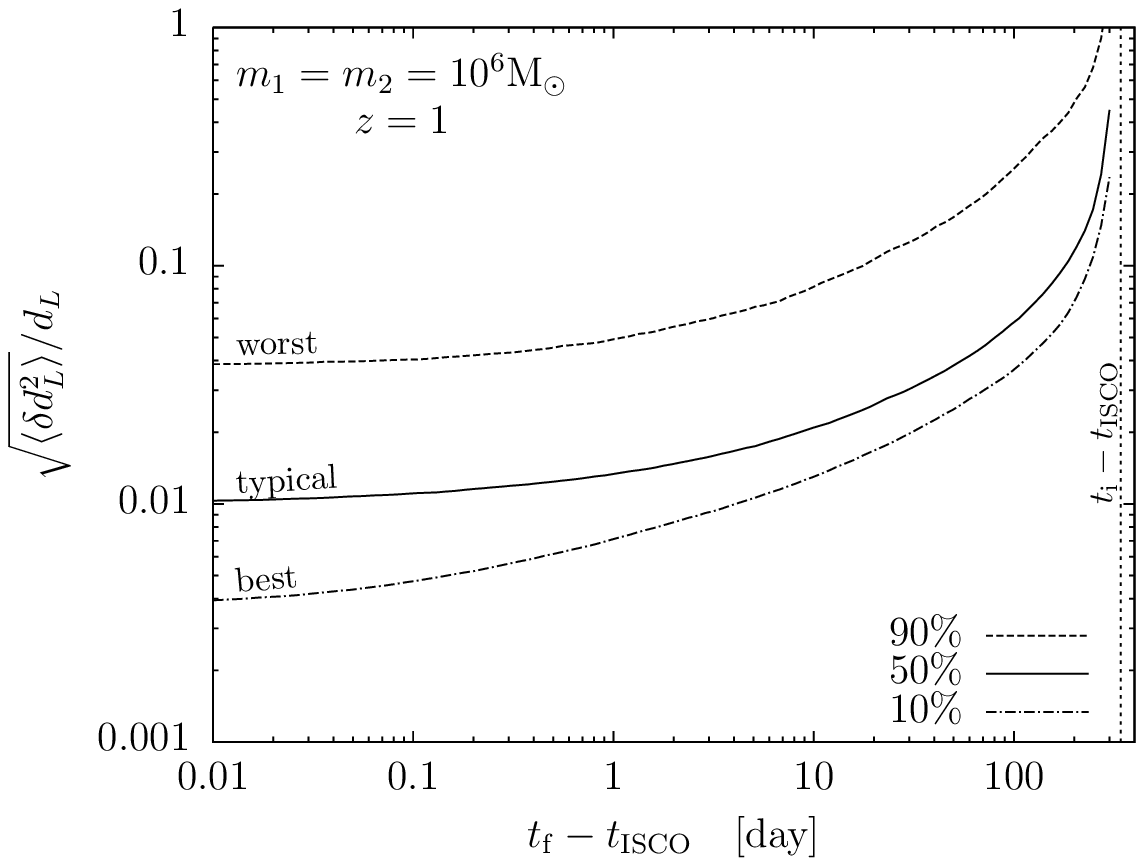}}\\
\bigskip
\mbox{\includegraphics[width=8.5cm]{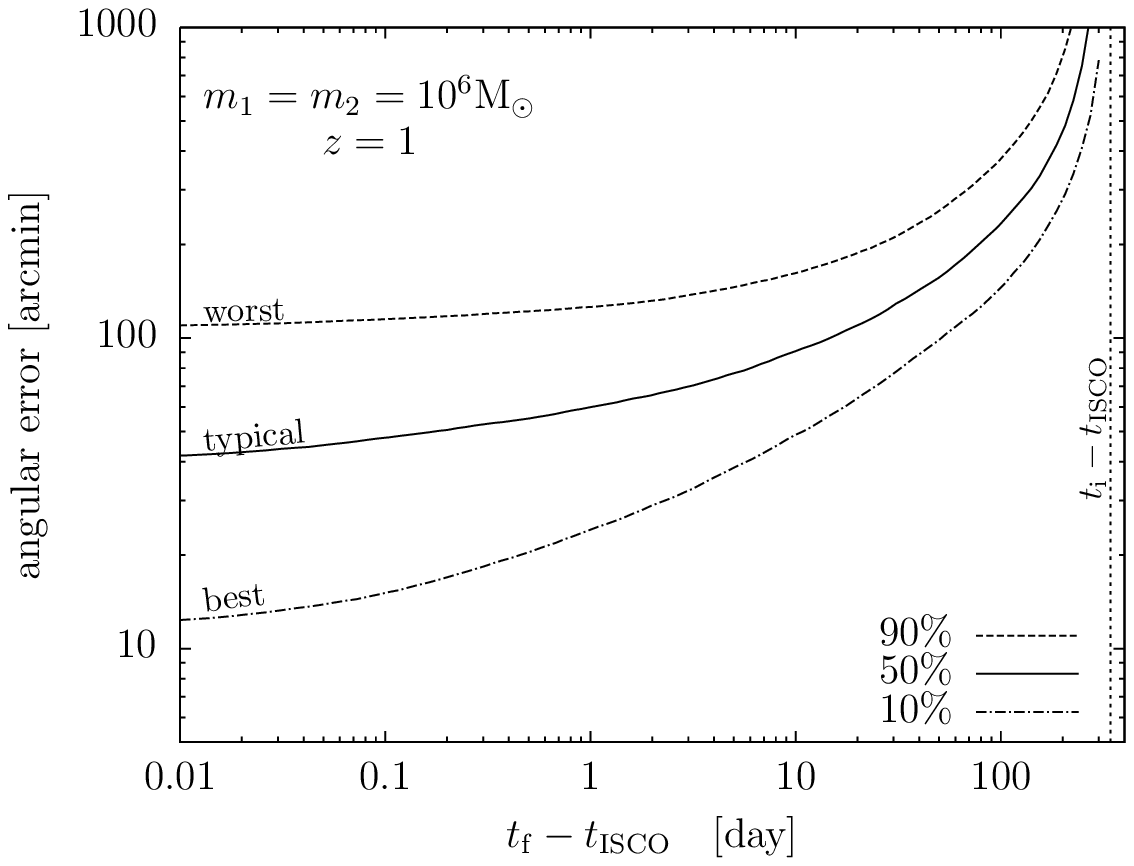}}}
\caption{\label{f:errortau} Evolution with pre-ISCO look--back time,
$\tf$, of LISA source localization errors, for $M=2\times 10^6\Msun$
and $z=1$. The top panel shows luminosity distance errors and the
bottom panel shows sky position angular errors (equivalent diameter,
$2\sqrt{ab}$, of the error ellipsoid). Best, typical, and worst cases
for random orientation events represent the $10\%$, $50\%$, and $90\%$
levels of cumulative error distributions, respectively.  Errors for
worst case events effectively stop improving at a finite time before
ISCO, even though the signal-to-noise ratio accumulates quickly at
late times. Errors for best case events (especially the minor axis)
follow the signal-to-noise ratio until the final few hours before
merger.}
\end{figure}

We calculate the variation with look--back time, $\tf$, of the
distribution of marginalized
parameter errors for a range of values of
$M,z,\theta_{N},\theta_{NL},\phi_{NL},\alpha,\gamma(0)$.
Figure~\ref{f:errortau} shows results for random angles and
$m_1=m_2=10^6\Msun$, at $z=1$.

The top panel shows the luminosity distance error, $\delta\dL$, while
the bottom panel describes the equivalent diameter, $2\sqrt{ab}$, of
the sky position error ellipsoid with minor and major axes $a$ and
$b$. The figure displays results for three separate cumulative
probability distribution levels, $90\%,50\%,10\%$, so that $10\%$
refers to the best $10\%$ of all events, as sampled by the random
distribution of angular parameters. The evolution of errors scales
steeply with look--back time for $\tf\gsim 40\dy$. In this regime, the
improvement of errors is proportional to $(S/N)^{-1}$. For smaller
look--back times, errors stop improving in the ``worst'' ($90\%$
level) case, improve with a much shallower slope than $(S/N)^{-1}$ for
the ``typical'' ($50\%$ level) case, and keep improving close to the
$(S/N)^{-1}$ scaling in the ``best'' case ($10\%$ level among the
realizations of fiducial angular parameters). Although
Figure~\ref{f:errortau} shows only the equivalent diameter of the 2D
sky localization error ellipsoid, we have also computed the evolution
of the distribution of the minor and major axes. We find that
$a\approx b\approx \sqrt{ab}$ initially (i.e. the ellipsoid is
circular), but the geometry changes significantly during the last two
weeks to merger. For example, in the typical case, the major axis $a$
stops improving at late times, while the minor axis $a$ maintains a
steep evolution. Therefore the eccentricity of the 2D angular error
ellipsoid changes quickly with look--back time. This is important
because large eccentricities can play a role in assessing
observational strategies for EM counterpart searches \cite{paper2}.

To map possible systematic effects with respect to source sky
position, we carried out MC computations with random $(\cos
\theta_{NL},\phi_{NL},\alpha)$ angles (sample size $N_{MC}=3\times
10^3$) but fixed source sky latitude and longitude relative to the
detector $(\theta_N,\gamma)$, for $m_1=m_2=10^6\Msun$ and $z=1$. We
find no systematic trends with sky position for $\delta\dL$, for any
value of the look--back time, $\tf$. Neither do we find systematic
trends with sky position for the distributions of minor and major axes
of the angular ellipsoid, for any value of the look--back time, $\tf$,
as long as $\theta_{N}$ is not along the equator. The case of
equatorial sources, with $\theta_{N}\approx 90^{\circ}$ and a short
look--back time $\tf$ before merger, is the only nontrivial one we have
identified. In that case, we find a minor systematic trend with
$\gamma$ longitude. The error distributions shift periodically up and
down, relative to the average, when changing $\gamma$ from $0$ to
$2\pi$.

In addition, to map dependencies with mass--redshift--look--back time of
localization errors, we carried out MC computations with arbitrary
$(\cos \theta_N,\cos\theta_{NL},\phi_{NL},\alpha,\gamma)$ angles, with
sample size $N_{MC}=3\times 10^3$, for several thousand pairs of
$(M,z)$ values. We find that the evolution with look--back time of error
distributions depends sensitively, and in a complicated way, on the
mass-redshift parameters.  Generally, localization errors increase
with redshift. Firstly, the $S/N$ is approximately proportional to the
instantaneous value $\sigma(t_{\rm ISCO})\propto
\eta^{3/4}[(1+z)M]^{5/4}/\dL(z)S_{n}(f_{\rm ISCO})^{-1}$ (eq.  [\ref{e:sigma}])
and, secondly, the beginning-of-observation time scales as $\ti\propto
\eta^{-1}[(1+z)M]^{-5/3}$ (eq. [\ref{e:t0(f)}]). For
$(1+z)M<4\times 10^6\Msun$, the total
observation time can exceed one year and the second effect is
unimportant. We further describe mass--redshift dependencies below, in
\S~\ref{s:results:warning}, in relation to advance warning times for
targeted electromagnetic counterpart searches.

The results on localization errors from our extensive exploration of
the parameter space of potential LISA sources can be summarized as
follows:
\begin{enumerate}
 \item Probability distributions
\begin{itemize}
\item The error distributions for $\delta\dL$, $2a$, and $2b$ all have
long tails: $1\%$--$99\%$ cumulative probability levels are separated
by factors of $\sim 100$, while the $10\%$--$90\%$ levels are
separated by factors of $\sim 10$.

\item The $\delta\dL$ distribution is skewed, with a median closer to
the best case, a median smaller than the mean, even on a logarithmic
scale. On the other hand, sky localization error distributions are
roughly symmetric on a logarithmic scale.
\end{itemize}

\item Fiducial parameter dependencies
\begin{itemize}
\item The $\delta\dL$ errors are roughly independent of fiducial
angles throughout the observation.

\item For non-equatorial sources, the distribution of sky localization
errors, $(2a,2b)$, is independent of sky position, i.e. the
distribution does not have a systematic dependence on $\theta_N$ and
$\gamma\equiv \Phi-\phi_{N}$ (for random
$\alpha,\theta_{NL},\phi_{NL}$).

\item There is a small systematic trend with $\gamma$ for equatorial
sources.

\item There is a complicated dependence of sky localization errors on
$\Mchirp$, $z$, and look--back time $\tf$. For $(1+z)(\eta/0.25)^{3/5}M\lsim 4\times 10^6$, and
long observation times, errors scale with $(S/N)^{-1}\approx
[(1+z)\Mchirp]^{5/4}\dL(z)^{-1}S_n(f_{\rm ISCO}(M,z))^{-1}f_{a}(\tf)$, where
$f_{a}(\tf)$ is the $\tf$-scaling shown in Fig~\ref{f:errortau}. For
larger redshifted masses, the scaling has a complicated structure in
the $M,z,\tf$ space that we did not analyze in detail (but see
eq. (\ref{e:cov3}) in the Appendix for scalings in terms of $\ti$ and
$\tf$.)
\end{itemize}

\item Time dependence
\begin{itemize}
\item Luminosity distance and sky localization errors roughly scale with
$(S/N)^{-1}$ until $2$ weeks before ISCO.

\item For the luminosity distance $\delta \dL$ and the major axis
$2a$, there is little improvement within the last week before ISCO for
the typical to worst cases (i.e. $50\%$--$90\%$ levels of cumulative
error distributions).

\item For the minor axis $2b$, only the worst case events stop
improving within the last week. The typical to best cases continuously
improve until the last hour.

\item The eccentricity of the sky localization error ellipsoid changes
with time during the first and last two weeks of observation.  The
eccentricities are smaller in between these two time intervals. For a
detailed discussion of the eccentricity and its impact on counterpart
searches, see Ref.~\cite{paper2}.

\item For the luminosity distance $\delta\dL$, the relative width of error
distributions does not change during observation and variations in the
difference between the $90\%$ and $10\%$ levels of the cumulative
distributions do not exceed $10\%$, except for the initial weeks,
when the distribution is much more spread out.

\item For the sky localization errors, the width of error
distributions increases during the final two weeks of observation, by
a factor $\sim2$ for the major axis and a factor $\sim 4$ for the
minor axis.
\end{itemize}
\end{enumerate}

\subsection{Advance warning times for EM searches}\label{s:results:warning}

From the astronomical point of view, being able to identify with
confidence, prior to merger, a small enough region in the sky where
any prompt electromagnetic (EM) counterpart to a LISA inspiral event
would be located, is of great interest. With sufficient ``warning time,''
it would then be possible to trigger efficient searches for EM
counterparts as the merger proceeds and during the most energetic
coalescence phase. In particular, an efficient strategy to catch such
a prompt EM counterpart would be to continuously monitor with a
wide-field instrument a single field-of-view (FOV), through
coalescence and beyond. Astronomical strategies for EM counterpart
searches are the focus of a second paper in this series
\cite{paper2}.

Given an angular scale, $\theta_{\rm FOV}$, corresponding to the
hypothetical FOV of a specific astronomical instrument, it is thus of
considerable interest to determine the value of the look--back time
$\tf$ at which the major axis, minor axis or equivalent diameter of
the sky localization error ellipsoid provided by LISA just reach the
relevant $\theta_{\rm FOV}$ scale. This would allow one to trigger an
efficient search for EM counterparts, in a well defined region of the
sky that can be monitored.  We will hereafter refer to this time as the
{\it advance warning time}. Note that it is important to
differentiate the sizes of the major and minor axes of the angular
error ellipsoid in this context because the eccentricity can be large,
and thus important in assessing optimal strategies for EM counterpart
searches \cite{paper2}.

\begin{figure}
\centering{
\mbox{\includegraphics[width=8.5cm]{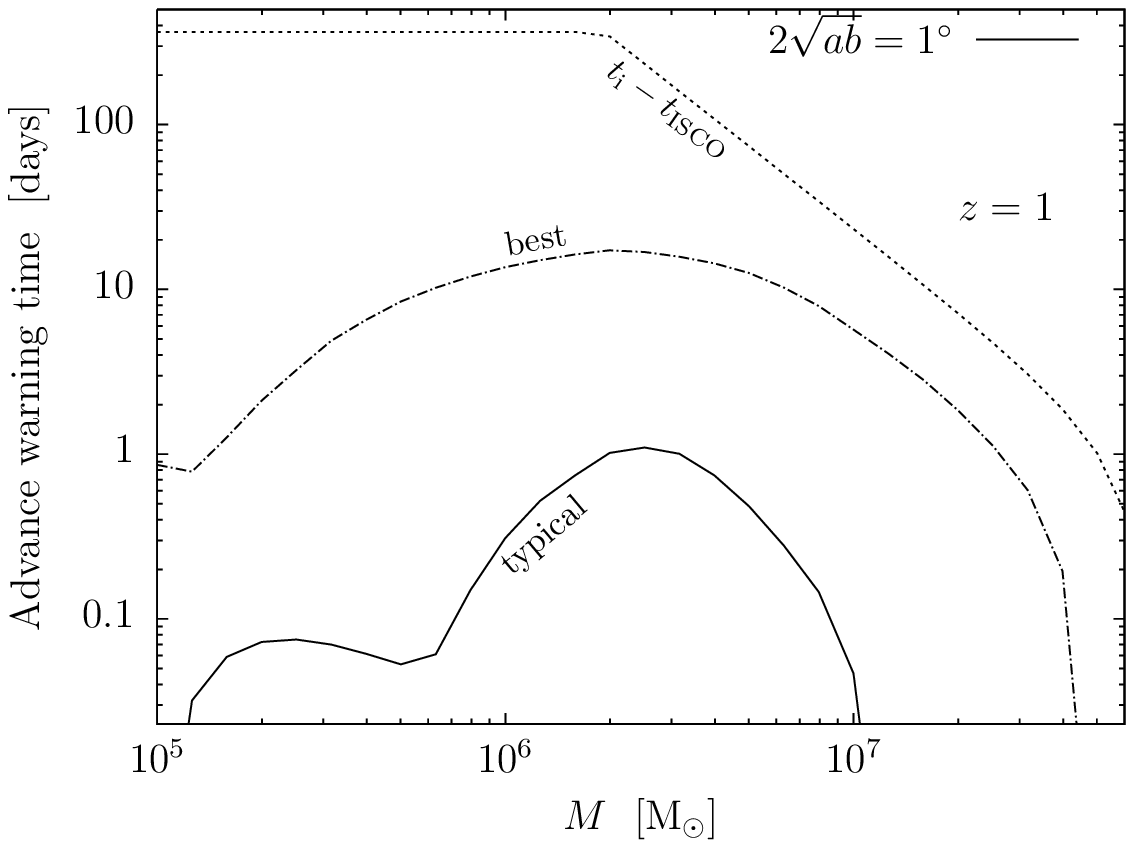}}\\
\bigskip
\mbox{\includegraphics[width=8.5cm]{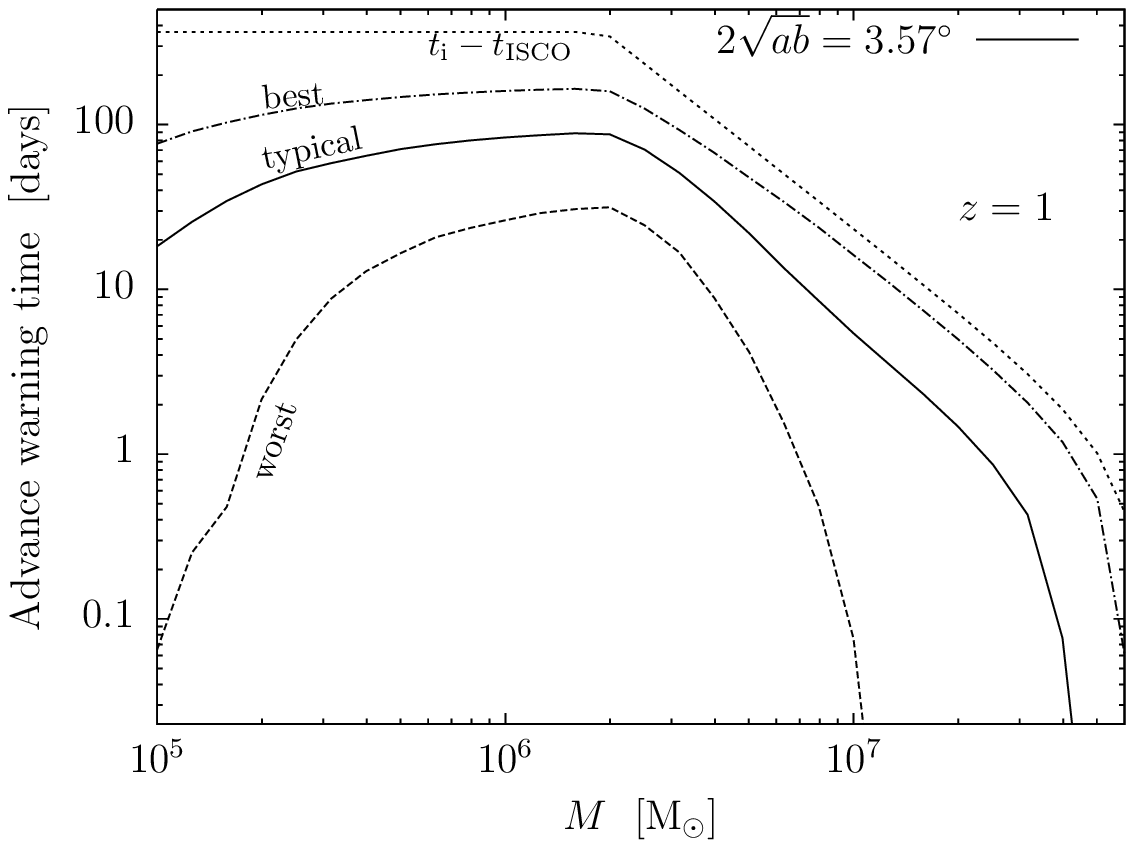}}}
 \caption{\label{f:warning_M} Advance warning times (in days) for
 equal mass binary inspirals at $z=1$, as a function of total mass,
 $M$ (in solar units). Best, typical, and worst cases refer to $10\%$,
 $50\%$, and $90\%$ levels of cumulative error distributions for
 random orientation events, as before. The advance warning times shown
 correspond to the values of look--back times when the equivalent
 diameter, $2\sqrt{ab}$, of the error ellipsoid first reaches
 $1^{\circ}$ (top panel) or an LSST-equivalent field-of-view
 ($3.57^{\circ}$, bottom panel). In the top panel, the worst case
 events are not shown because angular errors are too large even at
 ISCO. For the largest mass SMBHs, the maximum observation time (and
 thus $\ti$) is below one year.}
\end{figure}

\begin{figure}
\centering{
\mbox{\includegraphics[width=8.5cm]{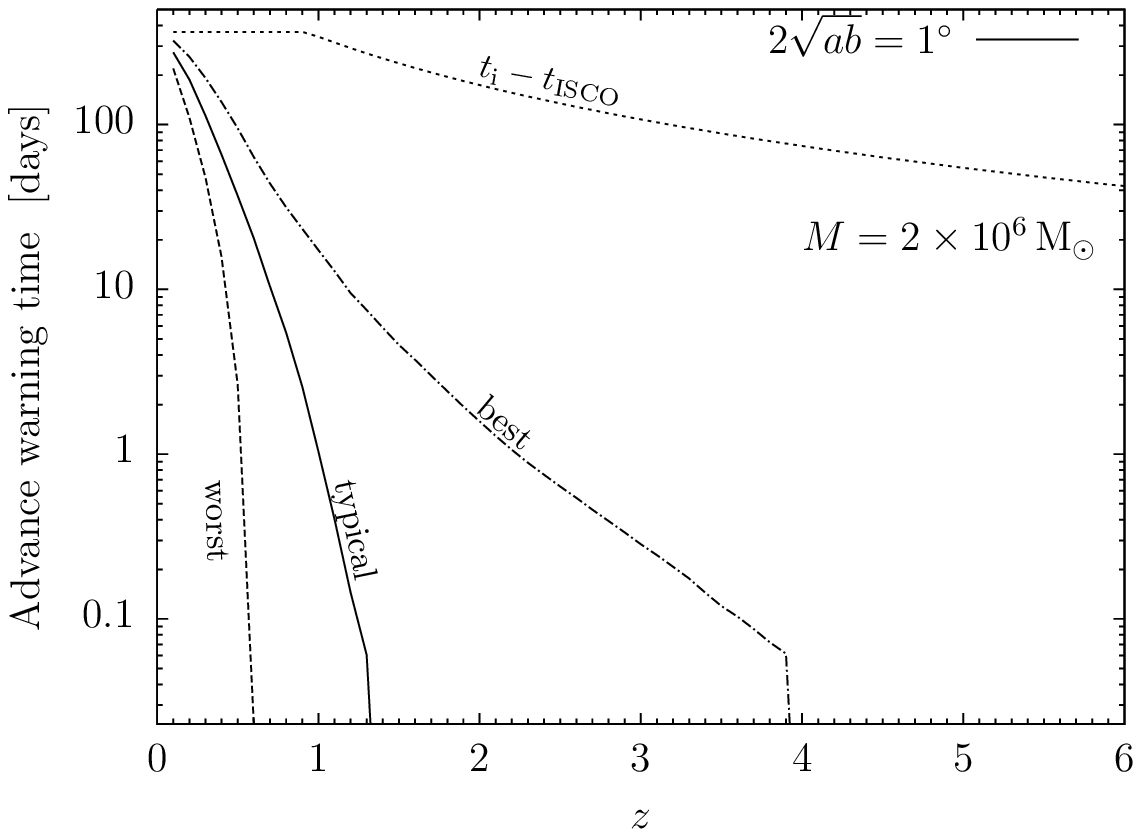}}\\
\bigskip
\mbox{\includegraphics[width=8.5cm]{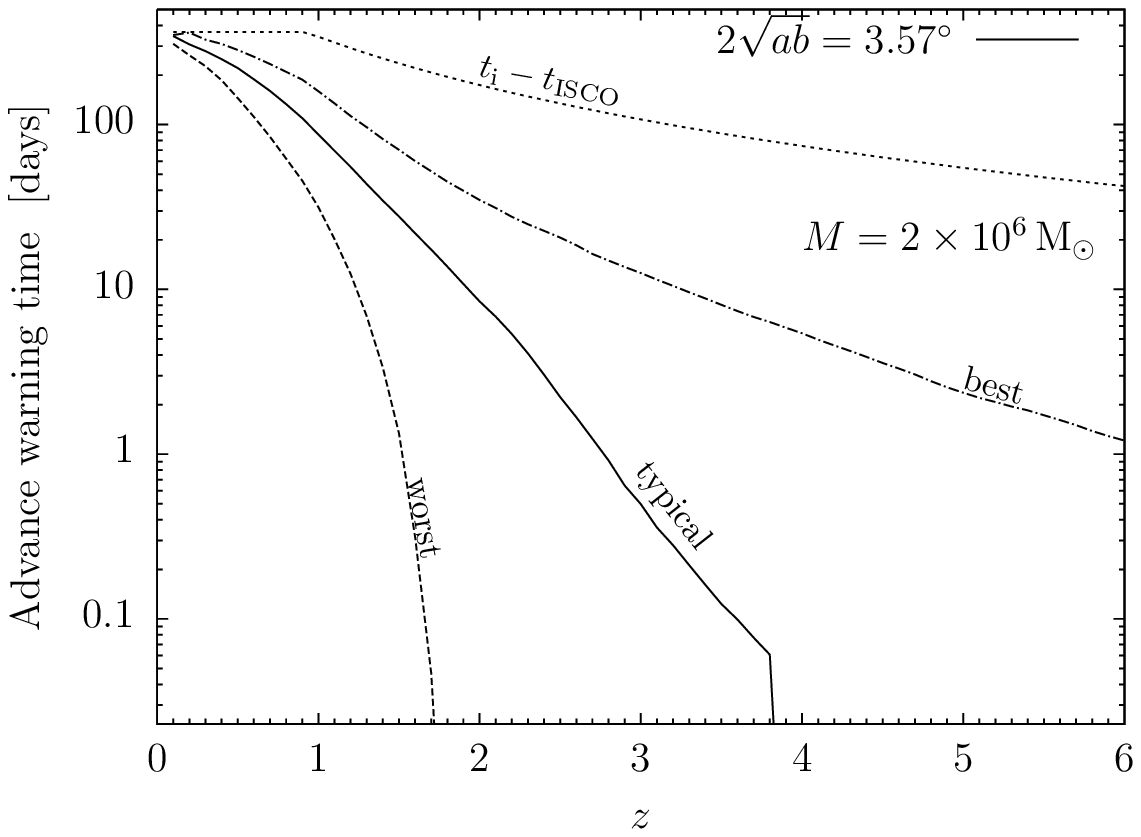}}}
 \caption{\label{f:warning_z} Same as Fig.~\ref{f:warning_M}, for a
 fixed total mass $M=2\times 10^{6}\Msun$ but various values of the
 source redshift, $z$.}
\end{figure}

For definiteness, we evaluate advance warning times for angular
diameters $\theta_{\rm FOV}=1^{\circ}$ and $3.57^{\circ}$ here but
generalizations to other $\theta_{\rm FOV}$ values are obviously
possible. The choice of the latter figure is motivated by the  $10\,{\rm deg}^2$ FOV
proposed for the future Large Synoptic Survey Telescope, or LSST
\cite{tys03}. Figure~\ref{f:warning_M} shows advance warning times
for a fixed source redshift at $z=1$ and various values of the total
SMBH mass, $M$. Figure~\ref{f:warning_z} shows similar results for
various source redshifts, at a fixed value of $M=2\times 10^6\Msun$.


In each case, we consider equal mass SMBH binaries and a maximum
observation time of $1\yr$ (or lower if set by the GW noise frequency
wall at $0.03\mHz$). Each panel in Figs.~\ref{f:warning_M}
and~\ref{f:warning_z} shows the values of advance warning times
at which the equivalent diameter $2\sqrt{ab}$ of the localization
error ellipsoid drops below the reference $\theta_{\rm FOV}$ value.
For each case, we show results for cumulative error distribution
levels of $10\%$, $50\%$, and $90\%$, labeled ``best'', ``typical,''
and ``worst'' cases, as before.
Figure~\ref{f:warning_M} shows that LISA can localize on the
sky events at $z=1$ to within an LSST FOV at least one month ahead of
merger, for $50\%$ of events with masses $2\times 10^5\Msun\leq M\leq
3\times 10^6\Msun$, and at least 4 days ahead of merger for $90\%$ of
events in the same mass range. Figure~\ref{f:warning_z} shows that
advance warning times decrease with redshift, leaving at least 1 day
ahead of merger for $50\%$ of events with $M=2\times 10^6\Msun$, as
long as $z\lsim 1$ for $\theta_{\rm FOV}=1^{\circ}$ and as long as
$z\lsim 3$ for an LSST FOV. For events with this mass scale and the
LSST FOV, there is a $10\%$ chance that a 1 day advance warning is
possible up to $z\sim5$--$6$.

\begin{figure}
\centering{
\mbox{\includegraphics[width=8.5cm]{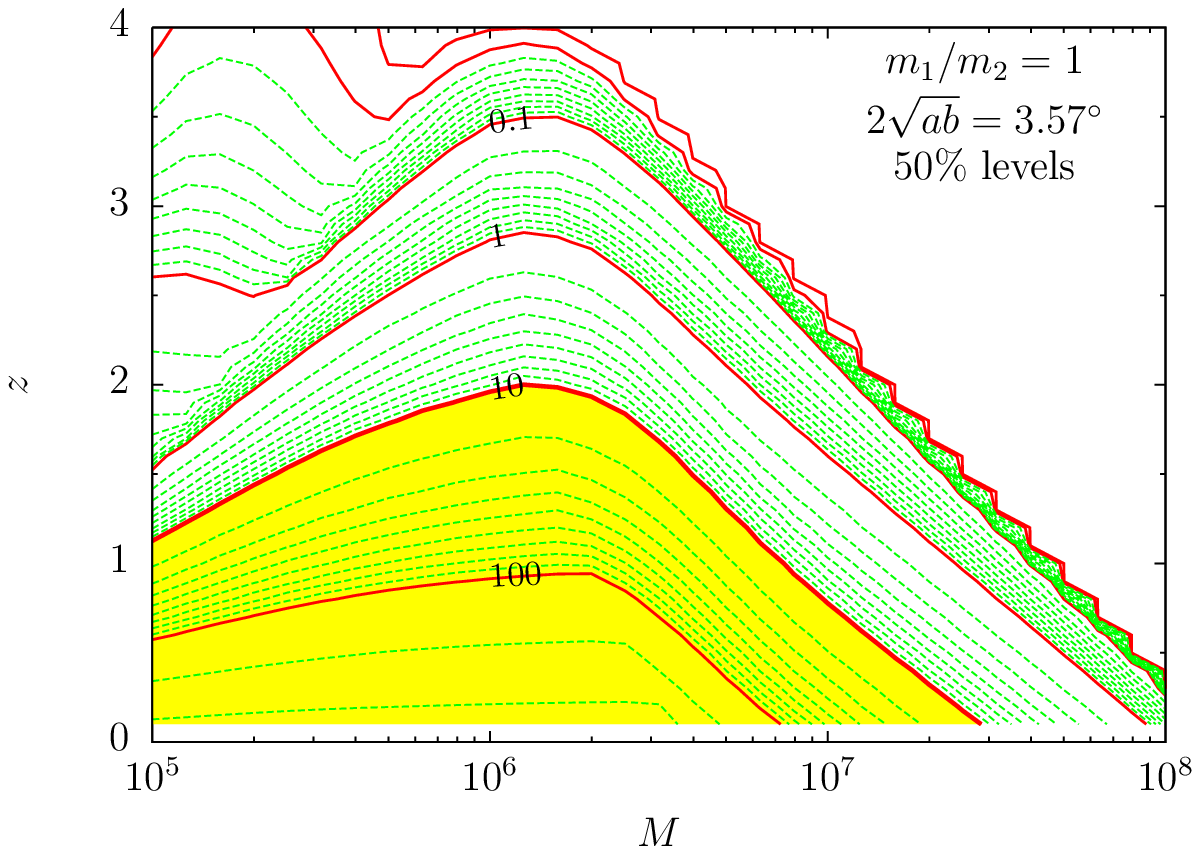}}\\
\bigskip
\mbox{\includegraphics[width=8.5cm]{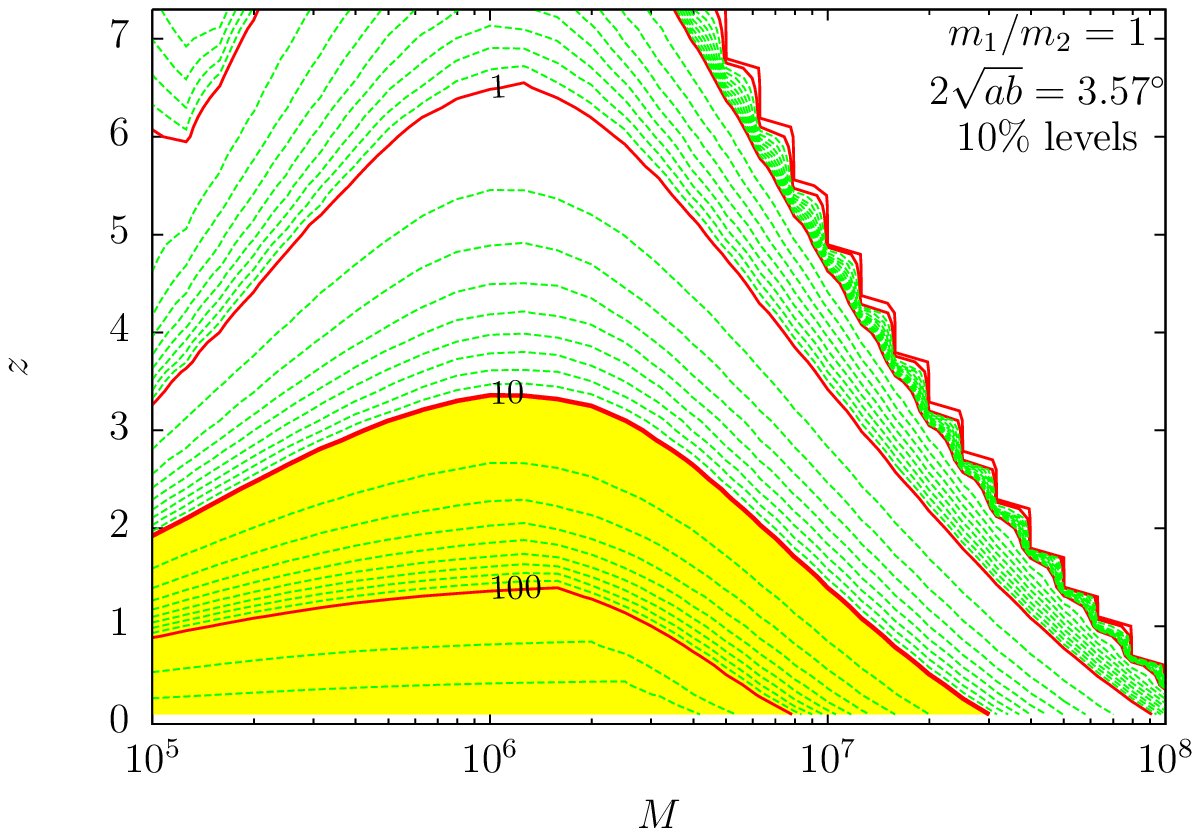}}
}
\caption{\label{f:warning_Mz50} Contours of advance warning times in
the total mass ($M$) and redshift ($z$) plane with SMBH mass ratio
$m_1/m_2=1$.  The contours trace the look--back times at which the
equivalent radius ($2\sqrt{ab}$) of the localization error ellipsoid
first reach an LSST-equivalent field-of-view ($3.57^{\circ}$). The
contours correspond to the $50\%$ {\it (top)} and $10\%$ {\it
(bottom)} level of cumulative distributions for random orientation
events.  The contours are logarithmically spaced in days and $10$ days
is highlighted with a thick line. }
\end{figure}

Figures~\ref{f:warning_M} and \ref{f:warning_z} display advance
warning times for single one dimensional slices of the full $(M,z)$
space of potential LISA events. With the HMD method, however, it is
possible to explore the entire parameter space of SMBH inspirals by
repeating the calculation on a dense grid of $(M,z)$ values. We
construct a uniform grid in the $(\log M,z)$ plane, with $\Delta
z=0.1$ and $\Delta \log M = 0.1$, and perform MC computations with
$3\times 10^3$ randomly oriented angles for each grid element. As a
result, we obtain a complete description of the time evolution of sky
localization errors in the large parameter space of potential LISA
sources.  Figure~\ref{f:warning_Mz50} displays advance warning time
contours from this extensive MC calculation, for typical ($50\%$)
and best case ($10\%$) events, adopting the LSST FOV as a reference.

Advance warning time contours are logarithmically spaced, with
solid-red contours every decade and a thick red line highlighting the
$10\,$day contour. Since advance warning times were computed on a
finite mesh, contour levels for arbitrary $M$ and $z$ values were
obtained by interpolation.  Our interpolated mesh is smooth if $\tf
\sim 0.1\,$day, but it gets edgy for short advance warning time
approaching ISCO. Figure~\ref{f:warning_Mz50} shows
that a $10$ day advance warning is possible with a unique LSST-type
pointing  for a large range of masses and
source redshifts, up to $M\sim 3\times 10^7\Msun$ and $z\sim 1.9$. The
bottom shows how far the advance warning concept can be stretched, by
focusing on the
$10\%$ best cases of random orientation events. In this case a 10 day
advance warning is possible up to $z\sim 3$ for masses around $M
\sim 10^6 \Msun$.  Note that, in both cases, allowing for a
warning of just one day would extend considerably the range of masses
and redshifts for which a unique LSST-type pointing is sufficient.

\begin{figure}
\centering{
\mbox{\includegraphics[width=8.5cm]{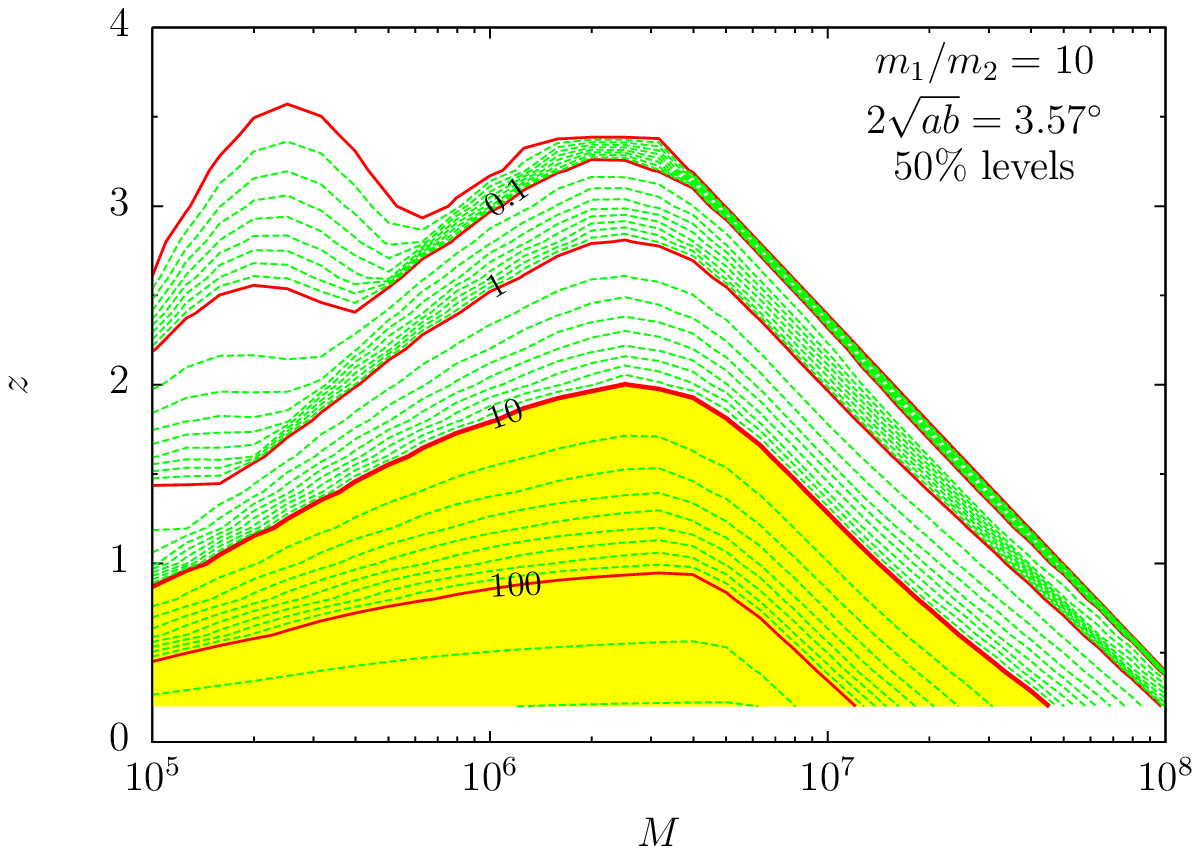}}\\
\bigskip
\mbox{\includegraphics[width=8.5cm]{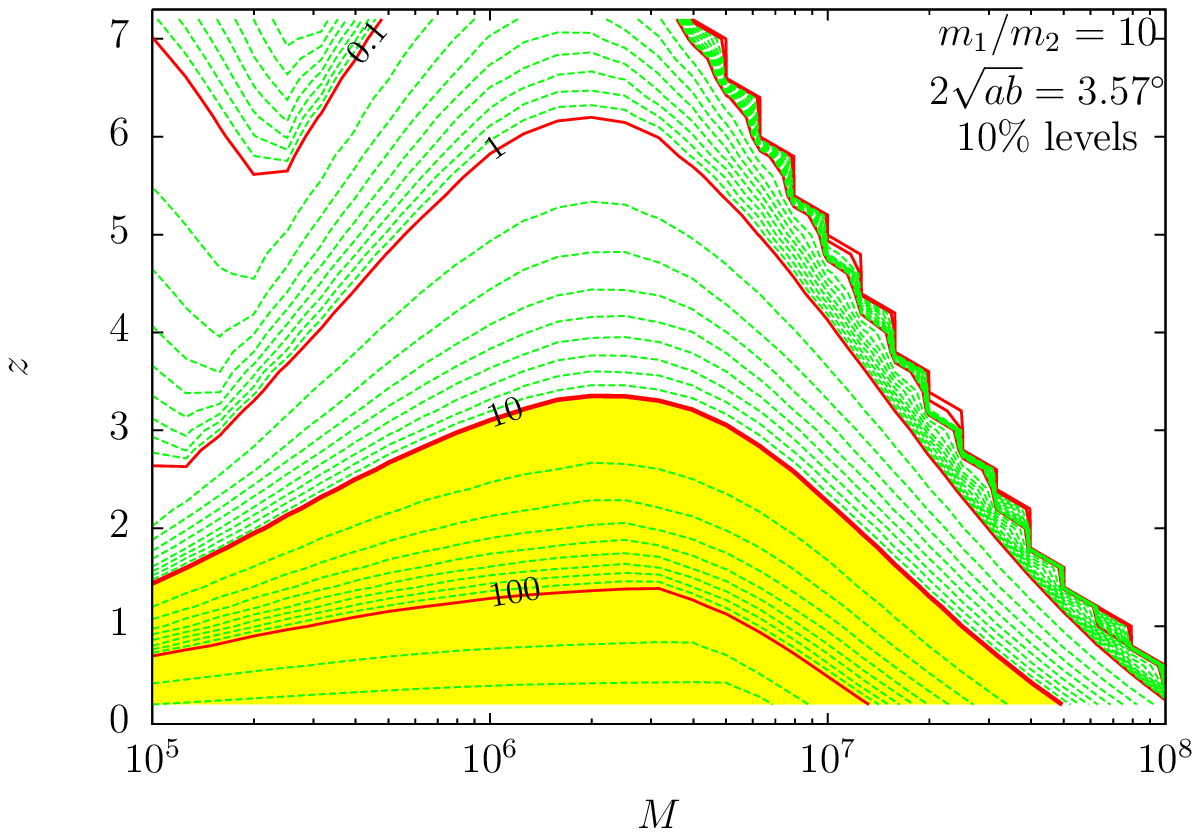}}
}
\caption{\label{f:warning_Mz50_y10} Same as Fig.~\ref{f:warning_Mz50},
except for a SMBH mass ratio of $m_1/m_2=10$.
}
\end{figure}

These results can also be generalized to unequal-mass SMBH binaries.
At fixed total mass, $M$, an unequal-mass binary has an instantaneous
signal-to-noise ratio that is reduced because of a lower $ \eta$
value, but it also has a total observation time that is potentially
longer. Localization errors for unequal-mass inspiral events with
total observation times longer than a month (i.e. with
$\eta_{0.25}^{2/5}(1+z)\,M<1.8 \times 10^7\Msun$) are degraded
relative to the equal-mass cases discussed so far. For larger total
mass, however, the worsening of errors is mitigated, or even reverted,
relative to the equal mass case, thanks to the longer observation
time. The error ellipsoid also becomes less eccentric thanks to this
additional observation time. Figure~\ref{f:warning_Mz50_y10}
summarizes results on advance warning
times from the same MC computations as in Fig.~\ref{f:warning_Mz50},
but this time for unequal-mass SMBH
binaries with mass ratio $m_1/m_2=10$. Despite a systematic
degradation in advance warning times (especially noticeable at low $M$
values), the main effect of introducing a mass ratio $m_1/m_2=10$ is
to shift advance warning time contours to somewhat larger values of
total mass, $M$. Our main conclusions on advance warning times are not
very strongly affected by the inequality of mass components in the
population of SMBH binaries considered.

\begin{figure}
\centering{
\mbox{\includegraphics[width=8.5cm]{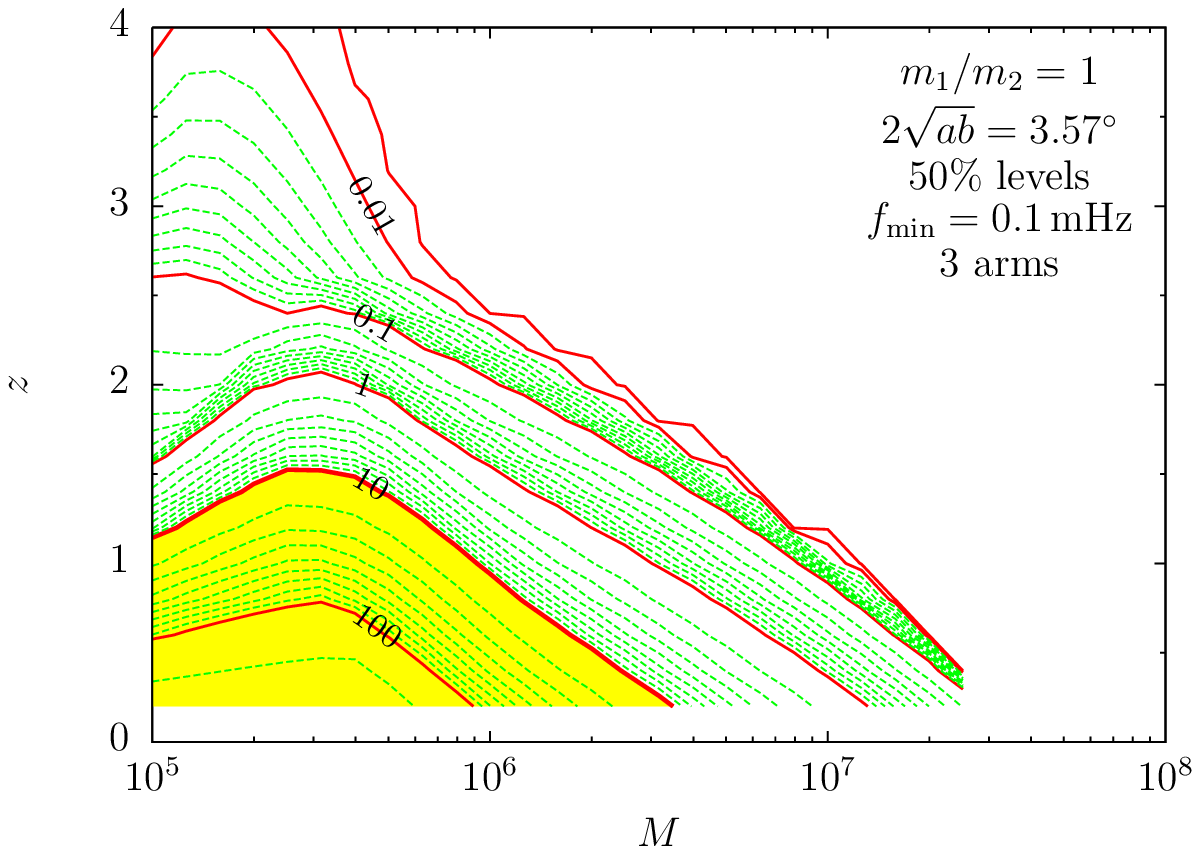}}\\
\bigskip
\mbox{\includegraphics[width=8.5cm]{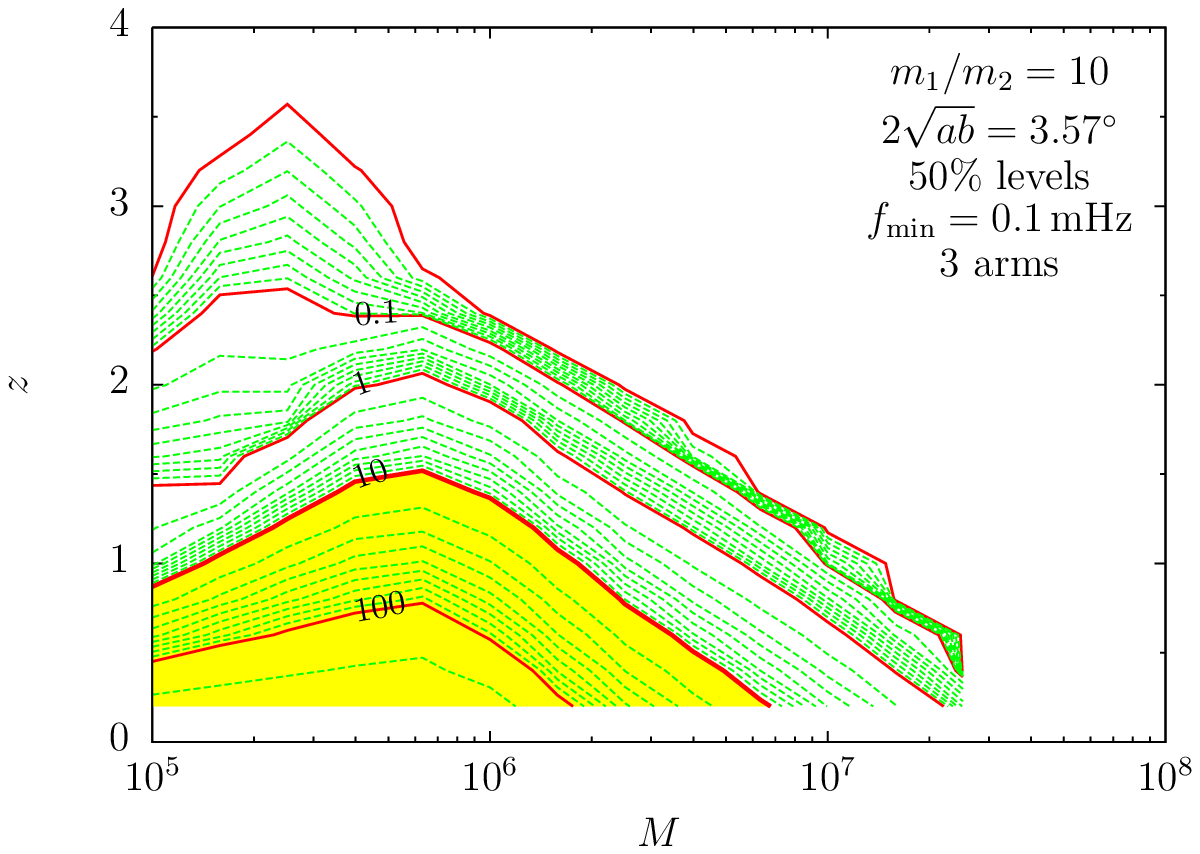}}
}
\caption{\label{f:warning_Mz50_fmin4} Same as the top panels of
Figs.~\ref{f:warning_Mz50} and \ref{f:warning_Mz50_y10}, except for a
degraded minimum detector frequency of $f_{\min}=0.1\mHz$.  }
\end{figure}

Finally, it is important to understand how sensitive the results
are to the LISA detector characteristics. In particular, we examined
how advance warning times are affected by increasing the minimum frequency
noise wall or by loosing one of the arms of the 3-arm constellation.
Figure~\ref{f:warning_Mz50_fmin4}
displays results for $f_{\min}=10^{-4}\Hz$, for $m_1/m_2=1$
and $m_1/m_2=10$.
Increasing $f_{\min}$ mostly reduces the total observation time
for high mass inspirals
($\ti \sim f_{\min}^{-8/3}M_z^{-5/3}$; see
eq.~[\ref{e:t0(f)}]) and reduces the signal-to-noise ratio
by a small factor. As a result, the advance warning
time contours primarily shift in the $(M,z)$ plane in the direction
of smaller total masses by a factor of $\sim 7$,
and secondly shift moderately
($30$--$50\%$) to smaller redshifts.
Loosing one LISA arm (i.e. using only one of the two interferometers)
most importantly removes the ability of the second datastream to break
correlations in localization errors and also reduces the
signal-to-noise by a small factor. As a result errors do not
improve much during the last $\sim 10\,$days before merger.
Compared to the case with two interferometers, contours
representing an advance warning of less than 10 days
are shifted to significantly smaller $z$ (especially for the
minor axis of the sky localization ellipsoid),
close to the $10$--day contour, but
warning times beyond $\sim 10\,$days worsen only moderately.
We conclude that even if $f_{\min}=0.1\mHz$ or if only one
of the two interferometers is used, LISA still admits $10$--day advance
localizations for a broad range of masses and redshifts, between
$10^5\lsim M \lsim 2\times 10^6$ and $z\lsim 1$.

\section{Discussion}\label{s:discussion}

We have introduced a novel technique, the HMD method, to compute time--dependent
GW inspiral signals for LISA. The method relies on the fact that LISA's orbital
motion induces a modulation on timescales that are long relative to
the inspiral GW frequency. Since this modulation is periodic, with a
fundamental frequency of $f_{\oplus}$, it can be expanded in a
discrete Fourier sum. In the HMD formalism, dependencies on sky
position, orbital angular momentum orientation, and detector
orientation in the LISA signal are inscribed in time-independent
coefficients, while time-dependent basis functions are independent of
these angles. This decomposition helps to reduce the computational
cost of Monte Carlo simulations exploring the time-dependence of
source localization errors by orders of magnitude.

Moreover, the HMD method can be used in conjunction with plausible
approximations to further decrease the computational cost of
explorations of the parameter space of localization errors for LISA
inspiral events. In our analysis, we identified two different
characteristic frequency constituents of the signal: the high
frequency restricted post-Newtonian GW inspiral waveform and the low
frequency amplitude modulation resulting from the detector's orbital
motion. In the HMD method, these two components separate and
parameters that depend only on the low frequency modulation (such as the
source position and the orbital angular momentum angle) can be
estimated independently of the other source parameters determined by
the high frequency carrier signal. Our working assumption was that
cross-correlations among these two sets of parameters must be much
smaller than parameter correlations within either set. This
hypothesis is valid very generally in the no spin limit for SMBHs, as
shown by full Fisher matrix calculations without such approximations
for general relativity \cite{hug02} and alternative theories of
gravity \cite{bbw05}.

In order to further examine the validity of our assumptions and the ultimate
boundaries of our models, and to understand our results, we have constructed
illustrative toy models that we now describe in some detail. These toy models show that the
separation of parameters into various subsets associated with
different characteristic frequencies of the signal is a rather general
property, which turns out to be an efficient way of reducing the
computational cost of error estimations for the LISA problem.

\subsection{Simple toy models}\label{s:advantages:simple}

In this section, we discuss very simple toy models which capture the
essence of the problem posed by the time-evolution of parameter error
estimations. We then use these models to answer general questions on
the LISA-specific parameter estimation problem.

Our harmonic decomposition technique is based on the simple intuition
that the angular information can be deduced from the slow periodic
modulation of the high frequency GW waveform. In \S~\ref{s:HMD}, we
have shown that modulation harmonics with frequencies larger than
$4f_{\oplus}$ vanish exactly. Here, we discuss the general properties
of such a modulation.  In the case of LISA, the high frequency carrier
signal has an effective, cycle-averaged signal-to-noise ratio which
monotonically increases with time as SMBH binaries approach merger. To
mimic such events, we also assume in all of our toy models that the
instantaneous signal-to-noise ratio continuously improves throughout
the observation.

We seek answers to the following questions:
\begin{enumerate}
\item How do mean errors evolve during the final days of observation?

On the one hand, in standard angle-averaged treatments
(e.g. \cite{bbw05,koc06,aru06}), an evolution of errors with the
inverse of the signal-to-noise ratio is generally assumed. This would
suggest a large improvement during the last day of inspiral. On the
other hand, the slow modulation picture suggests just the contrary:
not much improvement is expected at late times when there is
effectively very little modulation (Finn \& Larson 2005, private
communication).

\item Does the introduction of additional high frequency components in
the signal have any effect on the estimations of low frequency
parameters?

In the GW context, it is of general interest to determine under what
circumstances additional high frequency signal components, such as
higher order post-Newtonian corrections or spin-induced effects,
remain decoupled from the determination of angular and distance
information based on the signal amplitude modulation.

\item Are there combinations of signal parameters for which errors
improve rapidly in the last days of observation? If so, what are these
combinations? What determines how many such rapidly--improving combinations
there will be?

If the distance $\dL$ correlates with the angles, then in principle the
volume of the 3D error box can be much smaller than the product of the
marginalized errors $\delta \Omega\times \delta \dL$ would imply.
Unfortunately, in practice, this is unlikely to help to reduce the
number of false counterparts, because the $\delta z$ error will be
dominated by weak lensing \cite{koc06}.

\item How does the width of parameter error distributions evolve with
time? Are the best and worst cases approaching the typical case prior
to the final days of observation? How do we expect the eccentricity of
localization error ellipsoid to evolve with time for LISA?
\end{enumerate}

Here, we restrict our discussion to a brief summary of our findings
and direct the reader to Appendix~\ref{app:simple} for further details
on these toy models.

The parameter estimation uncertainties are defined by the correlation
error matrix. For $N_{p}$ parameters, this defines an
$N_{p}$-dimensional error-ellipsoid in the $N_{p}$-dimensional
parameter space, where parameters are constrained at a given
confidence level. Marginalized errors for a given parameter are then
related to the projection of this ellipsoid on the basis vector
corresponding to that parameter. Since the principal axes of this
error ellipsoid are generally not aligned with the original
parameters, the marginalized errors can be substantial even if the
volume of the error ellipsoid is close to zero. This happens if the
ellipsoid is very ``thin'' but has a large size in at least one
direction. Diagonal elements of the correlation matrix provide
marginalized squared errors on the parameters, while eigenvalues
provide squared errors along the principal axes.

We consider three versions of toy signals to understand how a
particular harmonic mode contributes to the time-dependence of
parameter uncertainties and to find answers to Questions 1--4
above. We start with the simplest toy model and refine this model by
adding more details and complexity in the successive models. In each
case, we discuss general implications for the model under
consideration.

\subsubsection{Basic toy model}

In our basic toy model, we assume that the true signal is comprised of
a constant carrier signal, which is modulated by a single
known-frequency cosine, $f_{\oplus}$:
\begin{equation}
\label{e:simple1} h(t)=c_0 + c_1\cos(2\pi f_{\oplus}t),
\end{equation}
where $c_0$ and $c_1$ are unknown parameters to be estimated. We
assume that the noise level is rapidly decreasing during the
observation, mimicking the gradual increase in the instantaneous
signal-to-noise ratio for LISA inspiral signals. The contradictory
statements made in relation to Question~1 above can be explored with
this model. We find that marginalized parameter errors scale with the
signal-to-noise ratio far away from merger (i.e. $\tf\gsim 0.1
f_{\oplus}^{-1}$) but they quickly converge to their final values at
late times, even though the signal-to-noise ratio keeps
accumulating. It is possible to derive analytical formulae for the
evolution of parameter errors to fully characterize this behavior (see
Appendix~\ref{app:simple}).  We find that, even though the error
ellipse rapidly decreases in volume, as the inverse of the
signal-to-noise ratio near merger, the error ellipse only shrinks
along one of its dimensions, the semi-minor axis, so that a
non-negligible residual uncertainty remains in the orthogonal subspace
(e.g. along the semi-major axis). This residual uncertainty carries
over to final marginalized errors for both parameters. Therefore, this
first toy model verifies the second option in relation to Question~1.:
there is no late improvement because there is very little effective
signal modulation, making the signal-to-noise argument largely
irrelevant. However, we find below that this model does not carry some
essential features of the LISA signal which modify somewhat our final
answer to Question~1 (see final toy model below).

\subsubsection{Second toy model}

In our toy second model, we modify the single frequency signal by
postulating two pairs of unknown amplitudes and phases for two
different {\it a priori} known frequencies, satisfying $f_2 \gg f_1$,
which modulate an otherwise constant signal:
\begin{eqnarray}
h(t)=&&c_0 + s_1 \sin(2\pi f_1 t) + c_1 \cos(2\pi
f_1 t)\nonumber\\
&+& s_{10} \sin(2\pi f_2 t) + c_{10} \cos(2\pi f_2 t).
\label{e:simple2}
\end{eqnarray}

The number of unknowns in this model is five:
$c_{0},s_{1},c_{1},s_{10}$ and $c_{10}$ are the coefficients of the
functions $1$, $\sin(2\pi f_1 t)$, $\cos(2\pi f_1 t)$, $\sin(2\pi f_2
t)$, and $\cos(2\pi f_2 t)$. Again, we assume that the noise decreases
quickly with time before merger, at $t=0$. This model is designed to
answer our Question~2 above.  In this case, we find that parameter
errors are correlated only with unique frequency components and the
constant signal, all the way to $\tf \gsim 0.1 f_{2}^{-1}$. The model
thus demonstrates how components associated with very different
variation timescales can decouple from each other. Moreover, as for
the first toy model, we find that marginalized parameter errors
effectively stop improving past a finite time before merger
(Question~1), which is simply related to their respective
frequencies. As a result, a nonzero residual error remains again, even
though the signal-to-noise ratio continuously increases near merger.

\subsubsection{Final toy model}

In our final toy model, we insert a few additional features essential
to a realistic LISA data-stream. Firstly, we assume 5 low-frequency
harmonics, $1$, $\sin(2\pi f_1 t)$, $\cos(2\pi f_1 t)$, $\sin(4\pi f_1
t)$, $(\sin 4\pi f_1 t)$, with unknown amplitudes. We also include a
high frequency carrier signal with known frequency, $f_2\gg f_1$, but
unknown amplitudes in $\sin(2\pi f_2 t)$ and $\cos(2\pi f_2 t)$, for a
total of seven free parameters. Secondly, we note that the LISA system
is equivalent to two orthogonal arm interferometers with both
detectors measuring polarization phases simultaneously (which
correspond to the real and imaginary parts of the amplitude
modulation, \S~\ref{s:HMDFisher}).  Therefore, the signal is comprised
of 4 simultaneous data-streams. We incorporate this feature by assuming
4 measurements (i.e. 4 corresponding Fisher matrices) of the signal
with 4 given phase shifts
($\varphi^{s_{1}}_{i},\varphi^{c_{1}}_{i},\varphi^{s_{2}}_{i},\varphi^{c_{2}}_{i}$;
$1\leq i\leq 4$) so that
\begin{eqnarray}
h(t) &=& c_0 + s_1 \sin(2\pi f_1 t + \varphi^{s_1}_{i}) + c_1 \cos(2\pi f_1 t + \varphi^{c_1}_{i})\nonumber\\
&&+ s_2 \sin(2\pi f_1 t + \varphi^{s_2}_{i}) + c_2 \cos(2\pi f_1 t + \varphi^{c_2}_{i}) \nonumber\\
&&+ s_{10} \sin(2\pi f_2 t) + c_{10} \cos(2\pi f_2 t),
\label{e:simple4}
\end{eqnarray}

In this case, we find that 4 principal components improve quickly at
late times. As in our second toy model, the high frequency parameters
decouple from the slow frequency ones, except at very late times when
$\tf\gsim 0.1f_2^{-1}$.

This final toy model allows us to answers all of Questions~1-4 as
follows.
\begin{itemize}
\item Answer~1: Four out of 5 slow principal components of the error
ellipsoid are quickly improving with time, while one of them stops
improving at $\tf\lsim 0.1 f_{1}^{-1}$. Therefore, any parameter with
a large projection along this one poor principal component will stop
improving, while parameters nearly orthogonal to it will keep
improving quickly. Thus, both statements made in relation to
Question~1 above can in fact be correct, depending on the connection
between a given parameter and the poor principal component. Typically,
we expect marginalized parameter uncertainties to evolve as
$(S/N)^{-1}$ for $\tf\gsim 0.1 f_{1}^{-1}$. For smaller $\tf$ values,
closer to merger, they would continue to improve, albeit with a
shallower slope.

\item Answer~2: We find that the introduction of additional high
frequency components does not change the evolution of original
parameter estimations as long as the time-to-merger is larger than a
fraction of the time period of the additional high frequency
components.

\item Answer~3: As the signal-to-noise ratio increases quickly at late
times, rapidly evolving parameter error combinations are given by the
principal components of the error ellipsoid corresponding to the final
situation at merger. With 4 data-streams, there are 4 such best
principal components. Analogously, for the LISA amplitude modulation
given by eq.~(\ref{e:amplitude_mod}), we expect that the 2 polarization phases for
the 2 beam patterns at ISCO can be best determined:
$(1+\cos^2\theta_{NL})F^{I,II}_{+}(\Omega_{\rm ISCO})$ and
$\cos\theta_{NL}F^{I,II}_{\times}(\Omega_{\rm ISCO})$. (In terms of
ecliptic angular variables, these are the real and imaginary parts of
the combination given by eq.~(\ref{e:hmod}).)

\item Answer~4: The widths of error distributions for slow parameters
do not change significantly as long as $\tf\gsim 0.1
f_{1}^{-1}$. During this final stretch of time before merger, however,
one of the principal components stops improving and the major axis of
the error ellipsoid freezes. Since the physical parameters can be
considered to be randomly oriented with respect to the ellipsoid axes,
distributions of marginalized errors suddenly start broadening for
$\tf\lsim 0.1 f_{1}^{-1}$, with a worst case relative orientation
leading to very little improvement and a best case relative
orientation corresponding to a scaling with $(S/N)^{-1}$.
\end{itemize}

\subsection{Implications for LISA}

These simple toy models offer a general interpretation of the time
dependence of LISA's parameter estimation errors for source
localization. The LISA data stream is described by $N_{p1}=5$ physical
parameters, ${\bm p}_{\slow}$, which are not the harmonic coefficients
themselves but determine these coefficients, $g_{j}$ (or conversely,
the mode expansion coefficients $g_{j}$ determine the physical
parameters ${\bm p}_{\slow}$; see \S~\ref{s:HMD}). Neglecting Doppler
phase and spin precession effects, $2J_{\max}+1=9$ modes determine the
signal by eqs. (\ref{e:hmodres},\ref{e:g_jres}). In principle, any
$N_{p1}=5$ of the $g_{j}$ mode amplitudes uniquely determine the
physical parameters, ${\bm p}_{\slow}$. However, in the presence of
noise, each of these modes are uncertain and the combination of all
modes helps in reducing the estimation errors of the ${\bm p}_{\slow}$
variables.

The key implication of our toy models for LISA is that the estimation
of low frequency $g_{j}$ modes with low $|j|$ are effectively
decoupled from the high frequency signal, unless the merger is within
$\sim 0.1$ times the cycle time of the fast-oscillating signal.  We
have shown that the HMD of the orbital modulation consists purely of
low-order harmonics, with $|j|\leq 4$. In comparison, the high
frequency GW phase has a much higher frequency, corresponding to
$j>1000$, and this high frequency signal's cycle time is greater than
the time to merger throughout $t\gsim t_{\rm ISCO}$. Hence, physical
parameters ${\bm p}_{\slow}$ will remain decoupled from parameters
${\bm p}_{\fast}$, all the way to ISCO. This finding is independent of
details of the waveform and the modulation, in agreement with the
results of Ref.~\cite{bbw05} which show that decoupling occurs
independently of the details of the $\hc(t)$ signals, including the
modified inspiral waveforms of alternative theories of gravity. In
terms of post-Newtonian expansions, only terms above second order have
cycle times as large as the cycle time of the amplitude
modulation. These terms are responsible for the small
cross-correlations of the two sets of parameters found by
Ref.~\cite{hug02}.

We have not considered spin precession effects, but Vecchio~\cite{vec04} and
Lang~\&~Hughes~\cite{lh06} find that spin precession effects can help improve the
final localization errors by a factor of $\sim 3$. Spin precession
cycle times decrease continuously, become of order a few days or less
during the last week prior to merger, and of order hours during the
last day of inspiral. Therefore, according to our simple models, we
expect spin precession effects to improve the source parameter
estimation errors especially during the final two weeks before
ISCO. During that period of time, in the absence of spin effects,
parameter uncertainties (especially the sky position major axis and
the luminosity distance) cease to improve when using only the
amplitude modulation.

The best-determined parameters at ISCO are, approximately, the
independent detector outputs at ISCO, i.e. the real and imaginary
parts of $h^{\rm I,II}_{1}(p_1)$:
$\dL^{-1}(1+\cos^2\theta_{NL})F^{I,II}_{+}(\Omega)$ and
$\dL^{-1}\cos\theta_{NL}F^{I,II}_{\times}(\Omega)$ (see
Appendix~\ref{app:simple:best}).  These are the 4 independent
combinations of 5 physical parameters $p_1$ which correspond to the
eigenvectors of the error covariance matrix following the steep
evolution $\propto (S/N)^{-1}$ all the way to ISCO.  We refer to the
fifth independent combination, which is orthogonal to these best
eigenvectors, the ``worst'' eigenvector, since for this combination,
the evolution ceases to improve as $(S/N)^{-1}$ within $\sim
0.1\times(\rm amplitude~modulation~cycle~time)$ of merger.  It is
straightforward to obtain this worst combination explicitly by using
the 4 other eigenvectors and Gram-Schmidt orthogonalization (but we
have not done this in practice).  Since the highest frequency harmonic
of the slow modulation is for $j=4$, the corresponding cycle time is
$\yr/4$. Thus, we expect errors will stop improving roughly $1$--$2$
weeks prior to merger. Distributions of errors will quickly broaden
during these final stages of observation before ISCO. Simply scaling
errors with $(S/N)^{-1}$, as in the angle-averaged formalism
(e.g. \cite{bbw05,koc06}), is acceptable if one studies the
evolution of parameter errors at $\tf\gsim 2$ weeks, or if one only
focuses on the best case parameter combinations. In general, the
exponent in the $(S/N)$ scaling decreases as one approaches merger
time depending on how close the particular combination of angles
considered is to the worst combination.

Our findings for the eccentricity evolution of LISA's sky localization
error ellipsoid can also be understood with the simple toy models. In
fact, we found this behavior to be expected for any model signal with
relative instantaneous signal amplitude increasing quickly with time,
e.g.  $t^{-\alpha}$, $\alpha \gsim 2$. In this case, the principal axes
of the general parameter error ellipsoid separate near $\tf=0$. There
are a limited number of principal errors which rapidly decrease to
zero near $\tf=0$, while others ``freeze out'' at a time related to a
fraction of the cycle time of the particular waveform ($\sim 0.1
T_{\rm cycle}$ if $\ti> T_{\rm cycle}$). For LISA, there are 5
variable parameters, ${\bm
p}_{\slow}=(\dL,\theta_N,\phi_N,\theta_L,\phi_L)$, and estimation
uncertainties of 4 combinations of these parameters,
$\dL^{-1}(1+\cos^2\theta_{NL})F^{I,II}_{+}(\Omega),\dL^{-1}\cos\theta_{NL}F^{I,II}_{\times}(\Omega)$,
improve quickly with $(S/N)^{-1}$. These combinations correspond to
the best 4 principal axes of the 5-dimensional error ellipsoid. The
remaining $5^{\rm th}$ principal axis does not improve as
$(S/N)^{-1}$, but rather stops improving at a fraction of the last
modulation cycle time. The two dimensional sky position error
ellipsoid is the projection of the general 5-dimensional error
ellipsoid on the $(\theta_N,\phi_N)$ plane. This plane will generally
not be aligned with the principal axes of the 5-dimensional
ellipsoid. In a typical case, therefore, there will be a nonzero
projection on the worst principal component and the sky position
ellipsoid will stop shrinking along the worst principal
component. This explains why the major axis, $2a$, ceases to improve
and the eccentricity increases close to merger.

According to this argument, it is somewhat surprising to find that the
minor axis, $2b$, can stop improving much before ISCO.
Figure~\ref{f:errortau} shows that this happens in the worst $10\%$ of
all cases for randomly chosen source angular parameters. The reason
for this is that, in some cases, not all rapidly improving ``best''
principal components have a small absolute error at ISCO.  For
example, consider an edge-on binary inspiral ($\cos\theta_{NL}\approx
0$). Since two of the quickly improving parameters are simply
proportional to $\cos \theta_{NL}$, the errors will be very large for
these parameters. Thus, depending on the relative orientation of the
detector and the source at ISCO, there can be large absolute errors in
some cases even for the best combinations of parameters. In short,
both axes of the sky position error ellipsoid can stop improving at
late times in those cases when LISA is oriented in its least favorable
direction at ISCO.

\section{Conclusions}\label{s:conclusions}

We have developed a new harmonic mode decomposition (HMD) method to
study the feasibility of using LISA inspiral signals to locate
coalescing SMBH binaries in the sky, as the mergers proceed.
According to our extensive HMD survey of potential LISA sources, it
will be possible to trigger large field-of-view searches for prompt
electromagnetic counterparts during the final stages of inspiral and
coalescence. Our results indicate, for instance, that
for a typical $z\sim 1$ merger event with total mass $M\sim
10^{5}-10^{7}\Msun$, a 10-day advance
notice will be available to localize the source to within a $10\deg^{2}$ region of the sky.
The advance notice to localize the source to a 10 times smaller area of  $1\deg^{2}$
is $<1$ day for the typical event, suggesting that
a wide--field instrument of the LSST class, with
a $10 \deg^2$ field-of-view, may offer significant advantages over a
smaller, $1 \deg^2$ field-of-view instrument for observational efforts
to catch prompt electromagnetic counterparts to SMBH binary inspirals.

The robust identification of such electromagnetic counterparts would
have multiple applications, from an alternative method to measure
cosmological parameters to precise measurements of merger geometries
in relation to host galaxy properties \cite{hh05,koc06}. If such
electromagnetic counterpart searches can be implemented effectively
and successfully, LISA could become an extremely valuable instrument
for astrophysics and cosmology, beyond the original general
relativistic measurement goals.  Given the advance warning time
capabilities established here, effective strategies for
electromagnetic counterpart searches, including the concept of
partially dedicating a $\gsim 10 \deg^2$ field-of-view fast survey
instrument of the LSST class, are considered in detail in a separate
investigation \cite{paper2}.

\acknowledgments
 We thank Samuel Finn and Shane Larson for influential early
 discussions on this problem and Scott Hughes and Tom Prince for
 valuable comments which improved our manuscript.
 BK acknowledges support from a Smithsonian Astrophysical
 Observatory Predoctoral Fellowship and from \"Oveges J\'ozsef
 Fellowship. ZH acknowledges partial support by NASA through grant
NNG04GI88G, by the NSF through grant AST-0307291, and
by the Hungarian Ministry of Education through a Gy\"orgy
B\'ek\'esy Fellowship. KM was  supported in part by the National
Science Foundation under Grant No. PHY05-51164 (at KITP). Z.F.
acknowledges support from OTKA through grant nos. T037548,
T047042, and T047244.

\appendix

\section{Simple Toy Models}
\label{app:simple}

\subsection{Single Frequency Model}\label{app:simple:1}

First, let us consider the following simple model with two unknowns,
$c_0$ and $c_1$,
\begin{equation}
\label{e:a:simple1} h(t)=c_0 + c_1\cos(2\pi f_{\oplus}t),
\end{equation}
where $f_{\oplus}\equiv \yr^{-1}$ is fixed and assumed to be known
prior to the observation. We call $t$ the ``look--back time'' before
merger. Let us assume that the relative noise continuously decreases
during the observation and that the differential squared
signal-to-noise ratio (without modulation) is given by
$\sigma^{-2}(t)=t^{-2}$ in eq.~(\ref{e:Gamma0}).  Here $t=0$ is a
proxy for the ``merger''. Close to merger, the signal-to-noise ratio
accumulates very rapidly. We assume that $h(t)$ is measured in the
time interval $\ti\geq t\geq \tf$, where $\ti$ is the start of
observation, $\tf$ is the end of observation (i.e.  $x=t_{\rm
merger}-t$, $x_{\min}=\tf$, and $x_{\max}=\ti$ in
eq.~[\ref{e:Gamma0}]). We fix $\ti$ and examine the dependence of
parameter estimation errors as a function of $\tf$, assuming $\tf \ll
\ti$.

Note that, for the signal (\ref{e:a:simple1}), the fiducial values
$(c_{0},c_{1})$ drop out when calculating the RMS parameter errors
$\Delta c_{0}$ and $\Delta c_{1}$ using eq.~(\ref{e:Gamma0}). More
generally, this is true for any signal which is a linear combination
of the unknown parameters. All our toy models will have this property
and the results presented in this section will be general in that
respect.

\begin{figure}
\centering{
\mbox{\includegraphics[width=8.5cm]{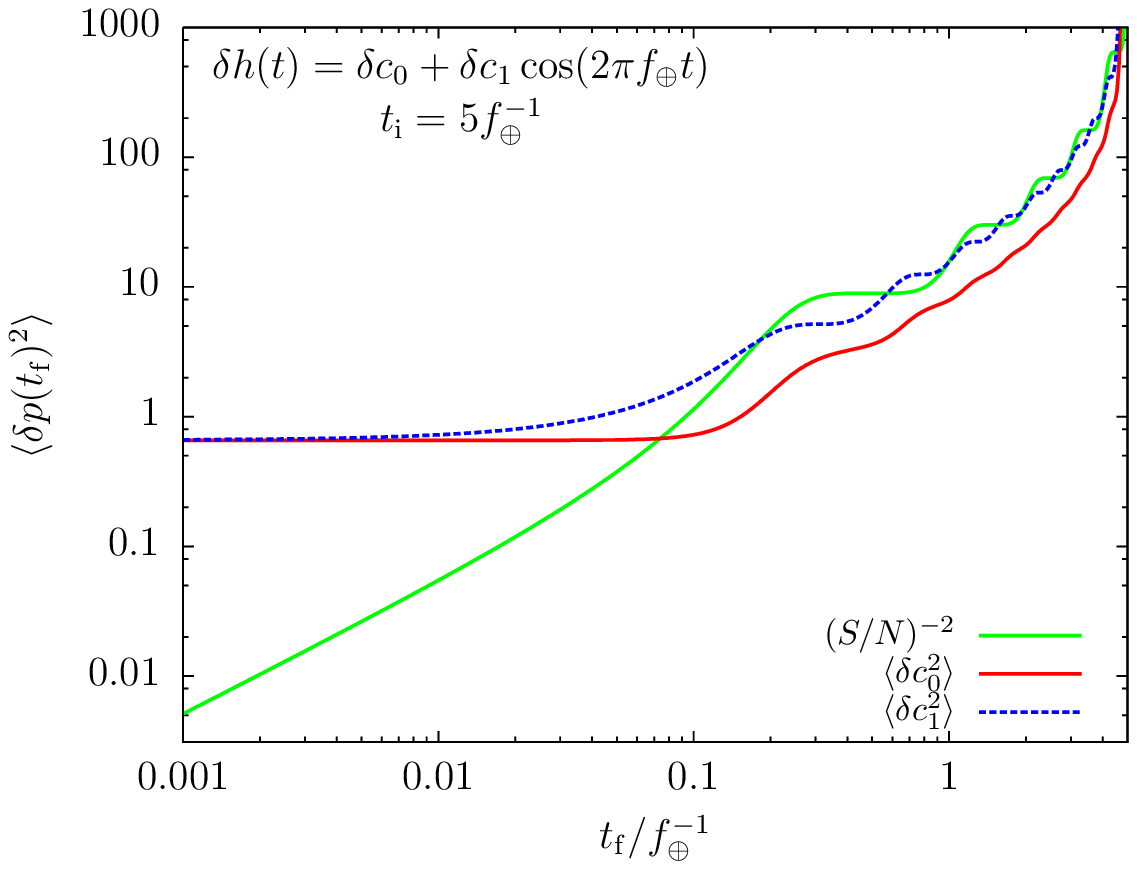}}\\ \bigskip
\mbox{\includegraphics[width=8.5cm]{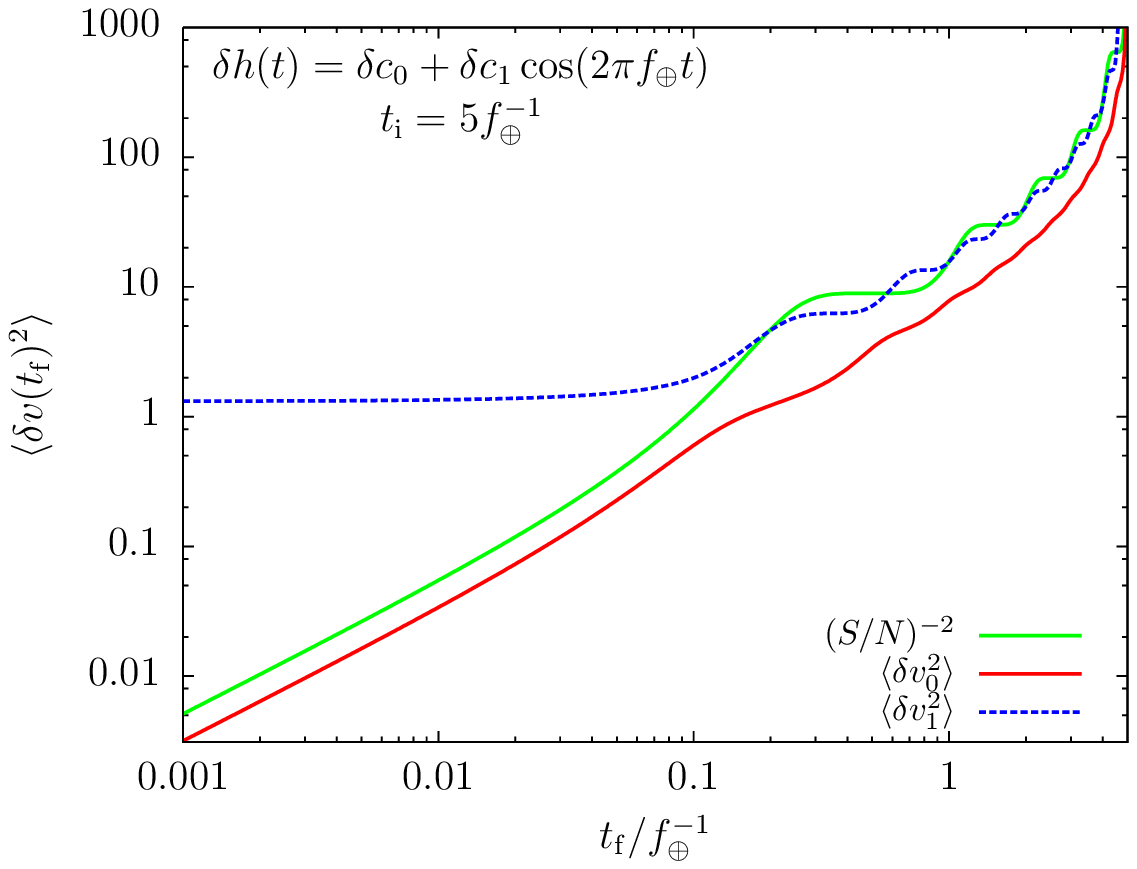}}
\caption{\label{f:simple1} Marginalized parameter errors {\it (top)}
and principal errors {\it (bottom)} for the single frequency
model. The green curve shows the scaling with inverse squared
signal-to-noise ratio, $(S/N)^{-2}$, for reference on both plots. A
total observation of $\ti=5\yr$ is assumed. Marginalized errors follow
the signal-to-noise ratio for large $\tf$, but they stop improving
within $\tf<\tc\sim 0.1 \yr$ from merger. Only one eigenvalue scales
with the signal-to-noise ratio near merger.\smallskip}}
\end{figure}

First, let us substitute (\ref{e:a:simple1}) in (\ref{e:papb}) and
(\ref{e:Gamma0}), and evaluate the expected covariance matrix
numerically. Figure~\ref{f:simple1} displays the time dependence of
marginalized parameter errors and principal errors. The plots show
that the parameter errors all decrease with the signal to noise ratio
when the look--back time before merger is large. However if the end of
the observation is within a certain critical time to merger, $\tf <
\tc$, only one principal component follows the signal-to-noise
ratio. Figure~\ref{f:simple1} shows that $\tc\sim 0.1\yr$. The start
of the observation in Figure~\ref{f:simple1} was fixed at $\ti=5\yr$.

\begin{figure}
\centering{ \mbox{\includegraphics[width=8.5cm]{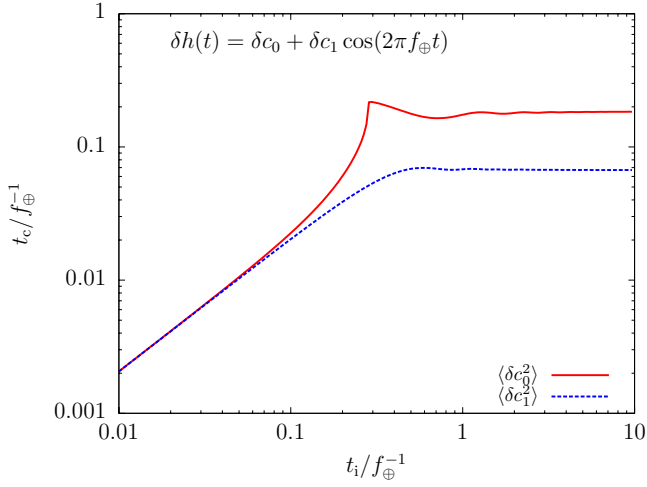}}}
 \caption{\label{f:simple1b} Critical look--back time, $\tc$, at which
parameter errors stop improving. Here $\tc$ is defined as the time at
which marginalized squared errors are within a factor of 2 of their
final values for the first time.\bigskip}
\end{figure}

It is also interesting to examine what happens for general total
observation times, do errors stop improving within some time $\tc$
before merger? If yes, how does $\tc$ depend on the two timescales
$\ti$ and $f_{\oplus}^{-1}$? We examine this question numerically,
substituting (\ref{e:a:simple1}) in (\ref{e:papb}) and
(\ref{e:Gamma0}) and now varying both $\tf/f_{\oplus}^{-1}$ and
$\ti/f_{\oplus}^{-1}$. Let us define the critical end-of-observation,
$\tc$, as the time when the marginalized squared parameter error is
first within a factor of 2 of its final value. Figure~\ref{f:simple1b}
plots the result for the two parameters. Figure~\ref{f:simple1b} shows
that $\tc$ is determined by $f_{\oplus}^{-1}$ for large $\ti$, but
becomes $\ti$-dependent for lower $\ti$ values. In the limit $\ti\ll
f_{\oplus}^{-1}$, the critical look--back time is independent of
$f_{\oplus}^{-1}$, it becomes a constant fraction of $\ti$.

Note that, in the limit of an observation extending up to merger, at
$t=0$, the signal becomes $h(0)=c_{0}+c_{1}$ and it has infinite
instantaneous signal-to-noise ratio. Therefore, this is the best
combination of parameters for which the scaling of errors can follow
$(S/N)^{-1}$ all the way to $t=0$. The worst combination is
$c_{0}-c_{1}$, which stops improving before $t=0$.

For this simple model, the origin of these features can be understood
by analyzing the principal errors and the marginalized errors in the
error covariance matrix. For this purpose, we present an analytical
algebraic solution to this problem. To simplify the equations, let us
set the time-scale to $f_{\oplus}^{-1}/(2\pi)$. In this case the
Fisher matrix (\ref{e:Gamma0}) is
\begin{equation}
\label{e:F1}
\Gamma_{ij}(\tf,\ti)=\left(
  \begin{array}{cc}
    \int_{\tf}^{\ti} t^{-2} \D t & \int_{\tf}^{\ti} \cos(t) t^{-2} \D t \\
    \int_{\tf}^{\ti} \cos(t) t^{-2} \D t & \int_{\tf}^{\ti} \cos^2(t) t^{-2} \D t \\
  \end{array}
\right).
\end{equation}
The integrals can be evaluated analytically,
\begin{equation}
\label{e:Gammasimple1}
\Gamma_{ij}(\tf,\ti)=\left.\left(
  \begin{array}{cc}
    \frac{1}{t} & \frac{\cos(t)}{t} + \Si(t) \\
    \frac{\cos(t)}{t} + \Si(t) & \frac{\cos(2t)+1}{2t} + \Si(2t)\\
  \end{array}
\right)\right]^{\tf}_{\ti},
\end{equation}
where $\Si(x)=\int_0^x \frac{\sin(x)}{x}\D x$ is the sine integral.

In the next two subsections, we find the limiting behavior of
marginalized and principal parameter errors in two different limits:
$f_{\oplus}^{-1}\ll \ti$ and $\ti \ll f_{\oplus}^{-1}$, respectively.

\subsubsection{Long Observations $(f_{\oplus}^{-1}\ll \ti)$}

Here, we assume that the signal has been measured for a very long
total time and we concentrate on the effects of changing the end of
the observation time, $\tf$, near merger.  Therefore, we take the
limit $\ti\rightarrow \infty$, for which
\begin{eqnarray}
\Gamma_{ij}(\tf)&=&\left(
  \begin{array}{cc}
    \frac{1}{\tf} & \frac{\cos(\tf)}{\tf} + \Si(\tf) \\
    \frac{\cos(\tf)}{\tf} + \Si(\tf) & \frac{\cos(2\tf)+1}{2\tf} + \Si(2\tf)\\
  \end{array}
\right)\nonumber\\&&- \left(
  \begin{array}{cc}
    0 & \pi/2 \\
    \pi/2 & \pi/2\\
  \end{array}
\right).\label{e:Gammasimple1a}
\end{eqnarray}
We consider the case of a total observation time which is not
negligible compared to a cycle time, $f_{\oplus}^{-1}$, i.e.  $\tf\ll
\ti$. We next examine two possible cases, $f_{\oplus}^{-1}\ll \tf \ll
\ti$ and $\tf\ll f_{\oplus}^{-1} \ll \ti$, separately.

First let us assume that the merger is still far away in time in units
of a cycle period $(f_{\oplus}\ll \tf \ll \ti)$. We substitute
(\ref{e:Gammasimple1a}) in (\ref{e:papb}) and expand
$\Gamma^{-1}(\tf)$ into a $\tf^{-1}$ series:
\begin{equation}
\label{e:cov2} (\Gamma^{-1})_{ij}\approx\frac{\tf}{1-\frac{\sin(2\tf)}{2\tf}+\frac{\cos(2\tf)-1}{\tf^2}}\left(
  \begin{array}{cc}
    1-\frac{\sin(2\tf)}{2\tf} & \frac{2\sin(\tf)}{\tf}\\
    \frac{2\sin(\tf)}{\tf} & 2\\
  \end{array}
\right).
\end{equation}
Equation~(\ref{e:cov2}) gives the large $\tf$ behavior of marginalized
errors and correlations, which can be compared to
Figure~\ref{f:simple1} in the appropriate regime, $\tf>1\yr$. In this
case, to leading order, {\it all of the squared errors} scale with
$\tf$, which is the scaling of the inverse squared signal-to-noise
ratio, $(S/N)^{-2}$, for our noise model.

Next, let us examine the case when the end-of-observation time is
close to merger, i.e. $\tf\ll f_{\oplus} \ll \ti$. Now, taking the
inverse of the matrix and expanding into a $\tf$ series around $\tf=0$
gives
\begin{equation}
\label{e:cov1} (\Gamma^{-1})_{ij}\approx\frac{2}{\pi}\left(
  \begin{array}{cc}
    1 + \frac{2}{3\pi}\tf^3 & -1 + \frac{\tf^2}{2\pi^2} \\
    -1 + \frac{\tf^2}{2\pi^2} & 1 + \frac{\pi}{2}\tf \\
  \end{array}
\right),
\end{equation}
which gives the short timescale behavior of marginalized errors and
correlations. The eigenvalues of $\Gamma^{-1}$ define the squared
length of the individual principal axes of the parameter error
ellipsoid, in this case
\begin{equation}
\label{e:eig1}
\left(
  \begin{array}{c}
    \langle\delta v^2_0\rangle \\
    \langle\delta v^2_1\rangle \\
  \end{array}
\right)\approx \left(
  \begin{array}{c}
    \frac{\tf}{2}+\frac{3\pi}{16}\tf^2\\
    \frac{4}{\pi} + \frac{\tf}{2} + \left(\frac{5\pi}{16}-\frac{2}{\pi}\right)\tf^2\\
  \end{array}
\right).
\end{equation}
Note that, in eqs.~(\ref{e:F1})-(\ref{e:eig1}), time is measured in
units of $f_{\oplus}^{-1}/(2\pi)$. In full units, the squared
marginalized parameter errors (i.e. diagonal elements) of
(\ref{e:cov1}) become
\begin{equation}
\label{e:err2}
\left(
  \begin{array}{c}
    \langle\delta c^2_0\rangle \\
    \langle\delta c^2_1\rangle \\
  \end{array}
\right)= \left(
  \begin{array}{c}
    \frac{2}{\pi}\left[ 1 + \left(\frac{\tf}{\sqrt[3]{3/(16\pi^2)}f_{\oplus}^{-1}}\right)^3\right] \\
    \frac{2}{\pi}\left[1 + \frac{\tf}{\frac{1}{\pi^2}f_{\oplus}^{-1}}\right] \\
  \end{array}
\right).
\end{equation}
For the eigenvalues (\ref{e:eig1}), we get
\begin{equation}
\label{e:eig2}
\left(
  \begin{array}{c}
    \langle\delta v^2_0\rangle \\
        \langle\delta v^2_1\rangle \\
  \end{array}
\right)= \left(
  \begin{array}{c}
    \tf/\left({\frac{1}{\pi}f_{\oplus}^{-1}}\right)\\
    \frac{4}{\pi}\left[1 + \frac{\tf}{\frac{4}{\pi^2}f_{\oplus}^{-1}}\right]\\
  \end{array}
\right).
\end{equation}
Equation (\ref{e:err2}) implies that the evolution of the marginalized
squared error on $c_0$ is very flat for small $\tf$, when the second
term is negligible, i.e. $\tf\ll
\sqrt[3]{\frac{3}{16\pi^2}}f_{\oplus}^{-1}=0.267\yr$, then rises
steeply ($\propto\tf^3$). The marginalized squared $c_1$ error is also
constant near merger, for $\tf\ll
\frac{1}{\pi^2}f_{\oplus}^{-1}\approx 0.1\yr$, and it increases
$\propto\tf\propsim (S/N)^{-2}$ for larger $\tf$.
Equation~(\ref{e:eig2}) shows that one of the principal errors has a
very different time-evolution: it has no constant term proportional to
$\tf^0$. Therefore the semi-minor axis of the error ellipsoid can
decrease continuously with the signal to noise ratio. On the other
hand, the semi-major axis becomes constant for $\tf\ll
\frac{4}{\pi^2}f_{\oplus}^{-1}=0.4\yr$. Since the marginalized errors
are nontrivial linear combinations of the principal errors, the
constant principal error carries over to both marginalized errors and
dominates their evolution.  All of these findings are in excellent
agreement with the numerical results shown in Fig.~\ref{f:simple1} for
$\tf\ll 1\yr$ and in Fig.~\ref{f:simple2} for $\ti/f_{\oplus}^{-1}>
1$.

It is worth emphasizing that, even if the total observation time had
been infinite, $\ti \to \infty$, the parameters could {\it not} have
been estimated to infinite precision in this model. It is not very
surprising if one recalls that in this model we defined errors to be
infinitely large at infinitely early times ($\sigma^2(t) \propto
t^2$). For stationary noise, the contribution of the last cycle to the
resultant RMS estimation error for a total observation of $N_{\rm
cyc}$ cycles is $1/\sqrt{N_{\rm cyc}}$. In contrast, {\it rather than
the total number of cycles, the typical error during the last cycle
dominates the determination of noise}, for the particular noise model
used here.

The main conclusion from this toy model analysis is that errors stop
improving close to merger, at $\tc\sim 0.1f_{\oplus}^{-1}$. It can be
extended to more general noise models, with
$\sigma^{-2}(t)=t^{-\alpha}$ and $\alpha\neq 2$. Repeating the
calculations for larger $\alpha$ values, we find that parameter
estimation errors become more and more insensitive to very early
times, $\tf\ll t\sim \ti$, and that marginalized parameter estimation
errors cease to improve at some $\tc$, which is now an
$\alpha$-dependent fraction of a single cycle time before merger. For
$\alpha>2$, we find that errors increase more abruptly at
$\tf\gsim\tc$, which is consistent with the signal-to-noise ratio
being a steeper function of time. On the other hand, for lower
$\alpha$ values, parameter estimation errors become more and more
sensitive to very early times, $\tf\ll t\sim \ti$. In this case, the
marginalized parameter estimation errors are again very slowly
changing for $0\sim\tf<\tc$, but the approximate time $\tc$ at which
parameter errors stop decreasing will be primarily determined by
$\ti$, rather than by the cycle period $f_{\oplus}^{-1}$. The
transition at $\tf\gsim \tc$ is not as abrupt, but extends to several
cycles. The $\alpha=0$ case corresponds to a stationary instantaneous
signal-to-noise ratio, with errors scaling slowly as
$1/\sqrt{\ti-\tf}$.  This case is irrelevant to LISA inspiral signals,
which have $\alpha\sim 2$ to a good approximation for $1$day $ <t<\ti$
in the relevant range of SMBH masses.

\subsubsection{Short observations $(\ti \ll f_{\oplus}^{-1})$}

Let us now examine the opposite limiting case, where the start of
observation time is already within the final cycle before merger. This
is relevant to LISA signals, since the observation time of SMBH
inspirals is often below a full year, especially for $(1+z)\,M\geq 4
\times 10^6 \Msun$.

We again restrict ourselves to the case with a total observation time
that is non-negligible, i.e. $\tf\ll \ti$. Using time units of
$f_{\oplus}^{-1}/(2\pi)$, expanding (\ref{e:Gammasimple1}) into a
series of both $\ti$ and $\tf/\ti$, we get
\begin{eqnarray}
(\Gamma^{-1})_{ij}&\approx&\frac{120}{\ti^3(10-\ti^2)}
\left[ \left(
  \begin{array}{cc}
    1 & -1 \\
    -1& 1  \\
  \end{array}
\right)\right.\nonumber\\&&\left. +\frac{\tf}{\ti}\left(
  \begin{array}{cc}
    \frac{30-10\ti^2}{10-\ti^2} & -\frac{30-5\ti^2}{10-\ti^2} \\
    -\frac{30-5\ti^2}{10-\ti^2} & \frac{30-\frac{5}{3}\ti^2}{10-\ti^2}\\
  \end{array}
\right)\right]\label{e:cov3}
\end{eqnarray}
Equation (\ref{e:cov3}) gives the parameter estimation covariance
during the final stages of observation before merger for small total
observation times. In this case, the final errors strongly depend on
the total observation time.  The errors reach their final values when
the second term becomes negligible in eq.~(\ref{e:cov3}). To leading
order, this happens at $\tc\sim \ti/3$ for both parameters,
independently of the cycle time, $f_{\oplus}^{-1}$.
Equation~(\ref{e:cov3}) approximates well the $\ti$ dependence of
$\tc$ shown in Fig.~\ref{f:simple1b} for $\ti/f_{\oplus}^{-1}<0.2$

\subsection{Double Frequency Model}\label{app:simple:2}

Now consider a more elaborate model with five unknowns $c_0$, $s_1$,
$c_1$, $s_{10}$, and $c_{10}$:
\begin{eqnarray}
\label{e:a:simple2} h(t)&=&c_0 + s_1 \sin(2\pi f_1 t) + c_1 \cos(2\pi
f_1 t)\nonumber\\ &&+ s_{10} \sin(2\pi f_2 t) + c_{10} \cos(2\pi f_2 t).
\end{eqnarray}
Here, the signal is comprised of two different characteristic
frequencies, $f_1$ and $f_2$, for which we assume $f_1\ll
f_2$. Moreover we assume that $f_1$ and $f_2$ are fixed and known
prior to the measurement, e.g. we take $f_1\equiv 1\yr^{-1}$ and
$f_{2}\equiv 10\yr^{-1}$. We again assume an observation in the
look--back time interval $\ti\geq t\geq \tf$ and take the average
instantaneous signal-to-noise ratio to increase as
$\sigma(t)^{-2}=t^{-2}$.

\begin{figure}
\centering{
\mbox{\includegraphics[width=8.5cm]{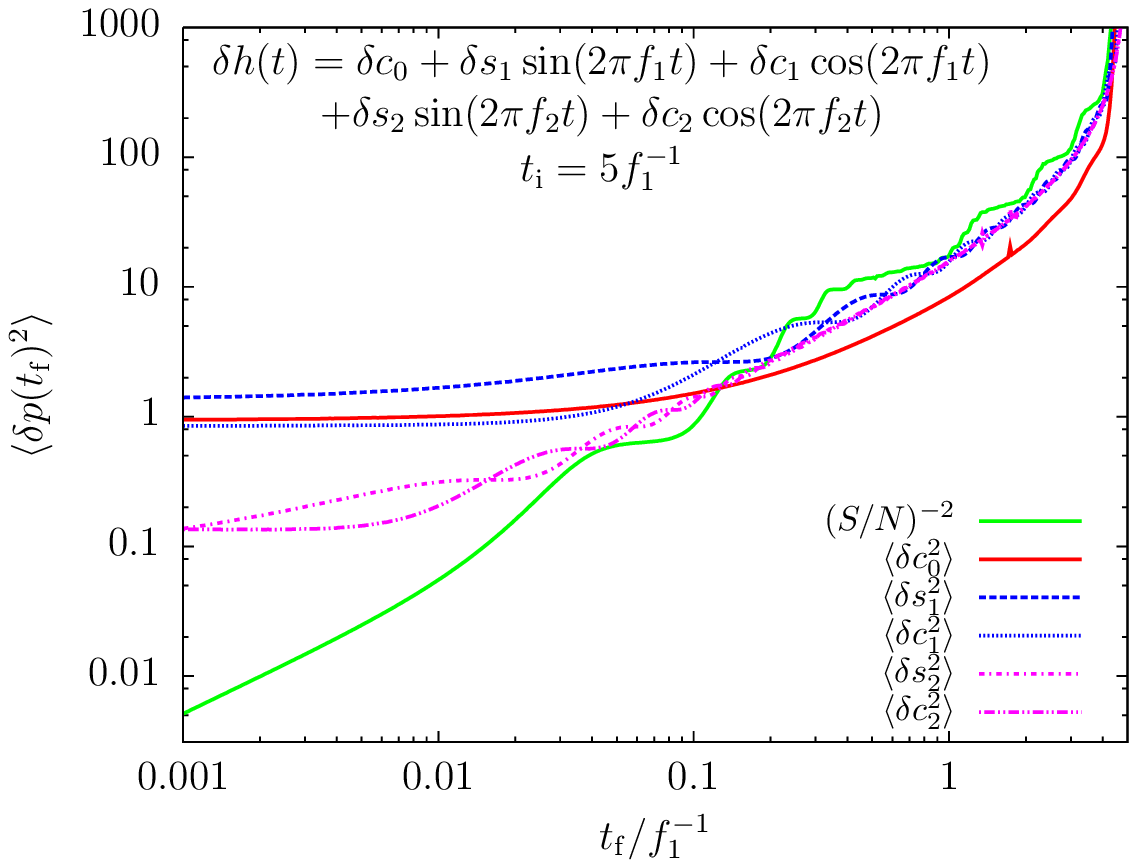}}\\ \bigskip
\mbox{\includegraphics[width=8.5cm]{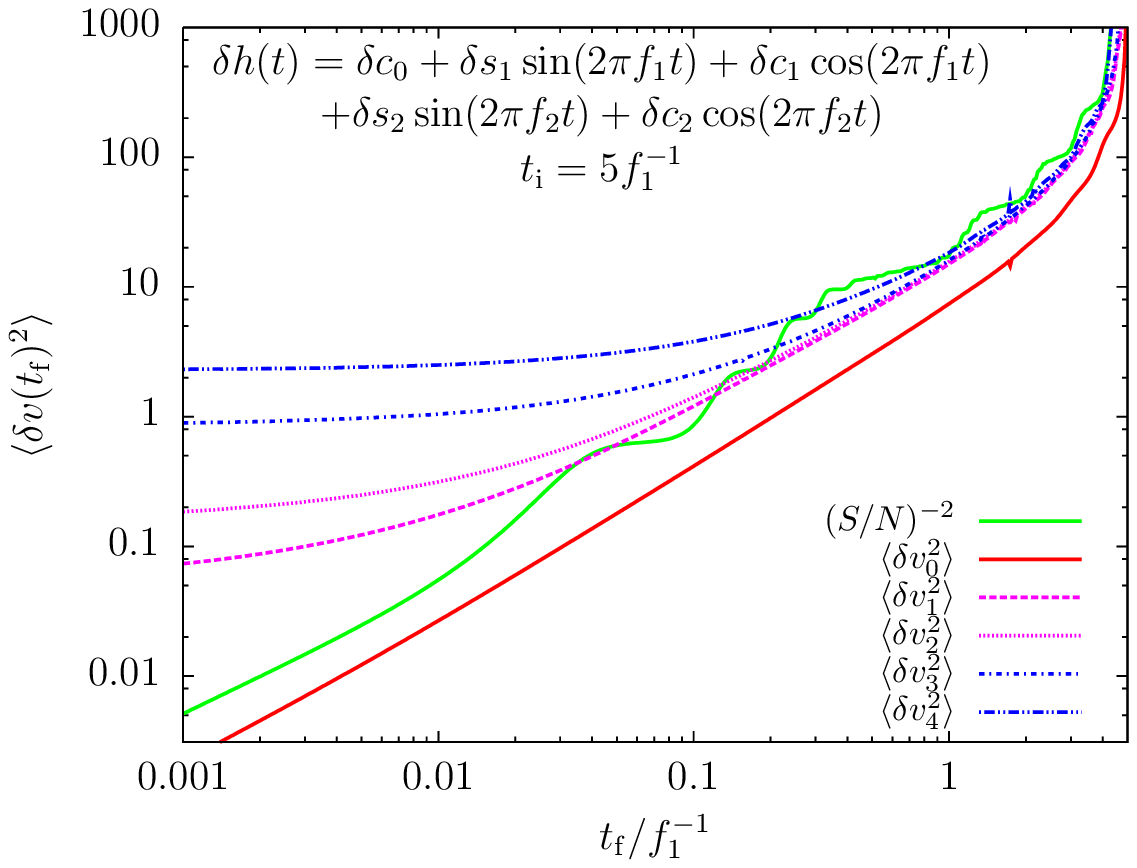}}
\caption{\label{f:simple2} Marginalized parameter errors (top) and
principal errors (bottom) for the double frequency model.  The green
curve shows the scaling with $(S/N)^{-2}$ for reference on both plots. A
total observation time $\ti=5\yr$ is assumed. Marginalized errors
follow the signal-to-noise ratio for large $\tf$ values, but they stop
improving after $\tf\lsim 0.1 f$, for both frequencies. By comparing
the two plots, it is clear that high frequency component errors
decouple and that they are determined by two corresponding eigenvalues
in the bottom panel.}}
\end{figure}

Let us substitute in (\ref{e:papb}) and (\ref{e:Gamma0}), and evaluate
the expected covariance matrix numerically. Figure~\ref{f:simple2}
displays the results. As in the previous model, these plots show that
{all parameter errors decrease with signal to noise ratio until the
last cycle} and {all marginalized errors stop improving beyond some
nonzero residual error} at late times. Thus, the general trends shown
in Fig.~\ref{f:simple2} are very much similar to the ones in the
previous simple model (Fig.~\ref{f:simple1}). Again, contrary to the
standard $1/\sqrt{N_{\rm cyc}}$ expectation, the {error during the
last cycle dominates} the total error of the accumulated
signal. Moreover, comparing Figs.~\ref{f:simple1} and \ref{f:simple2}
shows that the presence of additional independent high frequency
degrees of freedom practically does not modify the evolution of
marginalized parameter errors associated with low frequency
components, if $\ti> f_1^{-1}$. During the final cycle, the error
ellipsoid becomes ``thin'' and the narrow dimension will not be
aligned with any of the parameters. As a result, this bad principal
error dominates each of the marginalized parameter errors at late
times. (Note that the start-of-observation time in
Figure~\ref{f:simple2} is $\ti=5\yr$.)

\begin{figure}
\centering{ \mbox{\includegraphics[width=8.5cm]{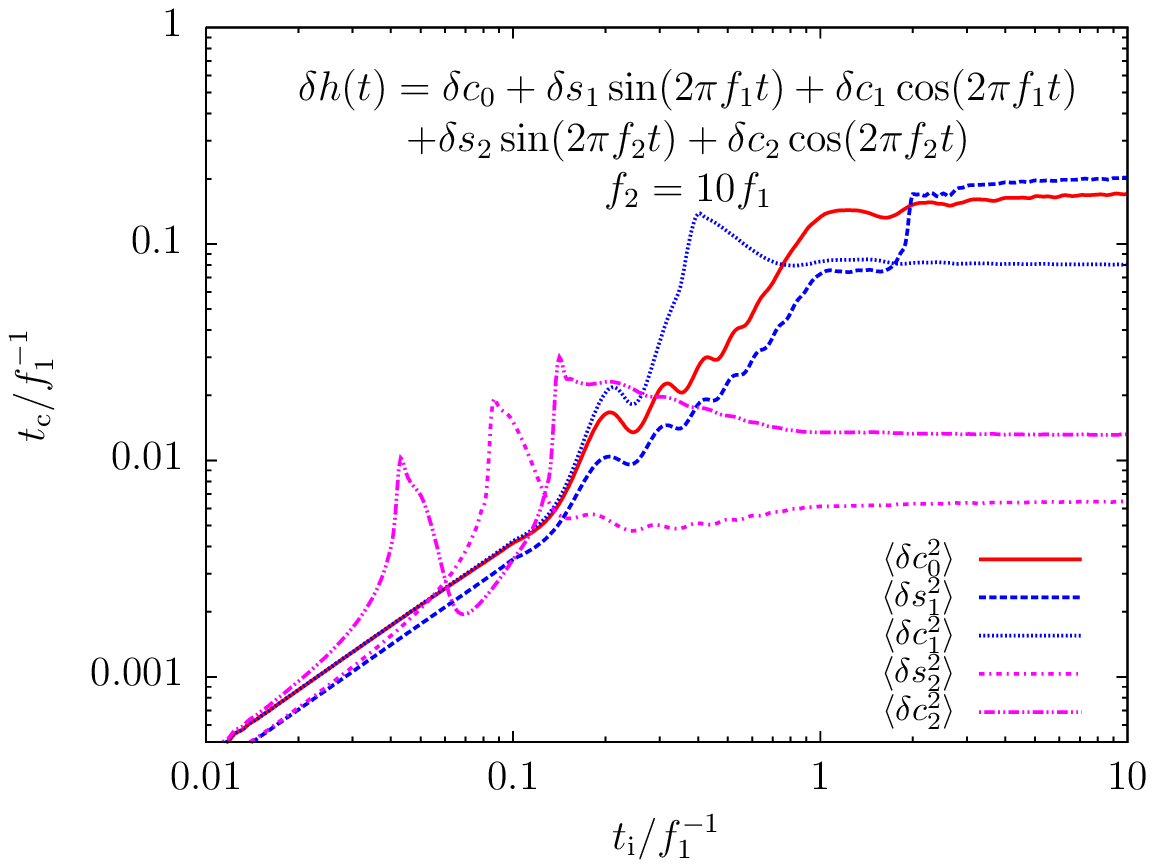}}\\
\bigskip \mbox{\includegraphics[width=8.5cm]{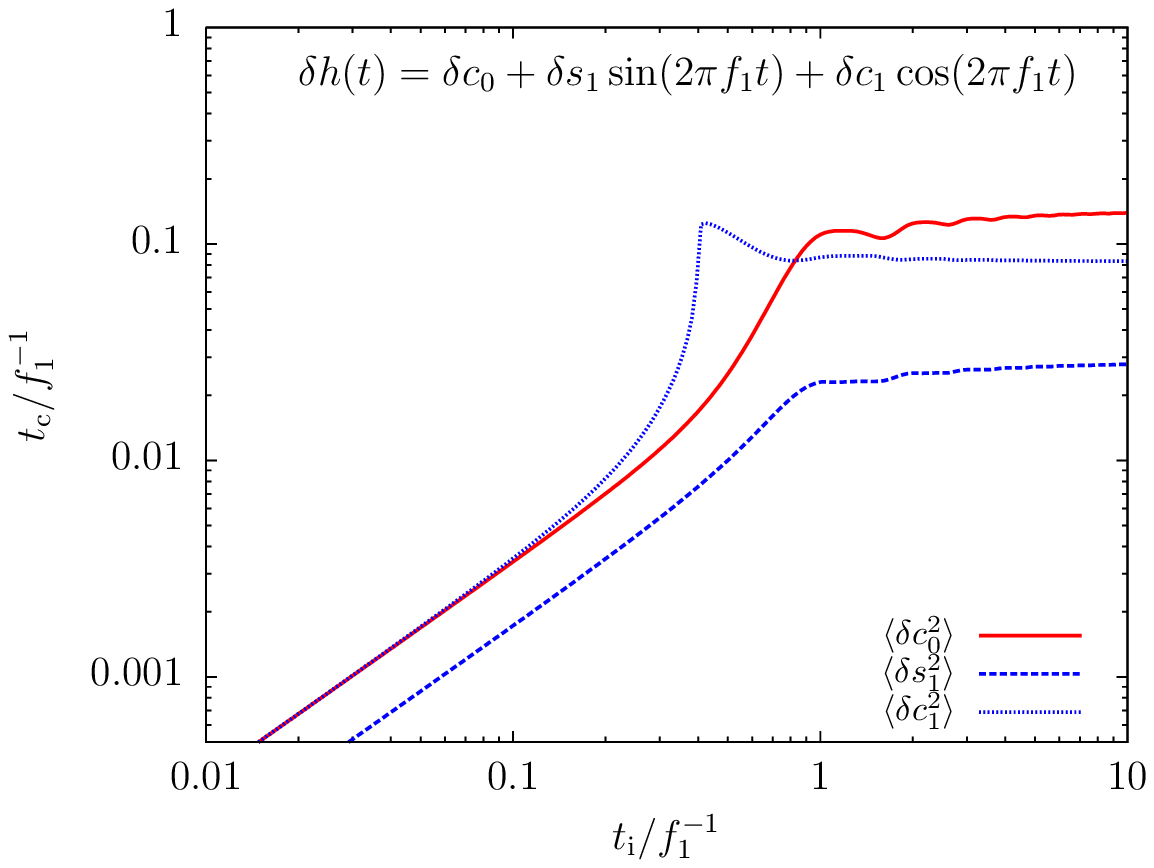}}
\caption{\label{f:simple2b} Critical look--back time, $\tc$ (as in
Fig.~\ref{f:simple1b}), at which marginalized parameter errors stop
improving. {\it Top:} Only $(c_0,s_1,c_1)$ are allowed to vary, using
the prior $(s_{10},c_{10})\equiv(0,0)$. {\it Bottom:} All 5 parameters
$(c_0,s_1,c_1,s_{10},c_{10})$ are determined from the observation. For
$\ti\gsim f_1^{-1}$, estimations of low frequency parameters
$(c_0,s_1,c_1)$ stop improving at $\tc\sim 0.1f_1^{-1}$, while
improvement for high frequency parameters occurs all the way to
$\tc\sim 0.1 f_2^{-1}$.} }
\end{figure}

The critical look--back time, $\tc$, at which this happens is different
for the different frequency components. The top panel in
Fig.~\ref{f:simple2} shows that ${\tc}_{i}\sim 0.1f_{i}$ approximately
for both sets of components $(s_1,c_1)$ and $(s_{10},c_{10})$, where
$f_i$ denotes the corresponding frequencies $f_1=1\yr^{-1}$ and
$f_2=10\yr^{-1}$, respectively. The bottom panel in
Fig.~\ref{f:simple2} shows that the principal errors separate in three
groups. There is one best eigenvector that improves continuously until
the end, two that stop improving near ${\tc}_{1}\sim 0.1f_1$ and two
that stop improving at ${\tc}_{2}\sim 0.1f_2$. The high frequency
parameters $(s_{10},c_{10})$ totally decouple from the two worst
principal components, $(v_0,v_1)$, and, as a result, decouple from the
low frequency parameters $(c_0,s_{1},c_{1})$ which are primarily
determined by $(v_0,v_1)$.

As for our previous model in \S~\ref{app:simple:1}, the critical
look--back time is generally different for different $\ti$ values.  The
bottom panel in Figure~\ref{f:simple2b} shows the time $\tc$ at which
the squared errors first double, as a function of $\ti/f_1^{-1}$, as
in Fig.~\ref{f:simple1b}. Fig.~\ref{f:simple2b} justifies the
rule-of-thumb scaling ${\tc}_{i}\sim 0.1f_{i}$ if the observation time
is at least one cycle period, $f_1^{-1}$.

The central question for the present analysis is how sensitive is the
time evolution of low frequency modulation errors to the presence of
high frequency components. We can examine this question by computing
the critical look--back time, $\tc$, when the high frequency terms are
totally neglected. The top panel in Fig.~\ref{f:simple1b} shows that,
if one limits the parameters to $(c_0,s_1,c_1)$, and the total
observation time is not smaller than the long-period cycle time, $\sim
f_{1}^{-1}$, the resulting $\tc$ value for parameters $c_0$ and $c_1$
is unchanged at the few percent level.  However, if the high frequency
components are introduced, the $s_1$ error evolves differently since
it asymptotes already at much larger $\tc$ values ($\sim 0.1f_{1}^{-1}$
rather than $\sim 0.03f_{-1}$). The reason is that, for small $t$,
with a noise level decreasing quickly, the corresponding function $s_1
\sin(2\pi f_1 t)\approx 2\pi f_1 s_1 t$ is linearly independent of,
and thus uncorrelated with, the functions $c_0$ and $c_1 \cos(2\pi f_1
t)$ which are both constant to first order. Hence, if there are no
more unknowns than $(c_0,s_1,c_1)$, then $c_0$ and $c_1$ are
correlated while $s_1$ is decoupled and can be determined
independently of the other parameters. However, if we add any
parameters which are {\it not} constant for $t\ll f_1^{-1}$, then
$s_1$ becomes correlated with those. This is exactly what happens in
the bottom panel of Fig.~\ref{f:simple1b}, when considering the high
frequency modulations: the estimation on $s_1$ becomes limited for
$t\lsim {\tc}_1 \sim 0.1 f_1^{-1}$ due to the correlations with
$s_{10}$ and $c_{10}$. Quite similarly, if one introduces any other
low-frequency function that is not constant to first order, like
$s_2\sin(4\pi f_1 t)$, then the correlations with this parameter will
limit the improvement of estimation errors for $s_1$ at $\tc\sim 0.1
f_1$, even when neglecting the high frequency components. As we shall
see, this is the case for LISA: there are generally more than one
$\sin$ and $\cos$ low-frequency modes. In this case, the evolution of
estimation errors for low frequency parameters can be obtained with
the high frequency modes (like $s_{10}$ and $c_{10}$) priored
out. This justifies our simple intuition: {once the signal is
decomposed into different time-scale components, the parameter
estimation problem becomes separable and the evolution of parameter
errors corresponding to different such time-scales can be estimated
independently from each other.}

Rather than going through an analytical derivation as in
\S~\ref{app:simple:1}, we answer one remaining question here: what
combination of the original parameters $(c_0,s_1,c_1,s_{10},c_{10})$
corresponds to the best principal component, $v_0$, which can be
determined extremely accurately at late times, $\tf \to 0$? At $t=0$,
the noise drops to zero.  Therefore, the quantity we can measure using
the $t=0$ information is simply $h(t=0)$. Looking back at
eq.~(\ref{e:a:simple2}), this is $c_0 + c_1 + c_2$. It will be
interesting to look for similar ``best determined combinations'' of
physical parameters for the case of the LISA's realistic signals.

\subsection{Four data-stream models}

For our final toy model, we insert additional features of a realistic
LISA data-stream. We consider five low frequency unknowns, $c_0$,
$s_1$, $c_1$,$s_2$, $c_2$, and a high frequency carrier signal with
additional unknowns $s_{10}$, and $c_{10}$. Moreover we consider the
simultaneous measurement of four data-streams.  The signal is
\begin{eqnarray}
\label{e:a:simple4} h(t)&=&c_0 + s_1 \sin(2\pi f_1 t + \varphi^{s_1}_{i}) + c_1 \cos(2\pi f_1 t + \varphi^{c_1}_{i}) \nonumber \\ & &
+ s_2 \sin(2\pi f_1 t + \varphi^{s_2}_{i}) + c_2 \cos(2\pi f_1 t +
\varphi^{c_2}_{i}) \nonumber \\ & & + s_{10} \sin(2\pi f_2 t) + c_{10} \cos(2\pi f_2
t),
\end{eqnarray}
where $\varphi^{c_1,s_1,c_2,s_2}_{i}$ ($i=1\dots 4$) are fixed at a
priori randomly chosen numbers defining the relative phases of the
various modes which are being simultaneously measured.  We compute
independent Fisher matrices for each four set of
$\varphi^{c_1,s_1,c_2,s_2}_{i}$.  We assume that $f_1\ll f_2$ and that
$f_1$ and $f_2$ are fixed and known prior to the measurement. We
choose $f_2=10f_1$ and find the evolution of marginalized errors and
principal errors in two limits:
\begin{enumerate}
 \item[(i)] neglecting cross-correlations with the high frequency parameters by assuming a prior $\delta s_{10}=\delta c_{10}=0$, and
 \item[(ii)] accounting for these high frequency parameters.
\end{enumerate}
We again assume an observation in the look--back time interval $\ti\geq
t\geq \tf$ and take the average instantaneous signal-to-noise ratio to
increase as $\sigma(t)^{-2}=t^{-2}$.

\begin{figure}
\centering{
\mbox{\includegraphics[width=8.5cm]{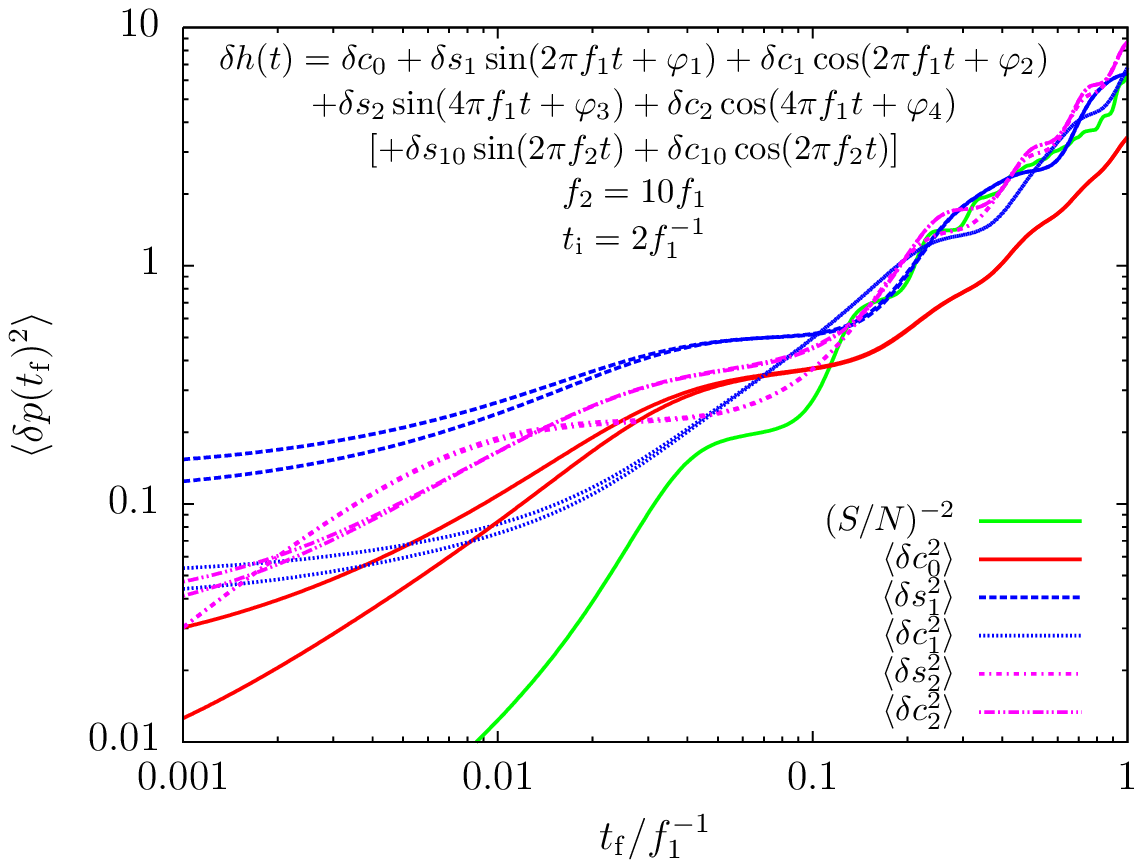}}\\
\bigskip
\mbox{\includegraphics[width=8.5cm]{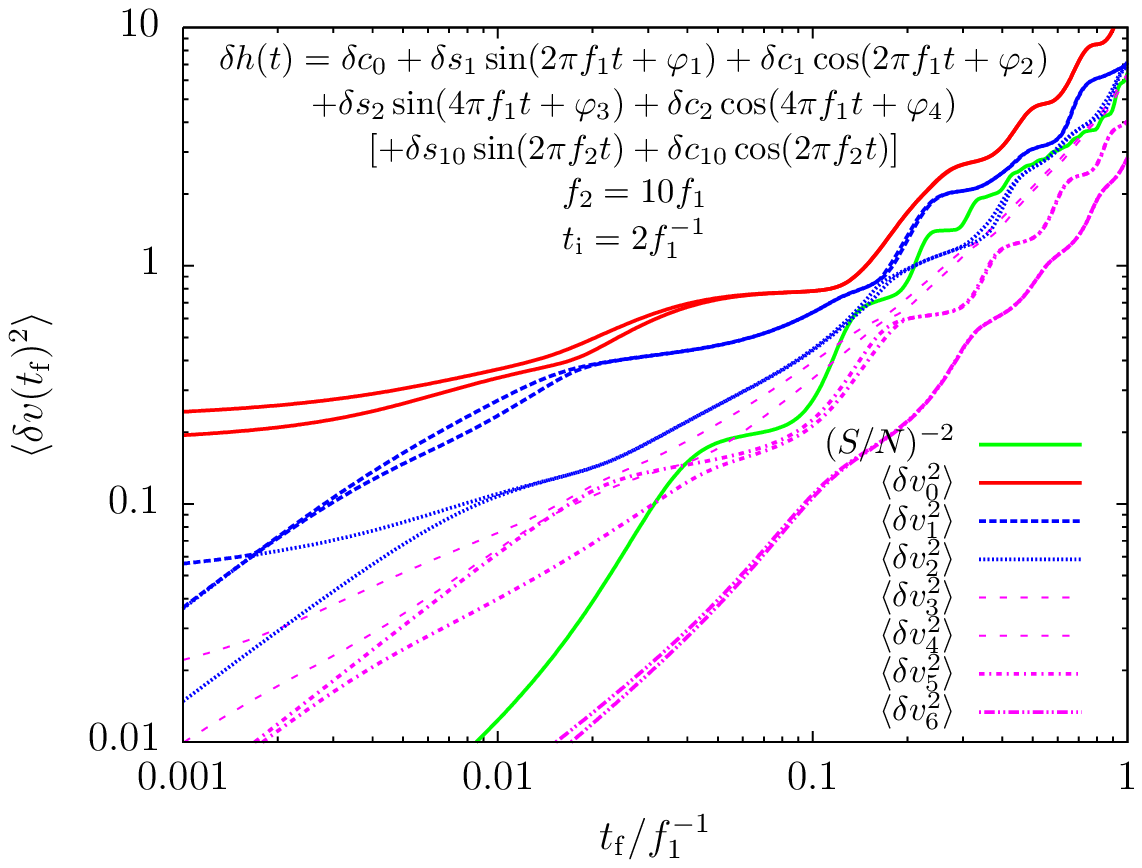}}
 \caption{\label{f:simple3} Marginalized parameter errors {\it (top)}
and principal errors {\it (bottom)} for the four data-stream
model. Pairs of curves with the same line style show results for cases
with five and seven parameters. The extra two parameters correspond to
high frequency ($f_{2}$) components, which affect errors on the other
parameters through correlations only slightly (factor of $\lsim 2$) if
$\tf\lsim 0.1f_{2}^{-1}$.  The green curve shows the scaling with
inverse squared signal-to-noise ratio, $(S/N)^{-2}$, for reference on
both plots. A total observation time $\ti=2\yr$ is assumed.
Marginalized errors follow the signal-to-noise ratio for large $\tf$
values. Four principal errors scale with the signal-to-noise ratio
near merger.}}
\end{figure}

The results for these models are shown in Figure~\ref{f:simple3}. Th
marginalized errors (top) and principal errors (bottom) are shown for
both cases (i) and (ii) above. The figures show that, in agreement
with our previous model, uncertainties on the low frequency parameters
are not affected by the high frequency parameters, except during the
final $0.1$ cycle time of the high frequency component,
$0.1f_2^{-1}$. The figures also show that the four principal
components of the error ellipsoid improve quickly at late times.

Marginalized parameter errors improve quickly if they have negligible
projection on the bad directions of the error ellipsoid. As a result,
our expectation is that errors will typically not stop improving
abruptly, but that there will be a shallower evolution in the final
two weeks. In the worst case for a given parameter, if it is aligned
with the bad ellipsoid principal component, it will stop improving
near merger. In the best case, if the parameter is orthogonal to the
bad ellipsoid principal component, it will improve quickly throughout
the final days of inspiral. Therefore, we understand that the
distribution of errors broadens for $\tf\ll 0.1f_{1}^{-1}$.

\subsection{Best Determined Parameters}\label{app:simple:best}

In the previous section, we have shown that, if the noise decreases
quickly like $t^2$ near merger (at $t=0$), the best-determined
parameters are the eigenvectors of the error covariance matrix that
improve with $(S/N)^{-1}$. Near merger, these are the independent
detector outputs at $t=0$.  In the case of LISA inspirals, the
observation only extends down to ISCO. In this case, the best
determined combination of physical parameters $p_1$ at ISCO are the
real and imaginary parts of $h^{\rm I,II}_{1}(p_1)$. To prove this, we
have to show that these are uncorrelated and decrease with
$(S/N)^{-1}$.  The functions $h^{\rm I}(t)$ and $h^{\rm II}(t)$ are
uncorrelated by construction, since they correspond to the two
independent Michelson detector outputs (see \S~\ref{s:waveform} and
Cutler~\cite{cut98}). The real and imaginary parts of one of the
detectors, $\Re h^{\rm I}_1(t)$ and $\Im h^{\rm I}_1$, are
uncorrelated since they are the coefficients of the high frequency
carrier, $\sin\phi_{GW}$ and $\cos\phi_{GW}$, for which correlation
over one $\phi_{GW}$ cycle (during which the detector noise is
approximately constant) is zero. Another way to see this is to focus
on the real part in the definition of the Fisher matrix
(\ref{e:GammaLISAt2}), which is expressed as the integral of
$\Re[\overline{\partial_a h^{\rm I,II}_{1}(t)}\partial_b h^{\rm
I,II}_{1}(t)]$.  The term in brackets is purely imaginary for the
cross correlation of $\Re h^{\rm I}_1(t)$ and $\Im h^{\rm I}_1$, hence
the real part is always zero. Therefore, the correlation matrix for
$\Re h^{\rm I, II}_1,\Im h^{\rm I, II}_1$ is diagonal. For diagonal
terms, the derivatives are $1$ and the integrals become simply $\int
\sigma^{-2} \D t$, which is exactly $(S/N)^2$.  The RMS estimation
uncertainty of $\Re h^{\rm I,II}_{1}(p_1)$ and $\Im h^{\rm
I,II}_{1}(p_1)$ follows the $(S/N)^{-1}$ all the way down to
ISCO. These best combinations are
$\dL^{-1}(1+\cos^2\theta_{NL})F^{I,II}_{+}(\Omega)$ and
$\dL^{-1}\cos\theta_{NL}F^{I,II}_{\times}(\Omega)$.

The evolution of an arbitrary combination of angles will be determined
by the projection of this combination on the covariance matrix
eigenvectors. A linear combination of good eigenvectors leads to
similarly quick improvement of errors with $(S/N)^{-1}$. However, as
soon as there is a nonzero projection on the fifth eigenvector, the
estimation uncertainty will stop improving at $\sim0.1 T_{\rm cycle}$
which, for the highest $j=4$ harmonic, is between $1$--$2$ weeks.

\section{Angular variables}\label{app:angles} Here we define the relative angles $\theta_{NL}$ and $\phi_{NL}$, using
the polar angles $(\theta_{N},\phi_{N})$ and $(\theta_{L},\phi_{L})$
and the corresponding unit vectors $\bf \hat N$ and $\bf \hat L$.

Let us write a rotation around $\bf \hat z$ and $\bf \hat y$ as
$O_z(\phi)$ and $O_y(\theta)$, respectively. Then, ${\bf \hat
z}=O_{y}(-\theta_N)O_{z}(-\phi_N){\bf \hat N}$ and we define
\begin{equation}
  \left(
    \begin{array}{c}
     \sin(\theta_{NL})\cos(\phi_{NL}) \\
     \sin(\theta_{NL})\cos(\phi_{NL}) \\
     \cos(\theta_{NL})\\
    \end{array}
  \right)\equiv O_{y}(-\theta_N)O_{z}(-\phi_N){\bf \hat L}.
\end{equation}
This uniquely defines $\theta_{NL}$ and $\phi_{NL}$, which
correspond to the relative latitude and longitude, respectively. More
explicitly, we get
\begin{align}
\theta_{NL} &= \arccos ({\bf \hat N}\cdot{\bf \hat L}) =\\
&=\arccos
\left[\sin\theta_N\sin\theta_L\cos(\phi_L-\phi_N)+\cos\theta_N\cos\theta_L\right],\nonumber\\
\phi_{NL} &= \left\{
\begin{array}{cl}
                   2\pi - \phi_0 &
                   \text{~if~}(\phi_L-\phi_N)/\pi\in[-1,0]\bigcup[1,2]\\
                   \phi_0 &\text{~otherwise}
                 \end{array}\right.,
\end{align} where
\begin{align}
\phi_0 &= \arccos
\left(\frac{\cos\theta_N\sin\theta_L\cos(\phi_L-\phi_N)-\sin\theta_N\cos\theta_L}{\sin\theta_{\rm NL}}\right).
\end{align}

\end{document}